\documentclass[fleqn,usenatbib]{mnras}

\usepackage{newtxtext,newtxmath}
\usepackage[T1]{fontenc}

\DeclareRobustCommand{\VAN}[3]{#2}
\let\VANthebibliography\thebibliography
\def\thebibliography{\DeclareRobustCommand{\VAN}[3]{##3}\VANthebibliography}

\usepackage{graphicx}	% Including Figure~files
\usepackage{amsmath}	% Advanced maths commands
\usepackage{makecell}
\usepackage{threeparttable}
\usepackage{cases}

\newcommand{\lya}{\hbox{Ly$\alpha$}}
\newcommand{\ha}{\hbox{H$\alpha$}}
\newcommand{\hb}{\hbox{H$\beta$}}
\newcommand{\hg}{\hbox{H$\gamma$}}
\newcommand{\oiii}{\hbox{O\,{\sc iii}}}
\newcommand{\oii}{\hbox{O\,{\sc ii}}}
\newcommand{\hi}{\hbox{H\,{\sc i}}}
\newcommand{\hii}{\hbox{H\,{\sc ii}}}

\newcommand{\fesc}{$f_{\rm esc}^{ {\rm Ly}\alpha}$}

\newcommand{\beagle}{\textsc{Beagle}}

\newcommand{\cloudy}{\textsc{Cloudy}}

\newcommand{\jwst}{\textit{JWST}}
\newcommand{\hst}{\textit{HST}}

\newcommand{\muv}{$M_{\rm UV}$}

\newcommand{\kms}{\,km~s$^{-1}$}
\newcommand{\um}{\,$\mu$m}

\defcitealias{Tang2023}{T23}

\title[\lya{} escape fraction in the reionization era]{JWST spectroscopy of $z\sim 5-8$ UV-selected galaxies: New constraints on the evolution of the \lya{} escape fraction in the reionization era} 

\author[Z. Chen et al.]{
Zuyi Chen$^{1}$\thanks{E-mail: zychen@arizona.edu},
Daniel P. Stark$^{1}$, 
Charlotte Mason$^{2,3}$, 
Michael W. Topping$^{1}$,
Lily Whitler$^{1}$,
Mengtao Tang$^{1}$,
\newauthor{Ryan Endsley$^{4}$,
\& Stéphane Charlot$^{5}$}
\\
\vspace{0in}\\
$^{1}$Steward Observatory, University of Arizona, 933 N Cherry Ave, Tucson, AZ 85721, USA \\
$^{2}$Cosmic Dawn Center (DAWN) \\
$^{3}$Niels Bohr Institute, University of Copenhagen, Jagtvej 128, 2200 Copenhagen N, Denmark\\
$^{4}$Department of Astronomy, University of Texas, Austin, TX 78712, USA\\
$^{5}$Sorbonne Université, UPMC-CNRS, UMR7095, Institut d’Astrophysique de Paris, F-75014, Paris, France
}

\date{Accepted XXX. Received YYY; in original form ZZZ}

\pubyear{2023}

\begin{document}
\label{firstpage}
\pagerange{\pageref{firstpage}--\pageref{lastpage}}
\maketitle

% Abstract of the paper
\begin{abstract}
We describe \jwst{}/NIRSpec prism measurements of \lya{} emission in $z\gtrsim 5$ galaxies.  
We identify \lya{} detections in 10 out of 69 galaxies with robust rest-optical emission line redshift measurements at $5\leq z<7$ in the CEERS and DDT-2750 observations of the EGS field.  
Galaxies at $z\simeq 6$ with faint continuum (F150W$=$27-29 mag) are found with extremely large rest-frame \lya{} equivalent widths (ranging up to 286~\AA{}).
Likely \lya{} detections are also seen in two new $z>7$ galaxies ($z=$ 7.49 and 7.17) from the second epoch of CEERS observations, both showing large \lya{} equivalent widths that likely indicate significant transmission through the IGM.
We measure high \lya{} escape fractions in the 12 \lya{} emitters in our sample (median 0.28), two of which show \fesc{} near unity ($>0.80$).
We find that $50_{-11}^{+11}$~\% of $z\simeq 6$ galaxies with [\oiii{}]+\hb{} EW$>$1000~\AA{} have \fesc{}~$>0.2$, consistent with the fractions found in lower-redshift samples with matched [\oiii{}]+\hb{} EWs. 
While uncertainties are still significant, we find that only $10_{-5}^{+9}$~\% of $z>7$ galaxies with similarly strong rest optical emission lines show such large \fesc{}, as may be expected if  IGM attenuation of \lya{} increases towards higher redshifts. 
We identify photometric galaxy overdensities near the $z\gtrsim 7$ \lya{} emitters, potentially providing the ionizing flux necessary to create large ionized sightlines that facilitate \lya{} transmission. 
Finally, we investigate the absence of \lya{} emission in a comparable (and spectroscopically confirmed) galaxy overdensity at $z=7.88$ in the Abell 2744 field, discussing new prism spectra of the field obtained with the UNCOVER program. 
\end{abstract}

% Select between one and six entries from the list of approved keywords.
% Don't make up new ones.
\begin{keywords}
dark ages, reionization, first stars - cosmology: observations - galaxies: evolution - galaxies: high-redshift
\end{keywords}

%%%%%%%%%%%%%%%%%%%%%%%%%%%%%%%%%%%%%%%%%%%%%%%%%%

%%%%%%%%%%%%%%%%% BODY OF PAPER %%%%%%%%%%%%%%%%%%

\section{Introduction}

The reionization of intergalactic hydrogen marks the important stage where early structure formation has impacted the vast majority of baryons in the universe.
Through a variety of observational efforts over the past two decades, substantial progress has been made to constrain the timeline of this reionization process and the nature of the early ionizing sources (see \citealt{Stark2016,fan2023,Robertson2022} for reviews).
Measurements of hydrogen absorption features presented in the quasar spectra indicate that the universe is partially neutral at $z\simeq7$ \citep{Greig2017,Davies2018,Wang2020,Yang2020} and becomes mostly ionized by $z\simeq$ 5.5--6 \citep[e.g.,][]{Fan2006,McGreer2015,Yang2020b,Qin2021,Zhu2022,Jin2023}.
Measurements of the Cosmic Microwave Background have suggested a similar picture, where constraints on the electron scattering optical depth imply a reionization mid-point at $z = 7.8\pm0.7$ \citep{PlanckCollaboration2016,PlanckCollaboration2020}.
With the rapidly declining quasar luminosity function at $z>3$ \citep[e.g.,][]{Matsuoka2018,Kulkarni2019,Jiang2022},  young star-forming galaxies are thought to be the likely dominant contributors to the ionizing photons necessary for this process \citep{Robertson2015,Stanway2016,Dayal2018,Finkelstein2019,Naidu2020}.

The \lya{} emission from early star-forming galaxies has been shown to be another useful probe in constraining the timeline of reionization \citep[e.g.,][]{Dijkstra2014,Ouchi2020}.
While strong \lya{} emission has been commonly observed in $z\simeq6$ galaxies \citep[e.g.,][]{Stark2011,Curtis-Lake2012,Cassata2015,DeBarros2017}, deep spectroscopic surveys show that it becomes increasingly rare at higher redshifts ($z\simeq$ 7--8; e.g., \citealt[]{Fontana2010,Stark2010,Treu2013,Caruana2014,Pentericci2014,Tilvi2014,Hoag2019,Mason2019,Jung2020}).
This decline of \lya{} visibility is consistent with a significant neutral hydrogen fraction of the intergalactic medium (IGM) at $z\gtrsim7$  (${\rm X}_{\rm HI} \gtrsim 0.6$; e.g., \citealt{Mesinger2015,Zheng2017,Mason2018,Mason2019,Hoag2019,Jung2020,Whitler2020,Bolan2022}).
 
\jwst{} has recently ushered in a new era of high-redshift \lya{} studies (e.g., \citealt{Jones2023,Jung2023,Nakane2023,Napolitano2024,Saxena2023,Saxena2023b,Tang2023,Witten2023}), with detections now being made out to $z\simeq 11$ \citep{Bunker2023}. 
\lya{} investigations with \jwst{} provide several advantages compared to earlier efforts.
The absence of atmospheric OH lines greatly increases \lya{} completeness and improves the reliability of upper limits on fluxes when lines are not detected. 
Access to strong rest-optical emission lines provides spectroscopic redshifts whether or not \lya{} is detected, allowing the environment around \lya{} emitters to be more efficiently characterized. 
Detection of \hb{} (or \ha{}) allows the intrinsic luminosity of \lya{} to be calculated, which when combined with the observed \lya{} flux yields estimates of the \lya{} escape fraction (e.g., \citealt{Ning2023,Roy2023,Saxena2023,Tang2023,Tang2024,Lin2024}).
The rest-optical emission lines also constrain the systemic redshifts for measurement of \lya{} velocity profiles, providing a key input to reionization calculations \citep[e.g.,][]{Bunker2023,Prieto-Lyon2023,Saxena2023,Tang2023}.

The Cosmic Evolution Early Release Science (CEERS\footnote{\url{https://ceers.github.io/}}; ERS-1345, PI: S. Finkelstein; Finkelstein et al. in prep, also see \citealt{Finkelstein2022,Finkelstein2023}) has recently obtained \jwst{} imaging and spectroscopy of the EGS field, providing one of first opportunities for a statistical investigation of \lya{} in the reionization era. 
This region is known to have a large number of $z\gtrsim 7$ \lya{} emitting galaxies, with previously confirmed structures at $z=7.48$, $z=7.73$, and $z=8.68$ \citep{Oesch2015,Zitrin2015,Roberts-Borsani2016,Stark2017,Tilvi2020,Jung2022,Larson2022,Cooper2023}. 
The CEERS spectra have revealed additional galaxies at these redshifts \citep{ArrabalHaro2023b,Fujimoto2023,Harikane2023,Sanders2023_directZ,Tang2023}, potentially suggesting the \lya{} emitters trace large-scale overdensities spanning several physical~Mpc.
Such overdensities of galaxies could generate significant amounts of ionizing photons, creating large ionized bubbles that facilitate the transmission of \lya{} photons from the member galaxies \citep[e.g.,][]{Barkana2004,Furlanetto2004,Wyithe2005,Iliev2006,Dayal2018,Weinberger2018}.

The CEERS spectra have also revealed insight into the nature of the $z\gtrsim 7$  \lya{} emitters in the field. 
\citet{Tang2023} demonstrated that the systems with \lya{} are atypical in their properties, with hard spectra and more efficient ionizing photon production than is common at this redshift. 
This may reflect very young stellar populations formed during a rapid burst \citep{Endsley2021_mmt,Saxena2023,Tang2023} or the presence of an AGN \citep{Larson2023}. 
Regardless of the origin, the ionizing properties will result in boosted intrinsic \lya{} luminosities, enhancing the likelihood of detecting the line in the face of IGM attenuation even within small ionized bubbles. 
The \lya{} detections at $z\gtrsim 7$ also appear to be redshifted significantly from systemic, with velocity offsets commonly in excess of 500 km~s$^{-1}$ \citep{Bunker2023,Tang2023}. 
Given that the IGM attenuation of \lya{} is strongest close to the line center, the large velocity offsets will further contribute to the visibility of \lya{}.

In spite of this progress, the size of the ionized bubbles around the $z\gtrsim 7$ \lya{} emitting galaxies in the EGS field remains poorly constrained, making it unclear the relative role of overdensities and intrinsic galaxy properties play in explaining their strong observed \lya{} \citep[e.g.,][]{Jung2022,Whitler2023c,Tang2023}.
If overdensities have carved out large ionized regions (i.e., $\gtrsim$1 physical~Mpc), we expect the transmission of \lya{} through the IGM to be enhanced \citep[e.g.,][]{Mason2018_transmission,Mason2020_bubble}. 
If the overdensities and bubble sizes are smaller, we would expect to see significant attenuation in the \lya{} line from the damping wing of the neutral IGM.
The CEERS \hb{} detections allow \lya{} escape fractions (\fesc{}) to be derived for the \lya{} emitters in the field. 
Typical escape fractions are low (\fesc{} = 0.03--0.09), suggesting the majority of \lya{} photons are not making their way through the NIRSpec slitlets \citep{Tang2023}. 
Galaxies with similar rest-optical spectral properties at lower redshifts ($z\simeq 0.3-3$) tend to have much stronger \lya{} \citep[e.g.,][]{Yang2017,Tang2021,Flury2022_I}, with \lya{} EWs that are 6--12$\times$ greater than the $z\gtrsim 7$ \lya{} emitters in CEERS. 
This may be indicating that the $z\gtrsim 7$ \lya{} emitters trace relatively small ionized bubbles, with excess attenuation from the neutral IGM on the outside of the bubbles. 
Alternatively, the interstellar medium (ISM) and circumstellar medium (CGM) of the $z\gtrsim 7$ galaxies may scatter the \lya{} more than the comparison samples at lower redshift.

In this paper, we seek to extend the \lya{} investigations presented in \citet{Tang2023} to a larger sample of galaxies. We have two primary goals. 
First, we seek to characterize the \lya{} properties of galaxies at slightly lower redshifts ($5\leq z < 7$) where the impact of the IGM on the line is reduced.
At these redshifts, we expect to see \lya{} with larger escape fractions than those studied to date at $z\gtrsim 7$. 
This comparison sample will be important in interpreting the emerging body of \lya{} detections deep in the reionization era. 
Second, we aim to use new measurements of \lya{} in faint $z\gtrsim 7$ galaxies to better characterize whether the previously-known \lya{} emitters in the EGS field are likely to trace large ionized regions. 
We will investigate whether there are significant overdensities of photometrically-selected galaxies in the EGS field, as may be expected if large bubbles are present. 

The connection between ionized bubbles and galaxy overdensities has been further tested in recent \jwst{} observations of the $z=7.88$ protocluster discovered in {GLASS (ERS-1324, PI: Treu; \citealt{Treu2022}), JWST Director's Discretionary Time (DDT-2756, PI: W. Chen; \citealt{Roberts-Borsani2023}), and GO-1840 (PI: Hashimoto; \citealt{Hashimoto2023}) observations of the Abell 2744 field \citep{Morishita2023}. 
The galaxies in this protocluster are significantly overdense (20$\times$) over a relatively small radius of 60 physical kpc. 
While such a region may be expected to have carved out a large ionized bubble, \jwst{} spectroscopy presented in \citet{Morishita2023} has revealed no \lya{} emission. 
In this paper, we use new \jwst{} spectra of galaxies in the $z=7.88$ protocluster  to understand why \lya{} does not appear to be enhanced in this overdense volume.

The organization of the paper is as follows. 
In \S \ref{sec:data_sample}, we first describe JWST/NIRSpec observations (\S~\ref{sec:spec}), our \lya{} emitting galaxy sample (\S~\ref{sec:lya_det}), and the photoionization modeling of their NIRCam spectral energy distributions (\S~\ref{sec:sed}).
In \S \ref{sec:fesc-z}, we then discuss the \lya{} properties inferred for these \lya{} emitting galaxies, comparing them to literature samples spanning a wide range of redshifts.
Based on the photometric samples identified from the NIRCam imaging, we characterize the large scale environments of the newly identified \lya{} emitting galaxies at $z>7$ in \S \ref{sec:bubble}.
In \S \ref{sec:discussion}, we discuss the dependence on galaxy local environment for \lya{} visibility, considering the detection of a large number of $z\sim$ 7--8 \lya{} emitting galaxies in the EGS field and the new \jwst{} observations of \lya{} emission in the $z=7.88$ protocluster presented in   \citet{Morishita2023}.
Finally, we summarize our findings in \S \ref{sec:summary}.
Throughout this paper, we adopt a flat $\Lambda$CDM cosmology with $H_0$ = 70 km~s$^{-1}$~Mpc$^{-1}$, $\Omega_{\rm m} = 0.3$, and $\Omega_\Lambda = 0.7$.
All magnitudes are measured in the AB system \citep{Oke1983}, and the emission line equivalent widths are calculated in the rest frame.

\section{Data and Sample Properties}\label{sec:data_sample}

The goal of this paper is to characterize the \lya{} emission strengths in galaxies observed with \jwst{}/NIRSpec \citep{Jakobsen2022,Boker2023}. We primarily focus on the CEERS program in the EGS field (Finkelstein et al. in prep, also see \citealt{Finkelstein2022,Finkelstein2023}).
We first describe the NIRSpec spectra and the spectroscopic sample in \S~\ref{sec:spec}.
We identify the $z\geq 5$ \lya{} emitting galaxies and constrain their \lya{} properties in \S~\ref{sec:lya_det}.
Based on broad-band SEDs of the \lya{} emitting galaxies, we then derive their galaxy physical properties through photoionization modeling in \S~\ref{sec:sed}.

\subsection{Spectroscopy of $z\gtrsim 5$ galaxies with \jwst{}/NIRSpec}\label{sec:spec}

\begin{figure*}
    \centering
    \includegraphics[width=\textwidth]{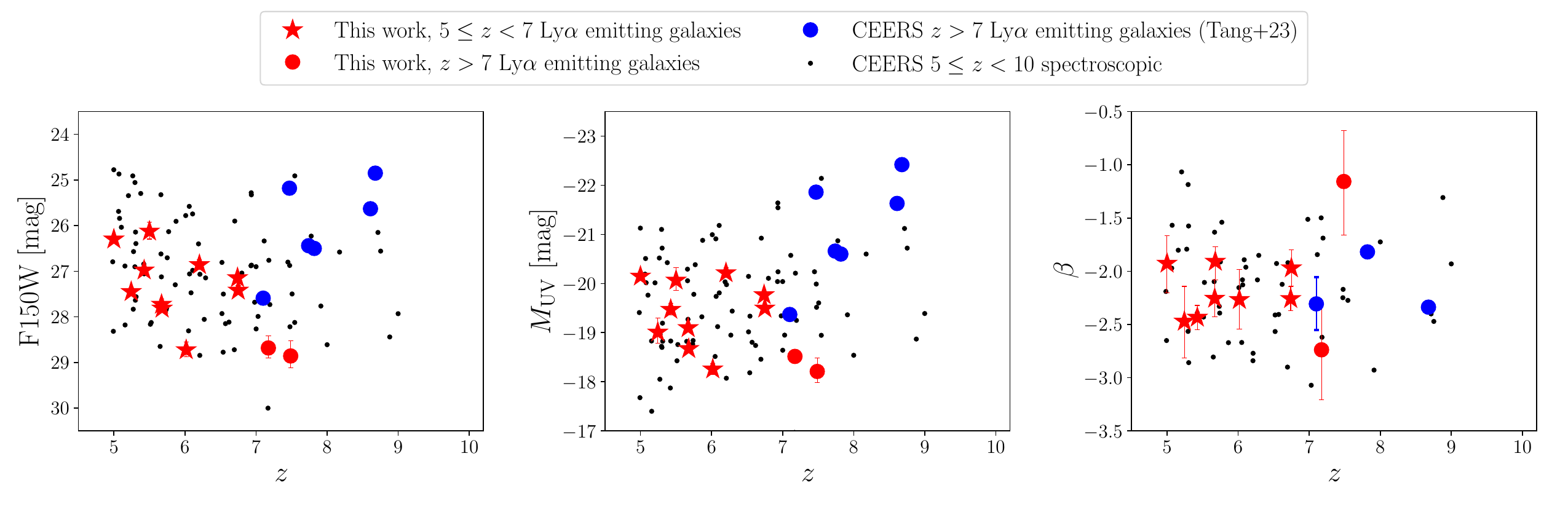}
    \caption{
    The distribution of basic photometric properties as a function of spectroscopic redshifts for the CEERS \lya{} emitting galaxies analyzed in this work.
    Shown are their F150W magnitudes (\hst{}/WFC3 F160W magnitudes when not covered by NIRCam) in the {\em left}, the absolute UV magnitudes (\muv{}) in the {\em middle}, and the UV continuum slope ($\beta$) in the {\em right} panel.
    We plot the $5\leq z < 7$ \lya{} emitting galaxies identified in this work as red stars and the new $z>7$ \lya{} emitting galaxies as red dots.
    For comparison, the CEERS spectroscopic sample spanning $5\leq z < 10$ identified in this work and from \protect\cite{Tang2023} are plotted as black dots.
    It is thus clear that the \lya{} emitting galaxies in this work occupy the fainter end of the full CEERS sample, reaching F150W $\approx$ 29 and \muv{} $\approx$ $-18$.
    They also show blue UV continuum slopes (median $\beta$ $= -2.2$) indicating little dust obscuration. }
    \label{fig:sample}
\end{figure*}

Our work builds on the previous efforts in characterizing the \lya{} emission at $z\gtrsim7$ from the CEERS NIRSpec spectra  \citep[e.g.,][]{Jung2023,Tang2023,Witten2023}. 
To date, six \lya{} emitting galaxies have been investigated at $z\gtrsim 7$ using the spectra taken in the CEERS epoch 2 (December 2022) observations. 
Two of these are newly identified \lya{} detections, whereas the other four were previously reported in the literature based on ground-based spectroscopy. Here we extend the \lya{} emitting galaxy search to slightly lower redshifts ($5\leq z<7$), where the IGM is expected to be significantly more ionized.
This requires us to focus on the prism observations, where the wavelength coverage extends blue enough to recover \lya{} at $z\geq 5$. 
To obtain a complete sample of $z\geq 5$ \lya{} emitters in CEERS, we augment this with an updated search for $z>7$ \lya{} emitting galaxies from the new NIRSpec prism spectra taken since the first \lya{} analysis in CEERS by \cite{Tang2023}. 
This includes the CEERS epoch 3 observations (taken in February 2023; see  Arrabal Haro et al. in prep) and the Director’s Discretionary Time (DDT) observations (DDT-2750, PI: P. Arrabal Haro; \citealt{ArrabalHaro2023a,ArrabalHaro2023b}), which focus on NIRCam-selected targets and allows us to detect fainter $z\geq$ 5 galaxies.
In \S \ref{sec:discussion}, we characterize the \lya{} properties within a spectroscopically-confirmed overdensity at $z\sim 7.88$. Here we utilize the NIRSpec prism observations for the overdensity recently taken as part of the Ultradeep NIRSpec and NIRCam ObserVations before the Epoch of Reionization (UNCOVER\footnote{\url{https://jwst-uncover.github.io/}}; GO-2561, PIs: I. Labb\'e \& R. Bezanson; \citealt{Bezanson2022,Goulding2023,Wang2023}; S. Price et al., in prep.) program.
Below we will briefly describe the spectroscopic observations and identification of $z\gtrsim 5$ galaxies in the CEERS field, but 
we note that the UNCOVER spectra are reduced and analyzed in the same manner as the CEERS data.

The CEERS NIRSpec multi-object spectroscopy (MOS) observations employ the Micro Shutter Array (MSA; \citealt{Ferruit2022}), with a detailed description of the configuration given in the CEERS phase-2 public PDF\footnote{\url{https://www.stsci.edu/jwst/phase2-public/1345.pdf}} and Arrabal Haro et al. in prep.
Briefly, the epoch 2 (December 2022) observations consist of six pointings with medium-resolution grating using the G140M/F100LP, G235M/F170LP, and G395M/F290LP grating/filter pairs and eight pointings with the low-resolution prism.
The prism observations that are the focus of this paper deliver simultaneous spectral coverage spanning  0.6--5.4 \um{}, with a spectral resolution of $R\sim 100$ ($\sigma\simeq$ 1300 \kms{}).
The observed targets on the prism pointings include 466 galaxies,  spanning a wide (photometric or spectroscopic) redshift range from $z$ = 0.1--12 \citep{ArrabalHaro2023a,ArrabalHaro2023b}.
Each of the targets was observed using the 3-shutter MSA slitlet (1.5\arcsec{} $\times$ 0.2\arcsec{}), with a total exposure time of 3107 s split into three exposures of 14 groups per disperser and per pointing.
We also take advantage of the two new prism pointings taken in CEERS epoch 3 (February 2023) observations and the additional prism pointing taken in the DDT observations (DD-2750; \citealt{ArrabalHaro2023b}).
The two CEERS epoch 3 pointings target in total 283 NIRCam-selected galaxies also across a wide range of redshift, with a total integration time of 6127s and 2042s, respectively.
The DDT observation employs a single 5.1 hr pointing, mainly targeting $z \sim$ 12--16 galaxy candidates selected with NIRCam but also with galaxies at lower redshifts in the MSA (145 galaxies in total).
All the NIRSpec spectra are reduced in the same manner as presented in \cite{Tang2023} using the \jwst{} data reduction pipeline\footnote{\url{https://jwst-pipeline.readthedocs.io/en/latest/}}.
In this step, we have applied slit loss corrections by assuming a point source given that the majority of sources analyzed in this work are not significantly extended.
More information on the data reduction is provided in \cite{Tang2023}.

Each extracted 2D spectrum was visually inspected by three co-authors (M. Tang, M. Topping, and ZC) in search of robust emission line detections indicating a redshift of $z\geq$ 5. Our redshift identification process is similar to that described in \citet{Tang2023}, requiring multiple emission line detections (often the [\oiii{}] doublet and \hb{} or H$\alpha$). 
After investigating the full database, we identify 69 emission line galaxies at $5\leq z < 7$ \footnote{Here, we have removed potential AGNs as indicated by very broad \ha{} (FWHM $\geq$ 1000 km$^{-1}$) and narrow [\oiii{}] emission lines observed in the medium resolution grating spectra \citep[e.g.,][]{Kocevski2023,Harikane2023_AGN}}.  
We further identify 10 new galaxies at $7\leq z \leq 9$ from the more recent CEERS observations conducted since \citet{Tang2023}.  
In the following sections, we will add these systems to the 21 $z\gtrsim 7$ galaxies in  \citet{Tang2023}, resulting in a total sample of 100 galaxies at $5\leq z \leq 9$. 
A significant subset of these sources overlap with those presented in the literature, and we have verified that our redshifts are in excellent agreement with those published elsewhere \citep[e.g.,][]{ArrabalHaro2023a,Fujimoto2023,Harikane2023,Nakajima2023,Sanders2023}.
The full prism spectroscopic catalog will be presented in a future work (Chen et al., in prep).

The 1D spectra are extracted with a boxcar aperture following the procedures in \cite{Tang2023}, which is set to match the emission line profile along the spatial direction for each target (typical aperture is 6 pixels).
We determine the systemic redshifts by fitting Gaussian profiles (plus a constant factor for the underlying continuum) to the brightest rest-optical lines (i.e., [\oiii{}]+\hb{} or \ha{}).
The derived redshifts for the 69 prism-selected galaxies at $5\leq z < 7$  range from $z=4.98$ to 6.98, with a median of $z=5.73$.
For the 10 $z>7$ galaxies identified with the new datasets taken in February--March 2023, their redshifts span from 7.00 to 8.75, with a median of 7.48.

We measure optical emission line fluxes through Gaussian profile fitting, following the method of \cite{Sanders2023}.
We allow the centroid of each line to vary from the determined redshift within half of the spectral velocity resolution ($\sim$650 \kms{} for the prism spectra).  To account for the wavelength dependence of the spectral resolution, we also restrict the line width to be within 50\% to 150\% of the line widths determined from the strongest optical lines (i.e., [\oiii{}]+\hb{} or \ha{}).  The uncertainties are derived by perturbing the fluxes according to the error spectra and repeating the measurements 500 times. Dust corrections will be conducted via the Balmer decrement where available (and the SED where the Balmer decrement is not 
measured). We will discuss the corrections in \S\ref{sec:sed} once SEDs have been characterized for our sample.

We measure  \hst{}/ACS+\jwst{}/NIRCam photometry for our spectroscopic sample of $z>5$ galaxies, following the photometric procedures detailed in \cite{Endsley2023_ceers} (also see \citealt{Tang2023,Whitler2023c}).
We utilize the NIRCam imaging in the EGS field observed as part of the CEERS program in two scheduling windows in June and December 2022 (see \citealt{Bagley2023}), and the \hst{}/ACS (WFC3/IR as well for those without NIRCam coverage) imaging assembled and reduced with {\sc Grizli} \citep{Brammer2022} as part of the Complete Hubble Archive for Galaxy Evolution (CHArGE) project (\citealt{Kokorev2022}; Kokorev et al. in prep.). 
Among the 79 new sources considered in this paper, 59 are covered by  CEERS NIRCam observations. We measure their Kron photometry \citep{Kron1980} in ACS F435W, F606W, and F814W filters, and in NIRCam F115W, F150W, F200W, F275W, F356W, F410M, and F444W filters.
For the remaining 20 sources, the flux density in the \hst{} WFC3/IR filters (F125W, F140W, and F160W) are measured instead. 
Figure~\ref{fig:sample} shows the NIRCam/F150W (or WFC3/F160W when not covered with NIRCam) magnitude distribution for the 79  galaxies we present in this paper, ranging from 24.8 to 28.9 with a median of 27.1 mag.

\subsection{\protect  \lya{} Detections at $z\gtrsim 5$ in CEERS spectra}\label{sec:lya_det}

\begin{figure*}
    \centering
    \includegraphics[width=1.0\textwidth]{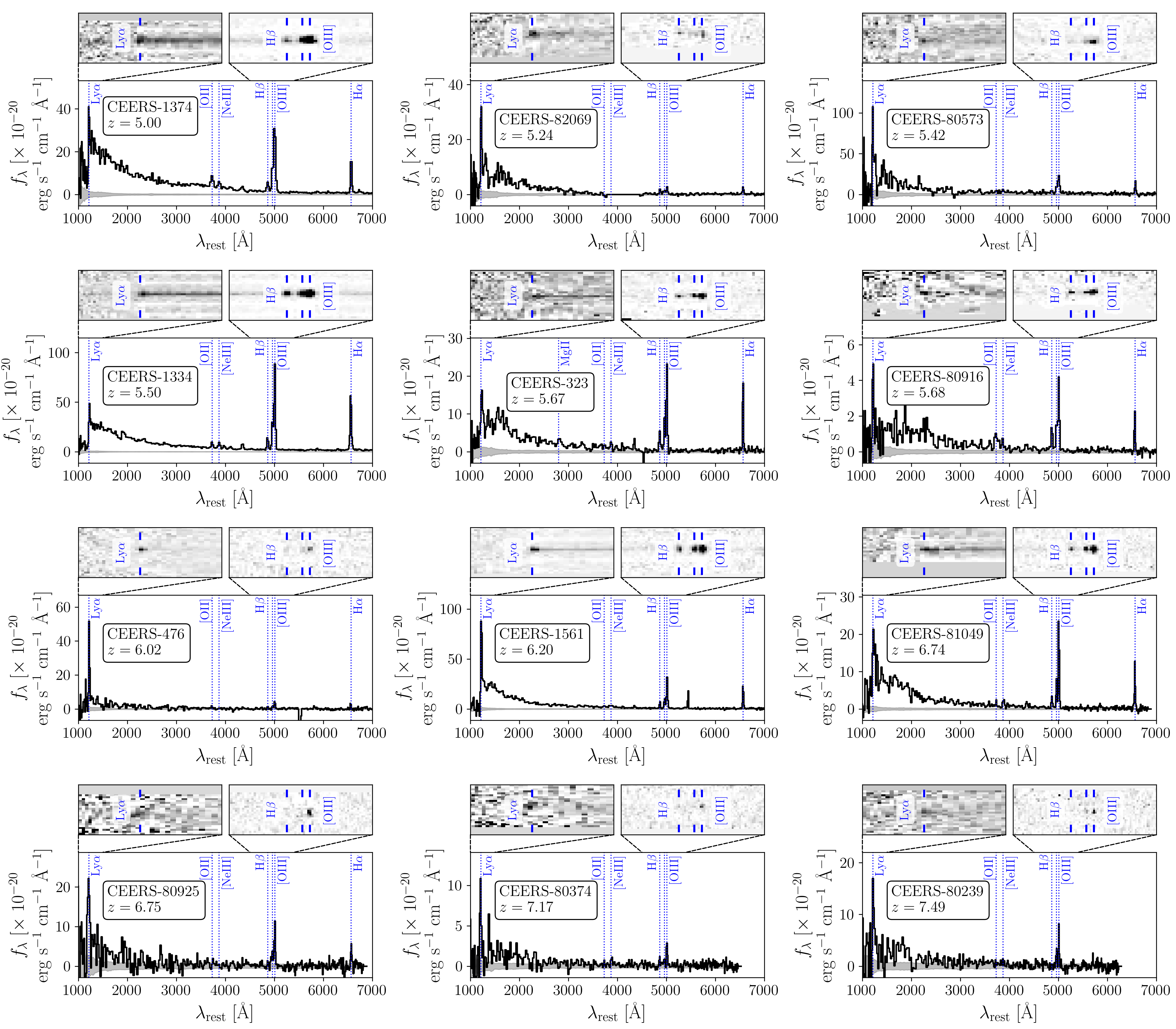}
    \caption{The 1D prism spectra for the 12 \lya{} emitting galaxies identified in our CEERS $z\gtrsim$ 5 spectroscopic sample.
    For each galaxy, we show the 1D spectrum and mark the key emission line features with blue lines and labels.
    The corresponding 2D spectrum for the \lya{} and optical [\oiii{}]+\hb{} emission lines is also shown on top of each 1D spectrum.
    The UV continuum for each \lya{} emitting galaxy is also detected at wide wavelength bins (i.e., 150~\AA{} per bin).
    }
    \label{fig:spec}
\end{figure*}

\begin{table*}
    \centering
    \begin{tabular}{cccccccc}
    \hline
    ID & $z_{\rm spec}$ & RA & Dec & NRC/F150W & WFC3/F160W & $M_{\rm UV}$ & O32 \\
    & & (deg) & (deg) & (mag) & (mag) & (mag) & \\ 
    \hline
CEERS-1374  & 5.00  &  214.9439108 &   52.8500417 & $26.3_{-0.1}^{+0.1}$  &                   --  & $-20.1_{-0.1}^{+0.1}$  & $ 7.3_{- 0.2}^{+ 0.3}$ \\[1pt]
CEERS-82069 & 5.24  &  214.7303221 &   52.7549722 & $27.4_{-0.1}^{+0.1}$  &                   --  & $-19.0_{-0.3}^{+0.2}$  & $>                2.6$ \\[1pt]
CEERS-80573 & 5.43  &  214.7739242 &   52.7806000 & $27.0_{-0.1}^{+0.1}$  &                   --  & $-19.5_{-0.1}^{+0.1}$  & $>                9.9$ \\[1pt]
CEERS-1334  & 5.50  &  214.7683562 &   52.7176417 &                  --  &  $26.1_{-0.2}^{+0.2}$  & $-20.1_{-0.3}^{+0.2}$  & $11.5_{- 0.3}^{+ 0.3}$ \\[1pt]
CEERS-323   & 5.67  &  214.8725558 &   52.8759500 & $27.7_{-0.1}^{+0.1}$  &                   --  & $-19.1_{-0.1}^{+0.1}$  & $>               58.2$ \\[1pt]
CEERS-80916 & 5.68  &  214.8916300 &   52.8159417 & $27.8_{-0.1}^{+0.1}$  &                   --  & $-18.7_{-0.1}^{+0.1}$  &                     -- \\[1pt]
CEERS-476   & 6.02  &  214.8055608 &   52.8363444 & $28.7_{-0.2}^{+0.2}$  &                   --  & $-18.3_{-0.1}^{+0.1}$  &   $4.8_{-1.0}^{+1.8}$  \\[1pt]
CEERS-1561  & 6.20  &  215.1660971 &   53.0707556 &                  --  &  $26.9_{-0.1}^{+0.1}$  & $-20.2_{-0.1}^{+0.1}$  & $>               38.0$ \\[1pt]
CEERS-81049 & 6.74  &  214.7898221 &   52.7307889 & $27.1_{-0.0}^{+0.0}$  &                   --  & $-19.8_{-0.0}^{+0.0}$  & $18.4_{- 2.4}^{+ 2.9}$ \\[1pt]
CEERS-80925 & 6.75  &  214.9486800 &   52.8532722 & $27.4_{-0.1}^{+0.1}$  &                   --  & $-19.5_{-0.1}^{+0.1}$  & $>                6.4$ \\[1pt]
CEERS-80374 & 7.17  &  214.8980742 &   52.8248944 & $28.7_{-0.3}^{+0.2}$  &                   --  & $-18.5_{-0.1}^{+0.1}$  & $>                5.2$ \\[1pt]
CEERS-80239 & 7.49  &  214.8960542 &   52.8698528 & $28.9_{-0.3}^{+0.3}$  &                   --  & $-18.2_{-0.3}^{+0.2}$  & $>                6.3$ \\[1pt]
    \hline
\end{tabular}
    \caption{The 12 $z>5$ CEERS \lya{} emitting galaxies newly identified in this work.
    We report the spectroscopic redshifts, coordinates, apparent F150W magnitudes (WFC3 F160W when not covered by NIRCam), the absolute UV magnitudes (\muv{}), as well as the O32 ratios.}
    \label{tb:lae}
\end{table*}

\begin{table*}
    \centering
    \begin{threeparttable}[t]
\begin{tabular}{ccccccc}
    \hline
    ID & $z_{\rm spec}$ & Flux \lya{} & EW \lya{} & \fesc{} & \fesc{} & \fesc{} \\
    & & ($\times 10^{-18}$ erg s$^{-1}$ cm$^{-2}$)  & (\AA{}) & (Case B) & (Case B, corrected\tnote{a}) & (Case A) \\ 
    \hline

CEERS-1374  & 5.00  & $ 5.4_{- 0.6}^{+ 0.5}$ & $ 23_{-  2}^{+  2}$ & $0.075_{-0.008}^{+0.007}$  & $0.123_{-0.013}^{+0.011}$ & $0.054_{-0.006}^{+0.005}$ \\[1pt]
CEERS-82069 & 5.24  & $ 6.7_{- 0.5}^{+ 0.6}$ & $ 70_{-  5}^{+  5}$ & $0.787_{-0.048}^{+0.053}$  & $1.004_{-0.061}^{+0.068}$ & $0.562_{-0.034}^{+0.038}$ \\[1pt]
CEERS-80573 & 5.43  & $37.0_{- 3.1}^{+ 3.0}$ & $286_{- 40}^{+ 46}$ & $0.832_{-0.046}^{+0.048}$  & $0.912_{-0.051}^{+0.052}$ & $0.594_{-0.033}^{+0.034}$ \\[1pt]
CEERS-1334  & 5.50  & $ 5.1_{- 0.4}^{+ 0.4}$ & $ 19_{-  1}^{+  1}$ & $0.047_{-0.003}^{+0.004}$  & $0.084_{-0.006}^{+0.006}$ & $0.034_{-0.002}^{+0.003}$ \\[1pt]
CEERS-323   & 5.67  & $ 4.8_{- 0.7}^{+ 0.7}$ & $ 85_{-  9}^{+  9}$ & $0.091_{-0.011}^{+0.011}$  & $0.110_{-0.014}^{+0.014}$ & $0.065_{-0.008}^{+0.008}$ \\[1pt]
CEERS-80916 & 5.68  & $ 1.6_{- 0.2}^{+ 0.2}$ & $ 72_{- 12}^{+ 18}$ & $0.304_{-0.035}^{+0.031}$  & $0.342_{-0.039}^{+0.035}$ & $0.217_{-0.025}^{+0.022}$ \\[1pt]
CEERS-476   & 6.02  & $11.4_{- 0.6}^{+ 0.6}$ & $215_{- 13}^{+ 16}$ & $1.505_{-0.078}^{+0.087}$  & $1.660_{-0.086}^{+0.096}$ & $1.075_{-0.056}^{+0.062}$ \\[1pt]
CEERS-1561  & 6.20  & $20.0_{- 0.7}^{+ 0.8}$ & $ 80_{-  2}^{+  2}$ & $0.254_{-0.007}^{+0.008}$  & $0.331_{-0.009}^{+0.010}$ & $0.181_{-0.005}^{+0.006}$ \\[1pt]
CEERS-81049 & 6.74  & $ 7.3_{- 0.3}^{+ 0.3}$ & $ 90_{-  3}^{+  3}$ & $0.254_{-0.010}^{+0.007}$  & $0.327_{-0.013}^{+0.008}$ & $0.181_{-0.007}^{+0.005}$ \\[1pt]
CEERS-80925 & 6.75  & $ 8.7_{- 1.5}^{+ 1.6}$ & $139_{- 20}^{+ 19}$ & $0.679_{-0.083}^{+0.088}$  & $0.825_{-0.101}^{+0.107}$ & $0.485_{-0.059}^{+0.063}$ \\[1pt]
CEERS-80374 & 7.17  & $ 3.3_{- 0.4}^{+ 0.4}$ & $205_{- 27}^{+ 48}$ & $0.472_{-0.048}^{+0.066}$  & $0.512_{-0.052}^{+0.072}$ & $0.337_{-0.034}^{+0.047}$ \\[1pt]
CEERS-80239 & 7.49  & $ 6.6_{- 0.8}^{+ 0.9}$ & $334_{- 62}^{+109}$ & $>0.060$        \tnote{b}  & $>0.065$                  & $>0.043$                  \\[1pt]
    \hline
\end{tabular}

\begin{tablenotes}
\item[a] The \fesc{} with prism \lya{} flux corrections, where the correction factors are computed using the \lya{} mock observations as shown in Figure \ref{fig:ew_simulation}.
\item[b] \fesc{} $7\sigma$ lower limits given the \hb{} non-detection.
\end{tablenotes}
    
\end{threeparttable}
    \caption{The \lya{} properties measured for the 12 $z>5$ CEERS \lya{} emitting galaxies newly identified in this work.
    We report the measured \lya{} fluxes, EWs, and the \lya{} escape fractions (\fesc{}) assuming both case B and case A recombination, as well as their 1$\sigma$ uncertainties.
    These uncertainties could be underestimated if the error spectra based on \jwst{} pipeline underpredict the flux uncertainties (see \citealt{ArrabalHaro2023b}), but we note that our analysis is not significantly affected if we inflate the error spectra by the factor of 1.75 as reported in \citealt{ArrabalHaro2023b}.}
    \label{tb:lya}
\end{table*}

\begin{figure}
    \centering
    \includegraphics[width=\columnwidth]{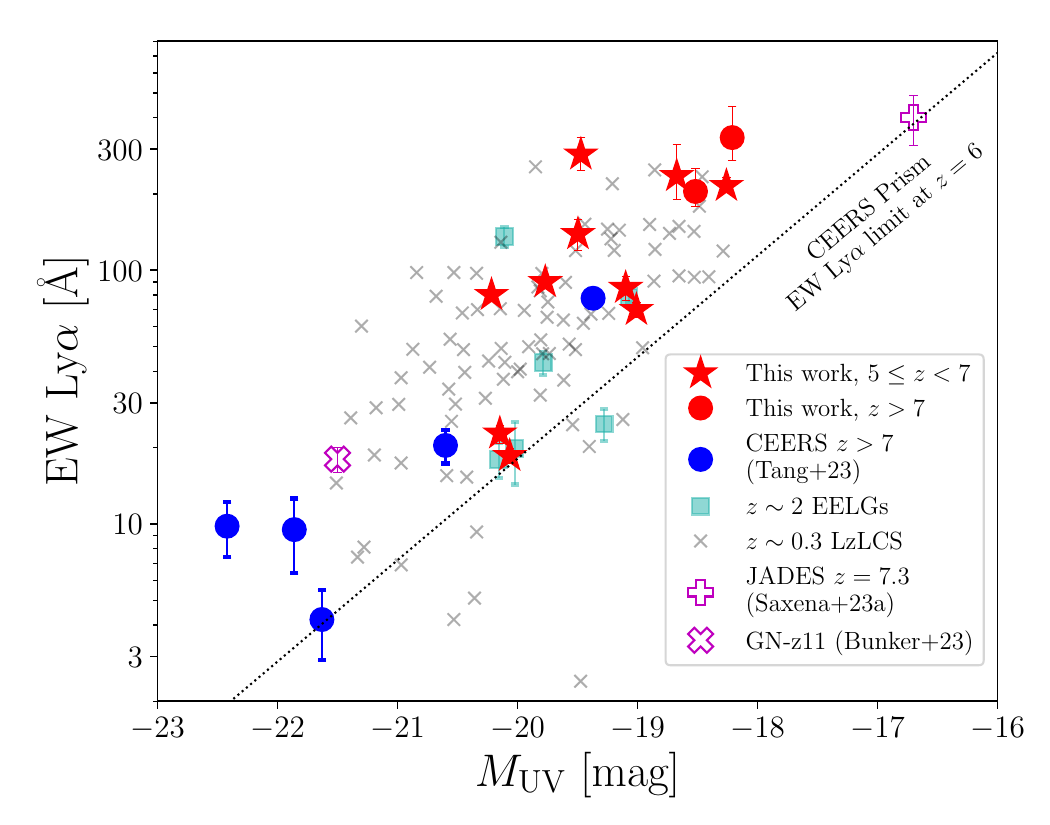}
    \caption{The \lya{} EW as a function of \muv{}.
    Similar to Figure~\ref{fig:sample}, the \lya{} emitting galaxies reported in this work are shown in red (stars for $5\leq z < 7$ \lya{} emitting galaxies and dots for $z>7$ \lya{} emitting galaxies), with the $z>7$ \lya{} emitting galaxies from \protect\cite{Tang2023} also shown as red stars.
    We also plot other NIRSpec spectroscopic confirmed \lya{} emitting galaxies at $z>7$, including JADES-GS-z7-LA at $z=7.3$ \citep{Saxena2023}, and the GN-z11 at $z=10.6$ \citep{Bunker2023}.
    For comparison, we show the measurements for the $z\sim2$ extreme emission line galaxies (EELGs) \citep{Tang2021} in cyan squares, and the $z\simeq$ 0.2--0.4 star-forming galaxies from the LzLCS survey in gray crosses \citep{Flury2022_I}.
    The black dotted line corresponds to the \lya{} EW 7$\sigma$ upper limit measured from the CEERS prism spectra (also see Figure \ref{fig:ew_simulation} below).}
    \label{fig:ew-muv}
\end{figure}

\begin{figure}
    \centering
    \includegraphics[width=1\columnwidth]{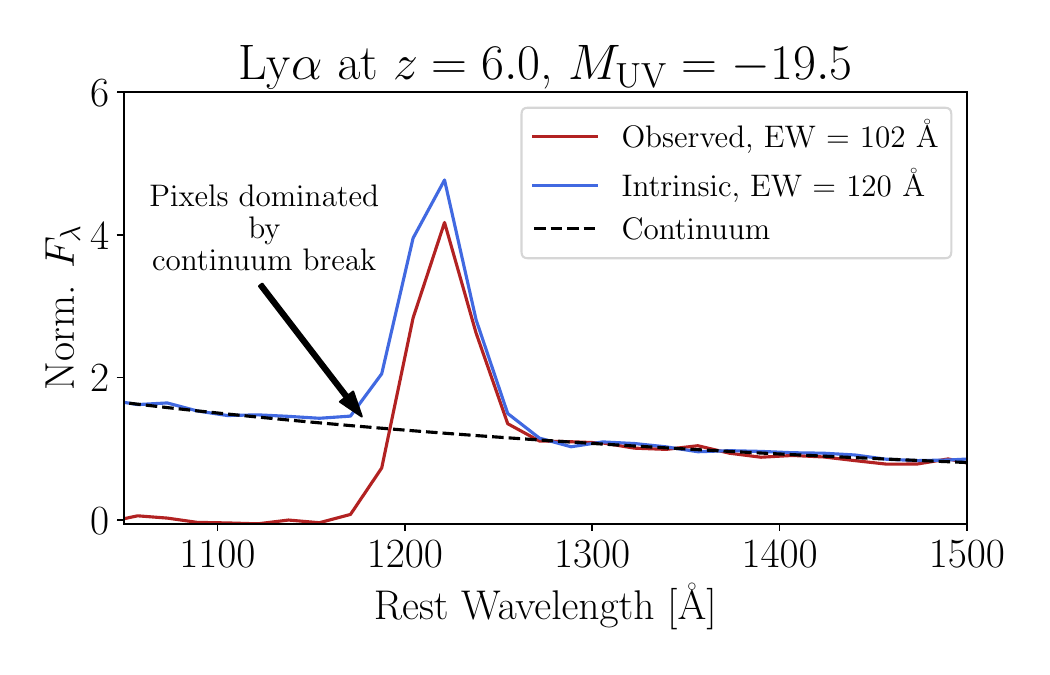} 
    \caption{The impact of the prism resolution on the recovery of \lya{} properties. 
    The red line corresponds to a mock observed \lya{} spectrum (intrinsic EW = 120 \AA{}) observed with the prism, taking into account the wavelength sampling, prism spectral resolution, and the IGM effect for a $z=6.0$ and \muv{}=$-19.5$ galaxy.
    The blue line is the same \lya{} spectrum but without the IGM attenuation.
    For this mock prism observation with IGM attenuation, a fraction of the \lya{} flux will fall in the pixels dominated by the continuum break, leading to a smaller measured \lya{} EW (102 \AA{}) than the true value (120 \AA{}).}
    \label{fig:mock_lya}
\end{figure}

\begin{figure*}
    \centering
    \includegraphics[width=1.0\textwidth]{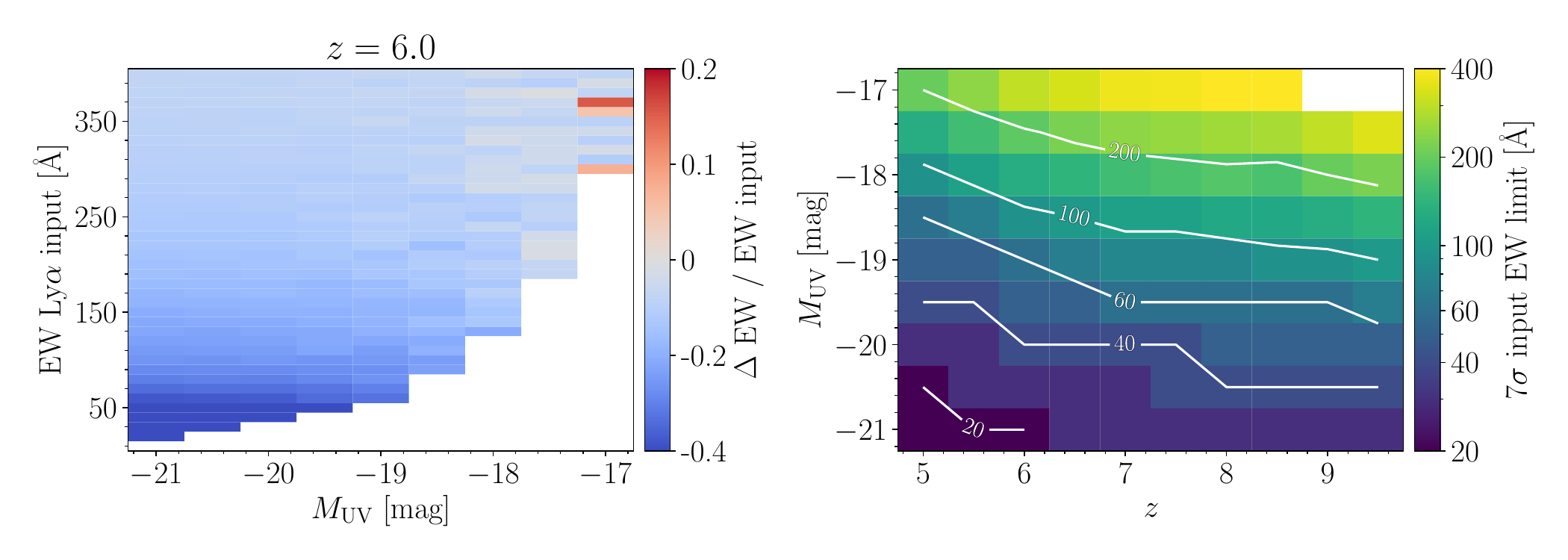}
    \caption{
    Our investigation of the robustness of \lya{} EW measurements from the prism spectra.
    In the {\em left} panel, we show the comparison between the recovered EW from the mock prism spectra and the input values: $\Delta$ EW / EW input, where $\Delta$ EW = EW recovered - EW input.
    We have chosen the redshift, absolute magnitudes (\muv{}), and input EW \lya{} grid spanning the full range of our \lya{} emitting galaxy sample.
    The EW differences shown here are for mock \lya{} observations at $z=6$, approximately the medium redshift of our \lya{} emitting galaxy sample.
    In the {\em right} panel, we present the input \lya{} EW limit that could be measured at 7$\sigma$ from the prism spectra for sources at different redshifts and \muv{} magnitudes.
    }
    \label{fig:ew_simulation}
\end{figure*}

We now seek to identify and characterize \lya{} emission from the CEERS spectroscopic sample established in the previous section, building on our earlier work characterizing  $z\gtrsim7$ \lya{} emitting galaxies in CEERS  \citep{Tang2023}. 
We first will focus on \lya{} emitting galaxies at $5\leq z < 7$ in the prism spectra before discussing new systems at $z>7$ from data obtained since \cite{Tang2023}.

We visually search for \lya{} in each of the 69 galaxies with prism redshifts at $5\leq z\leq 7$. 
We detect emission features at the wavelength of \lya{} in 10 galaxies (see Table~\ref{tb:lae}), with redshifts spanning $z = 4.999$ to 6.749 (median 5.67) and SNR = 5.6 to 26.9 (median 11.9). 
We have verified that these emission features are not due to cosmic rays by visually inspecting the individual exposures, thus most likely corresponding to \lya{} detection.
For one source (CEERS-80925), the \lya{} SNR is below 7 (5.6), and we consider it to be a tentative detection.
We will include this in our analysis below, but our results are not significantly altered if we remove this source.
We show the prism spectra in Figure~\ref{fig:spec}, highlighting the rest-optical lines and \lya{} detection.
The UV continuum is also detected for the 10 \lya{} emitters, at SNR = 5.8--132 (median 13.6) per rest-frame 150~\AA{} bin.
The continuum (measured in consistent aperture as the \lya{} flux) allows direct constraints to be placed on the \lya{} EW.  
We first fit the UV continuum with a linear function over the window of [100, 400]~\AA{} redwards of \lya{} in the rest frame.
We then compute the total \lya{} flux by directly integrating the continuum-subtracted line profile over the window of [-50, 80]~\AA{} (in the rest frame) around the line center. This aperture is chosen to match the observed width of the line. 
The measured  \lya{} fluxes range from $1.59\times10^{-18}$ to $3.69\times10^{-17}$ erg s$^{-1}$ cm$^{-2}$, corresponding to luminosities of $5.56\times10^{41}$ to $1.17\times10^{43}$ erg s$^{-1}$.
The resulting \lya{} EWs are presented in Table~\ref{tb:lya},  ranging from 19 to 286~\AA{} with a median EW = 134~\AA{}.

For the 10 galaxies with \lya{} at $5 \leq z < 7$, their NIRCam H-band (F150W) magnitudes (or ACS F160W if not covered by NIRCam) range from 26.1 to 28.7 (median 27.3, see Table~\ref{tb:lae}), which are on average fainter than the full CEERS spectroscopic sample (Figure~\ref{fig:sample}).
Following \cite{Tang2023}, we estimate their absolute UV magnitudes from the NIRCam (or WFC3) photometry in the filter where the central wavelength is closest to the rest-frame 1500~\AA{}. 
The resulting absolute UV magnitudes span \muv{} $= -18.3$ to $-20.2$ (median $-19.5$). These values overlap with a subset of the CEERS $z>7$ \lya{} emitting galaxies recently identified in \cite{Tang2023}, but are fainter than the $z>7$ \lya{} emitting galaxies identified with ground-based observations in this field \citep[\muv{} $< -20.0$; e.g.,][]{Oesch2015,Zitrin2015,Stark2017,Jung2022,Larson2022}.
We note that given the flux limit of the CEERS prism spectra, only large EW \lya\ ($>$ 100~\AA{}) will be detected in galaxies with faint continuum (\muv{} $>-20.0$; see Figure~\ref{fig:ew-muv}).
The sources with high \lya{} EW identified in this work are similar to the strong \lya{} emitting galaxies found at lower redshifts \citep[e.g.,][]{Malhotra2002,Cowie2011,Nakajima2012,Yang2017,Izotov2018_higho32,Tang2021,Flury2022_I,Naidu2022}.
For context, the CEERS \lya{} emitting galaxies are up to $\approx$3 mag fainter in \muv{} than GN-z11 (\muv{} = $-21.5$ and \lya{}~EW = 18~\AA{}; \citealt{Oesch2015,Bunker2023}), powering up to 10 times larger \lya{} EW.
Their large \lya{} EWs and low UV luminosities approach what is measured for the ultra-faint \lya{} emitting galaxy at $z=7.3$, JADES-GS-z7-LA (\muv{} = $-16.7$ and \lya{} EW = 400~\AA{}; \citealt{Saxena2023}).

We now investigate the CEERS spectra obtained after the first epoch of observations with the goal of identifying \lya{} emitters at $z\gtrsim 7$. 
The data reveal two likely \lya{} detections, CEERS-80374 at $z$ = 7.17, and CEERS-80239 at $z$ = 7.48.
Both \lya{} emitting galaxies are identified in the CEERS epoch 3 observations taken in February 2023. The spectra show multiple rest-optical emission lines (i.e., \hb{} and [\oiii{}]) as well as an emission line feature at the position of \lya{} at SNR = 7.8 and 8.2 that supports our identification (see Figure~\ref{fig:spec}).  
Both \lya{} emitting galaxies show the detection of UV continuum in the spectra.  
We compute \lya {} EWs in the same manner as above. 
The derived values are $205_{-27}^{+48}$~\AA{} for CEERS-80374 and $334_{-62}^{+109}$~\AA{} for CEERS-80239, among the highest EWs detected at $z>7$ \citep[e.g.,][]{Jung2020,Bunker2023,Saxena2023,Tang2023}.  
We measure \muv{} = $-18.5$ for CEERS-80374 and $-18.2$ for CEERS-80239 as above, which are among the faintest UV continuum magnitudes in our sample (see Figure~\ref{fig:ew-muv}).
The UV magnitudes are also fainter than the majority of the $z>7$ \lya{} emitting galaxies found previously, approaching the values measured for the strongest \lya{} emitting galaxies found in other surveys \citep[\muv{} $\sim$ $-15$ to $-17$; e.g.,][]{Maseda2018,Saxena2023,Maseda2023}.
We will come back to discuss this population in later sections.

To assess the robustness of our \lya{} EW measurements from the prism spectra, we simulate the observed \lya{} spectra for galaxies spanning the redshift and \muv{} range relevant to our \lya{} emitting galaxy sample.
We create mock 1D \lya{} spectra for galaxies spanning $z = 5$--10 in steps of $\Delta$z = 0.2 and \muv{} = $-16$ to $-22$ in steps of $\Delta$\muv{} = 0.5.
Each \lya{} spectrum consists of a skewed Gaussian profile as the intrinsic \lya{} emission line before IGM attenuation. 
We also adopt a flat continuum that matches the assumed \muv{}, where the flat continuum is consistent with the median UV slope ($\beta = -2.3$, see \S~\ref{sec:sed}) measured in our sample of \lya{} emitting galaxies.
For the \lya{} line, we assume a narrow (FWHM = 300~\kms{}) profile with the peak redshifted by 200~\kms{} relative to the systemic redshift, consistent with observations of Green Pea galaxies in the local universe where the IGM effect is negligible  \citep[e.g.][]{Orlitova2018}. However, we note that the results do not strongly depend on the assumed FWHM and velocity offset given the much poorer prism spectral resolution ($\sim$ 6,000--10,000~\kms{} at the observed \lya{} wavelength).
By adjusting the amplitude of the line, we obtain the intrinsic \lya{} profile. We choose amplitudes that correspond to EW = 10--400~\AA{} in steps of 5~\AA{}.
In this process, we have assumed that emission bluewards of 1216~\AA{} is completely attenuated by IGM.
Next, we convolve the intrinsic \lya{} profiles to match the wavelength-dependent resolution of the prism observations and rebin the spectra to the observed wavelength grid \footnote{\url{https://jwst-docs.stsci.edu/jwst-near-infrared-spectrograph/nirspec-instrumentation/nirspec-dispersers-and-filters}}.
We also add wavelength-dependent noise that is estimated from the median noise of our spectroscopic sample at the same observed wavelength grid.
An example of the resulting mock spectra is shown in Figurer \ref{fig:mock_lya}.
For each simulated observation of \lya{} spectrum, we then compute the observed \lya{} EW using the same method as we described above, adopting the same integration window of [$-50$, $80$]~\AA{} relative to the line center in the rest-frame to compute the line flux.
We repeat this process 100 times to obtain the median observed \lya{} EW and SNR at each redshift,  \muv{}, and input \lya{} EW.
Considering only the spectra where the \lya{} is observed at SNR $>$ 7 (similar to our adopted threshold for identifying \lya{} in CEERS observations), we measure the observed \lya{} EW from simulated spectra and compare them with the input values (Figure~\ref{fig:ew_simulation}).

At $z=6$, for mock spectra with input \lya{} EW $>90$~\AA{} and \muv{} $<-18.5$, we recover similarly high EW values, which in general are only slightly smaller (by $\lesssim$ 20\%) than the input EWs.
The smaller ``observed'' EWs are expected if a fraction of the \lya{} photons fall on pixels dominated by the continuum break at the prism resolution, reducing the recovered line flux (Figure \ref{fig:mock_lya}; also see \citealt{Jones2023}).
The percentage difference between the measured and input \lya{} EW increases towards smaller input EW, with the measured value potentially underestimated by  30--50\% for more moderate EW ($\lesssim 70$~\AA) lines. Among the 8 strongest  \lya{} emitting galaxies  ($>$80~\AA{}) in our sample, we find that the 
EWs may be underestimated by a small fraction ($\lesssim$ 25\%) due to the instrumental effects described above.
For the remaining 4 sources with weaker \lya{} (23--80~\AA{}), the left panel of Figure~\ref{fig:ew_simulation} suggests the measured values may be underestimated by up to 50\%. 
We note that the left panel of Figure 4 indicates that the EW could also be slightly overestimated (by $\lesssim$15\%) at extremely faint magnitudes (\muv{}$=-17.0$).
This is due to fluctuations of EW measurements with the low SNR continuum at such faint magnitudes, and all galaxies in our sample are brighter than this magnitude and thus have their \lya{} EWs underestimated.
In what follows, we will report the measured values, but we will also consider how our results may change if the \lya{} EWs are underestimated. 

As the majority of galaxies in our spectroscopic sample do not show \lya{}, we now consider what EW limits we can place on them given the prism non-detections.
The right panel of Figure~\ref{fig:ew_simulation} shows the limiting input \lya{} EW that leads to a $>$ 7$\sigma$ emission line in the mock spectra.
For bright (\muv{} $< -20$) sources with intrinsic EW = 40~\AA{}, we are able to detect \lya{} at $z\sim 5-6$. 
This translates to an observed EW $\approx$ 20~\AA{} that corresponds to the smallest \lya{} EW value measured in our CEERS \lya{} emitting galaxies. 
We also see in Figure~\ref{fig:ew_simulation} that the  \lya{} EW limit increases rapidly at lower luminosities.  At $z =$ 5--8, these tests suggest that a \lya{} non-detection at \muv{} = $-18$ would correspond to EW $<$ 90--180~\AA{}, whereas at 
\muv{} = $-20$, the non-detection would correspond to EW $<$ 30--50~\AA{}.
We note that these tests are consistent with  \lya{} being detected in GN-z11 at moderate resolution ($R=1000$) but not with the prism \citep{Bunker2023}:
given its absolute magnitude (\muv{} = $-21.5$) and redshift ($z=10.6$), we find that the $R=100$ prism can only recover \lya{} with EW$>$35~\AA\ (assuming continuum per pixel SNR = 20 and no damping wing effect of IGM absorption), above the value recovered at higher resolution (EW = $18\pm2$~\AA{}). 

For individual galaxies with non-detections of \lya{} in our sample, we have derived the \lya{} EW upper limits by direct integration of the error spectra.
We use the same integration window (i.e., [-50, 80]~\AA{} relative to \lya{} in the rest frame) used for individual \lya{} emitting galaxies in our sample.
The resulting 7$\sigma$ upper limits on \lya{} fluxes range from $4.2\times 10^{-18}$ to $6.0\times 10^{-17}$ erg s$^{-1}$ cm$^{-2}$, translating into EW spanning from 15~\AA{} to 560~\AA{}.
However, as already shown above with the mock \lya{} observations, when the \lya{} is detected, the observed EW could be underestimated (by up to 50\%) compared to the input \lya{} EW. As a result, the \lya{} upper limits derived from the observed spectra could still be underestimated given the poor resolution of the prism observations.
In our following analysis, we will include these non-detections of \lya{} and consider the impact of underestimated upper limits on our results. 
By combining these non-detections with the 12 \lya{} emitting galaxies, we will be able to assess how common strong \lya{} emission lines are at different redshift ranges, which we will discuss in the following sections.

We note again that the \lya{} measurements reported in this section may be missing diffuse flux scattered from gas surrounding the UV-bright components.
Studies with the VLT/MUSE Integral Field Spectrograph have characterized the surface brightness profile of \lya{} halos for large numbers of faint \lya{} emitting galaxies up to $z \sim$ 6 \citep[e.g.,][]{Wisotzki2016,Leclercq2017}.
To assess roughly how much of the total \lya{} flux we may be missing in these NIRSpec microshutter observations, we compare the three compact $z\sim$ 6 \lya{} emitting galaxies that have been observed by both NIRSpec prism and VLT/MUSE IFS \citep{Saxena2023b}.
This analysis will be presented in Tang et al., in prep.
The NIRSpec line flux is taken from \cite{Saxena2023b}, and the MUSE \lya{} flux is from \cite{Bacon2017,Bacon2023} extracted with the ``ORIGIN'' software \citep{Mary2020}, and more details will also be described in Tang et al. in prep..
For the three systems, we find that the NIRSpec measured \lya{} fluxes are 70--80\% of the fluxes measured from VLT/MUSE.
If we assume the $z\sim 5$ \lya{} surface brightness profile from \cite{Leclercq2017}, the MUSE ``ORIGIN'' aperture in turn recovers $\sim$ 75\% of the total \lya{} after accounting for the diffuse flux in the \lya{} halo.
This suggests that, for the $z\leq5$ galaxies in our sample, the total \lya{} flux could be up to a factor of 1.67 larger than our NIRSpec measurements.
In this paper, we are mostly focused on a comparison of \lya{} properties in galaxies observed with NIRSpec MSA, somewhat mitigating the impact of the slit loss associated with \lya{} halos.
Future studies with NIRSpec Integral Field Spectroscopy are required to assess slit loss fractions in larger samples of \lya{} emitters in the reionization era.

\subsection{SED Fitting and Photoionization Modeling}\label{sec:sed}

\begin{table*}
    \centering
    \begin{threeparttable}[t]
\begin{tabular}{cccccccc}
    \hline
    ID & $z_{\rm spec}$ & $\beta$ & log ($M_*/M_\odot$) & Age & EW [\oiii{}]+\hb{} & log ($\xi_{\rm ion}$/erg$^{-1}$Hz) & $\tau_{\rm V}$ \\
    & & & & (Myr, CSFH) &(\AA{}) &  & \\ 
    \hline

CEERS- 1374 & 5.00  & $-1.93_{-0.28}^{+0.26}$  & $7.91_{-0.19}^{+0.28}$  & $ 10_{-  4}^{+ 16}$  & $1982_{- 333}^{+ 462}$           & $25.76_{-0.14}^{+0.13}$  & $0.013_{-0.010}^{+0.033}$ \\[1pt]
CEERS-82069 & 5.24  & $-2.47_{-0.34}^{+0.33}$  & $7.67_{-0.13}^{+0.16}$  & $  2_{-  1}^{+  9}$  & $ 682_{- 315}^{+ 344}$           & $25.86_{-0.23}^{+0.04}$  & $0.077_{-0.049}^{+0.067}$ \\[1pt]
CEERS-80573 & 5.42  & $-2.43_{-0.12}^{+0.11}$  & $8.14_{-0.21}^{+0.22}$  & $ 54_{- 24}^{+ 52}$  & $ 835_{- 181}^{+ 253}$           & $25.53_{-0.07}^{+0.10}$  & $0.005_{-0.003}^{+0.012}$ \\[1pt]
CEERS- 1334 & 5.50  &                      --  &                   --    &                  --  & $2949_{-  54}^{+  50}$ \tnote{a} &                      --  &                        -- \\[1pt]
CEERS-  323 & 5.67  & $-2.26_{-0.17}^{+0.17}$  & $7.11_{-0.06}^{+0.29}$  & $  3_{-  1}^{+  8}$  & $4300_{-1006}^{+ 693}$           & $25.83_{-0.12}^{+0.06}$  & $0.008_{-0.006}^{+0.027}$ \\[1pt]
CEERS-80916 & 5.68  & $-1.91_{-0.18}^{+0.14}$  & $7.29_{-0.18}^{+0.29}$  & $  7_{-  3}^{+  8}$  & $2266_{- 552}^{+1063}$           & $25.79_{-0.10}^{+0.11}$  & $0.106_{-0.068}^{+0.039}$ \\[1pt]
CEERS-  476 & 6.02  & $-2.27_{-0.28}^{+0.29}$  & $7.28_{-0.34}^{+0.38}$  & $ 25_{- 17}^{+ 59}$  & $1094_{- 308}^{+ 550}$           & $25.60_{-0.13}^{+0.19}$  & $0.005_{-0.004}^{+0.014}$ \\[1pt]
CEERS- 1561 & 6.20  &                      --  &                   --    &                  --  & $2301_{- 224}^{+ 252}$ \tnote{a} &                      --  &                        -- \\[1pt]
CEERS-81049 & 6.74  & $-2.26_{-0.11}^{+0.11}$  & $7.50_{-0.16}^{+0.63}$  & $  5_{-  2}^{+ 34}$  & $1475_{- 315}^{+ 565}$           & $25.73_{-0.16}^{+0.08}$  & $0.013_{-0.011}^{+0.041}$ \\[1pt]
CEERS-80925 & 6.75  & $-1.97_{-0.17}^{+0.17}$  & $7.39_{-0.13}^{+0.55}$  & $  3_{-  2}^{+ 18}$  & $3865_{-1329}^{+ 883}$           & $25.82_{-0.16}^{+0.08}$  & $0.012_{-0.010}^{+0.044}$ \\[1pt]
CEERS-80374 & 7.17  & $-2.74_{-0.47}^{+0.42}$  & $7.19_{-0.36}^{+0.43}$  & $ 16_{- 13}^{+ 38}$  & $1416_{- 348}^{+1298}$           & $25.65_{-0.12}^{+0.16}$  & $0.006_{-0.004}^{+0.019}$ \\[1pt]
CEERS-80239 & 7.49  & $-1.16_{-0.50}^{+0.48}$  & $7.32_{-0.20}^{+0.48}$  & $  5_{-  3}^{+ 16}$  & $2289_{- 832}^{+1674}$           & $25.80_{-0.16}^{+0.15}$  & $0.201_{-0.116}^{+0.093}$ \\[1pt]
    \hline
\end{tabular}

\begin{tablenotes}
\item[a] \textbf{These two galaxies are not covered by NIRCam imaging, and we infer their EW [\oiii{}]+\hb{} directly from the prism spectra given the optical continuum also detected.}
\end{tablenotes}    

\end{threeparttable}
    \caption{The NIRCam SED properties for the $z>5$ CEERS \lya{} emitting galaxies inferred with \beagle{} photoionization modelling.
    We provide the UV continuum slope ($\beta$, measured from photometry), stellar mass and stellar population ages (both assuming a CSFH), [\oiii{}]+\hb{} EW, the intrinsic ionizing photon production efficiency ($\xi_{\rm ion}$) corrected for dust attenuation, and the V-band optical depth ($\tau_V$), as well as their inner 68 percent confidence interval uncertainties.}
    \label{tb:sed}
\end{table*}

\begin{figure*}
    \centering    \includegraphics[width=1.0\textwidth]{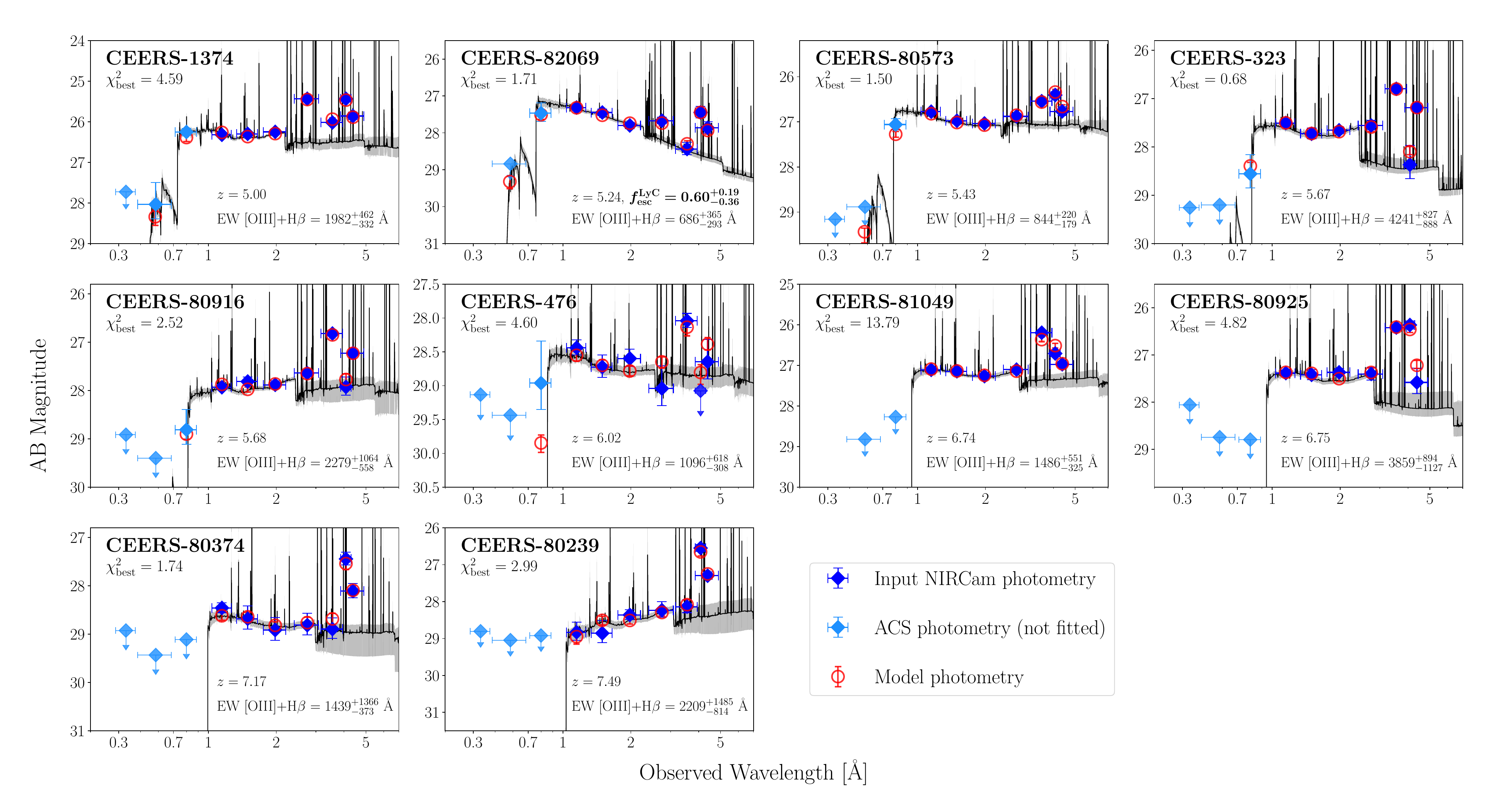}
    \caption{
    The BEAGLE SED fits of the \jwst{}/NIRCam photometry for the CEERS \lya{} emitting galaxies at $z\sim$ 5--7.5.
    We show the SEDs for 10 out of the 12 galaxies within the CEERS NIRCam coverage.
    In each panel, the median fit SED model is shown in black with the gray shading corresponding to the inner 68 percent confidence interval of the \beagle{} posterior.
    The dark blue diamonds are the input NIRCam photometry, while the red circles indicate the median model photometry.
    We also show in light blue the observed photometry covering or bluewards of \lya{}, which is not included in our BEAGLE modeling. 
    }
    \label{fig:beagle}
\end{figure*}

\begin{figure}
    \centering
    \includegraphics[width=\columnwidth]{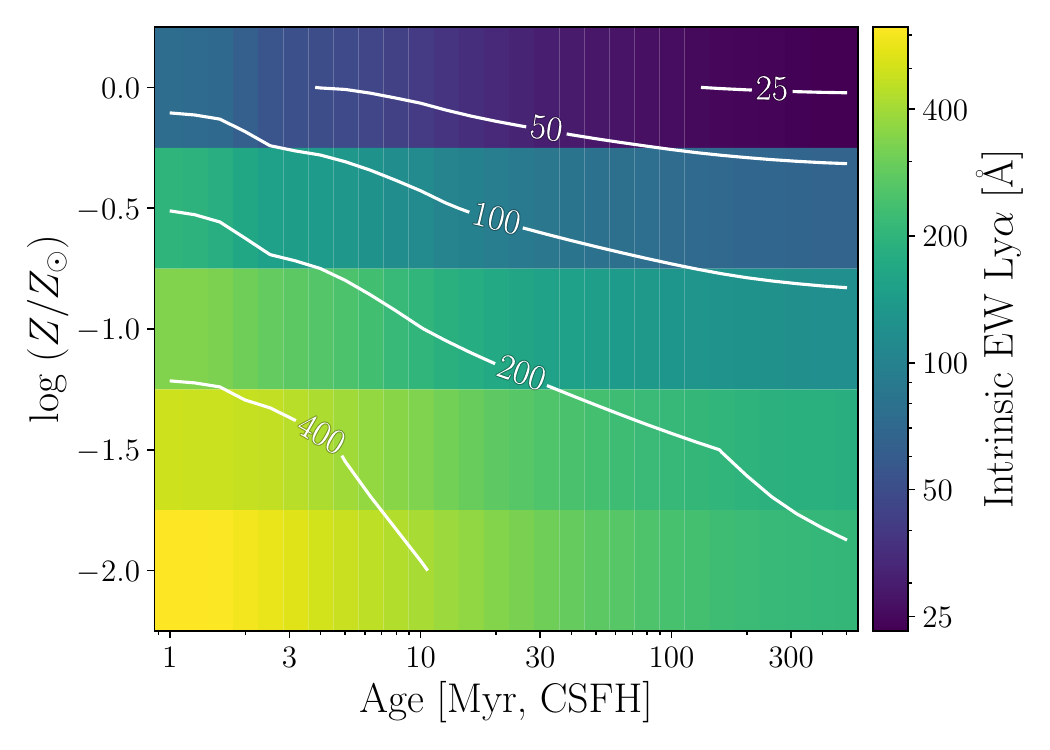}
    \caption{The intrinsic \lya{} EW as a function of stellar population age (assuming CSFH) and the stellar metallicity.
    We use \beagle{} photoionization modeling to predict the intrinsic \lya{} emission that is not attenuated by dust or scattered by neutral hydrogen.
    Our modeling assumes a \protect\cite{Chabrier2003} IMF (0.1--300 $M_\odot$), an ionization parameter of log $U$ = $-2.5$, with the gas-phase metallicity coupled with the stellar metallicity through the depletion factor $\xi_{\rm d}$ = 0.3.}
    \label{fig:ew_model}
\end{figure}

The stellar population properties of our spectroscopic sample, including the new $z\geq$ 5 \lya{} emitting galaxies are derived from spectral energy distributions (SEDs). 
The CEERS NIRCam footprint covers 10 of the 12  \lya{} emitters (49 of the 67 galaxies not detected in \lya{}) in our new prism sample. 
We will focus primarily on these sources in this section given the significant improvement NIRCam provides relative to \hst{} and {\it Spitzer}. 
We show the NIRCam SEDs of the 10 newly identified \lya{} emitting galaxies in Figure~\ref{fig:beagle}. 
Their SEDs show characteristic flux excesses (0.3--1.6 mag) from [\oiii{}]+\hb{} and H$\alpha$ in the NIRCam long wavelength filters that indicate the presence of young massive stars.
Through power law fitting to the broad-band SED sampling the rest-frame UV (see \citealt{Topping2022_ceers,Cullen2023,Topping2023}), we estimate that their UV slopes ($\beta$) range from $-1.2$ to $-2.7$, with the median $-2.3$ (see Table \ref{tb:sed}).
The bluest UV slopes are also consistent with substantial leakage of ionizing photons \citep[e.g.,][]{Bouwens2010,Ono2012,Topping2022_ceers,Furtak2023,Topping2023}, which we will discuss below.
We infer the physical properties of the galaxies in our sample by fitting the NIRCam photometry with the BayEsian Analysis of GaLaxy sEds (\beagle{}; \citealt{Chevallard2016}) code.
We largely follow the modeling process described in \cite{Endsley2023_ceers} and \cite{Tang2023}, which we summarize below. 
For the 10 \lya{} emitting galaxies, we also consider a set of models allowing the escape of ionizing photons from the host galaxies, which will also be described below.
In both cases, we employ the \cite{Gutkin2016} photoionization models that self-consistently combine the stellar emission from the latest \cite{Bruzual2003} stellar population synthesis models with the nebular emission computed with the photoionization code \cloudy{} (\citealt{Ferland2013}).
We fix the redshift to the spectroscopic values determined in \S~\ref{sec:spec}, and we only consider the NIRCam filters redwards of the \lya{} emission line.

We assume a constant star formation history (CSFH), adopting the \cite{Chabrier2003} initial mass function with the mass range of 0.1--300 $M_{\odot}$.
We place a log-uniform prior on the galaxy age (defined as the duration of the constant star formation) from 1 Myr to the age of the universe at the given redshift, and on the stellar mass over the range of $5 \leq \log (M_*/M_\odot) \leq 12$. 
However, we note that in cases where the star formation history has undergone a recent upturn, the stellar masses inferred from CSFH could substantially underestimate the total stellar mass if there is a hidden older stellar population \citep[e.g.,][]{Tacchella2023,Topping2022_rebels,Endsley2023,Whitler2023_cosmos,Whitler2023_ceers}.
We assume log-uniform priors on both stellar metallicities and ionization parameters over a broad range of the parameter space allowed by the models (i.e, $-2.2 \leq \log (Z/Z_\odot) \leq -0.3$; $-4.0 \leq \log U \leq -1.0$).
The interstellar (the gas-phase + dust) metallicity is kept the same as the stellar metallicities through a fixed depletion factor ($\xi_{\rm d}$ = 0.3) of metals into dust grains, and we assume the gas density $n_{\rm H} = 100$ cm$^{-3}$.
We use the SMC  curve \citep{Pei1992} with V-band optical depth ($\tau_{\rm V}$) adjusted from $0.001\leq \tau_{\rm V} \leq5$
Finally, we assume the \cite{Inoue2014} model to account for the IGM attenuation.

Our default BEAGLE models are ionization-bounded, assuming that all ionizing photons emitted from stars have been reprocessed by the interstellar medium into nebular emission lines or absorbed by dust grains.
Previous work has shown that extremely strong \lya{} emitting galaxies (i.e., $>$ 50--100~\AA) could leak a significant amount of Lyman continuum (LyC) photons \citep[e.g.,][]{Flury2022_II,Dijkstra2016,Verhamme2017,Izotov2020,Pahl2021}, diminishing the nebular continuum and emission line flux. 
This can result in very blue UV slopes ($\beta<-2.8$) that are not seen in the 
standard ionization-bounded models \citep[e.g.,][]{Raiter2010,Zackrisson2017,Yamanaka2020,Chisholm2022,Topping2022_ceers,Topping2023,Kim2023}.
Given the presence of very blue sources in our sample, we also consider a set of BEAGLE models that account for the escape of LyC photons ($f_{\rm esc}^{\rm LyC}$ models). The models we adopt use a ``picket fence'' geometry where a fraction of LyC photons escape through low \hi{} density sightlines from the host galaxy. \citep[e.g.,][]{Heckman2001,Heckman2011,Reddy2016,Gazagnes2020}.
We place a uniform prior on the LyC escape fraction ($f_{\rm esc}^{\rm LyC}$) over [0,1], with the remaining BEAGLE parameters kept the same as in our fiducial ionization-bounded models. 

For the majority (9/10) of the \lya{} emitting galaxies, the $f_{\rm esc}^{\rm LyC}$ models provide equally accurate fits to the NIRCam SEDs, with the implied $f_{\rm esc}^{\rm LyC}$ ranging from 0.19--0.47.
In both sets of \beagle{} fits, the minimum $\chi^2$ values are found to have small differences between the two models.
Their fitted parameters from the two models are mostly consistent, except that the $f_{\rm esc}^{\rm LyC}$ models prefer slightly younger CSFH ages ($\sim$ 0.36 dex) and higher stellar masses ($\sim$ 0.43 dex), similar to what is found in \cite{Tang2023}.
However, for CEERS-82069, where the observed SEDs cannot be fitted well with the fiducial BEAGLE models (best $\chi^2$ = 36.0), the $f_{\rm esc}^{\rm LyC}$ models improve the fitted SED significantly (best $\chi^2$ = 1.71, see its SED in Figure~\ref{fig:beagle}).
Its potential high escape fraction of LyC photons is also consistent with the very blue UV continuum slope ($\beta = -2.47_{-0.34}^{+0.33}$) and the relatively weak [\oiii{}]+\hb{} ($682_{-315}^{+344}$ \AA{}, the smallest value in our sample), both of which may be expected when a fraction of the ionizing photons escape rather than being used to produce nebular continuum and emission lines by the ISM \citep[e.g.,][]{Topping2022_ceers,Topping2023}.
We also obtain an O32 (=[\oiii{}]$\lambda\lambda$4959,5007 / [\oii{}]$\lambda$3728) lower limit of $>2.6$ (3$\sigma$) from the weakly detected [\oiii{}] and nondetection of [\oii{}] (Figure \ref{fig:spec}), but deeper spectroscopy is required for a more robust measure of O32.
Furthermore, as we will show in \S~\ref{sec:fesc-z}, we also infer a high \lya{} escape fraction (\fesc{}=$0.787_{-0.048}^{+0.053}$) that is also potentially consistent with a high escape fraction of LyC photons \citep[e.g.][]{Dijkstra2016,Flury2022_II}.
Therefore, the SED fitting results derived from the $f_{\rm esc}^{\rm LyC}$ models are preferred over those from the default ionization-bounded models for CEERS-82069.
In the following analysis, we will primarily focus on the results from the $f_{\rm esc}^{\rm LyC}$ models for CEERS-82069, and the results from our default models for the remaining 9 galaxies, which are reported in Table~\ref{tb:sed}.

The \beagle{} models of the \lya{} emitters show large flux excesses in the NIRCam SEDs at  3--5 \um{}, resulting in the models preferring large EW [\oiii{}]+\hb{} (682~\AA{} up to 4300~\AA{} in the rest frame).
The median EW [\oiii{}]+\hb{} is 1982~\AA{}, significantly higher than the typical values inferred for $z\sim$ 6--8 galaxies (median $\approx$ 780~\AA{}; \citealt{Endsley2021,Endsley2023_ceers,Endsley2023}).
The very high [\oiii{}]+\hb{} EWs also translate into extremely young stellar population ages (assuming CSFH) ranging from 2 to 54 Myr (median 6 Myr), younger than the more general populations at these redshifts (median 30--69 Myr; \citealt{Whitler2023_cosmos,Endsley2023}) but comparable to the $z>7$ \lya{} emitting galaxies found in \cite{Tang2023}.
The stellar masses associated with these young stellar populations are estimated to be $1.3\times 10^7$ $M_\odot$--$1.4\times 10^8$ $M_\odot$ (median $2.3\times 10^7$ $M_\odot$).
The corresponding specific star formation rates range between 19--473 Gyr$^{-1}$ and occupy the high end of the sSFR distribution of UV-bright galaxies at this redshift (median $18_{-5}^{+7}$ Gyr$^{-1}$; \citealt{Topping2022_rebels}), consistent with rapidly rising SFHs occurring in these \lya{} emitting galaxies \citep{Endsley2023_ceers,Endsley2023}.  
The rest-optical spectra also reveal O32 (=[\oiii{}]$\lambda\lambda$4959,5007 / [\oii{}]$\lambda$3728) ratios up to $>58.2$  (median $>$7.3), suggesting gas under extreme ionization conditions. 
We also note that when the rest-optical continuum is also well-detected (SNR$>$7 over a 1200~\AA{} window in the rest-frame) in the prism spectra, we derive values of EW [\oiii{}]+\hb{} that are consistent with the \beagle{}-based values within the 1--2 $\sigma$ uncertainties.

With the constraints on the dust attenuation from SED modeling, we now reassess the dust correction to the optical emission line fluxes measured from prism spectra.
We perform dust correction considering both the Balmer decrement derived directly from the spectra and the effective optical depth estimated with \beagle{}.
Among the 12 \lya{} emitting galaxies, we detect both \ha{} and \hb{} at SNR(\hb{}) = 16--106 for 6 systems, allowing us to estimate the dust attenuation using the \ha{}/\hb{} flux ratios.
In one case (CEERS-1334), we use the \hg{}/\hb{} ratio (\hg{} detected at SNR = 45) owing to an anomalously high \ha{} flux.
Assuming the case B recombination with the gas temperature $T_e=10^4$ K and density $n_e=250$ cm$^{-3}$, we expect an intrinsic \ha{}/\hb{} (\hg{}/\hb{}) = 2.86 (0.47; \citealt{Draine2011}).
Comparing this with the observed ratios (median \ha{}/\hb{} = 3.20, and \hg{}/\hb{} consistent with zero dust for CEERS-1334), we derive the corresponding optical depth in the V band ($\tau_{\rm V}$), assuming the SMC attenuation curve \citep{Pei1992}.
We find that inferred $\tau_{\rm V}$ range from 0.0 (CEERS-1334 and CEERS-80916) to 0.56 (CEERS-1561), corresponding to small correction factors to the \ha{} flux of 1.00--1.55 (median 1.24) to the six galaxies.

However, for the remaining 6 galaxies, the Balmer emission lines (\ha{} and \hb{} at $z<6.8$, and \hb{} and \hg{} at $z>6.8$) are poorly detected (SNR$<$7 for at least one line), leading to large uncertainties when deriving the dust attenuation based on the Balmer decrement.
In order to estimate their dust attenuation, we rely on the optical depths in the V band derived with NIRCam SED modeling above, which span $\tau_{\rm V}$ = 0.005--0.20 (median 0.009).
To investigate whether these SED-based attenuation factors are applicable to emission lines, we compare the $\tau_V$ derived from the continuum SED and that from the Balmer decrement for four galaxies where both measurements are available.
We find that the continuum SED suggests a $\tau_V$ that is 3.1--4.9 $\sigma$ smaller than the value inferred from Balmer emission lines.
Since the SED-based attenuation is mostly driven by the reddening of the continuum, it is often dominated by the fraction of dust in the diffuse ISM, affecting stars that have freed themselves from their birth clouds.
If the age of the stellar population is very young throughout the galaxy, we may expect the stars and nebular-emitting gas to face similar attenuation.
It is unclear whether the offset of $\tau_V$ we find is due to \hii{} regions facing greater attenuation than stellar continuum \citep[e.g.,][]{Calzetti2000,ForsterSchreiber2009,Wuyts2011,Price2014,Reddy2015,Reddy2020,Shivaei2020}, or due to uncertainty in estimating attenuation from SEDs of very young and blue galaxies.
Based on these results, when computing attenuation facing emission lines, we will adopt the \cite{Reddy2020} relation, where the nebular attenuation is 4.331$\times$ (assuming the SMC curve) of that faced by the stellar continuum.
However, we note that our analysis in the following is not significantly impacted if we adopt the original  $\tau_{\rm V}$ inferred with \beagle{} fits to NIRCam SEDs.
This requires small correction factors to the \ha{} (\hb{}) flux of 1.02--1.96 (median 1.03) for these six $z<7$ galaxies without robust Balmer decrement measurements.

In the following section, we will quantify the escape fraction of \lya{} for the \lya{} emitting galaxies in our sample, using the \ha{} or \hb{} detections to compute the intrinsic \lya{} luminosity. 
The \beagle{} models described above provide an independent estimate of the ionizing photon production rate, which also translates into an intrinsic \lya{} luminosity and EW.
In this process, \beagle{} takes the \lya{} luminosity output by \cloudy{} \citep{Ferland2013}, which is supposed to account for the absorption of ionizing photons by dust before they ionize hydrogen \citep{Charlot2001,Gutkin2016}.
In Figure~\ref{fig:ew_model}, we show the intrinsic \lya{} EW implied by \beagle{} as a function of both stellar population age (assuming CSFH) and stellar metallicity.
Here, we have assumed the default ionization bounded \hii{} regions with an ionization parameter of log $U$ = $-2.5$, and adopted the same \cite{Chabrier2003} IMF  and treatment to the gas-phase metallicity as described for our SED fitting.
It is clear that the intrinsic \lya{} EW are largest at the youngest stellar population ages and lowest metallicity, reaching $\geq$ 200~\AA{} at age $\lesssim$ 30 Myr and $Z<0.1 Z_\odot$.
At the young stellar population ages that are common for our \lya{} emitting galaxy sample (2--54 Myr), the \beagle{} models infer very high ionizing photon production efficiency log ($\xi_{\rm ion}$ / erg$^{-1}$ Hz) = 25.5--25.9, which corresponds to the hydrogen ionizing photon production rate per unit intrinsic UV luminosity at rest-frame 1500~\AA{} including nebular and stellar continuum \citep[e.g.,][]{Chevallard2018,Tang2019}.
The efficient production of ionizing photons results in the production of very large \lya{} EWs in the sample, with values ranging from 113 to 601~\AA{} (median 256~\AA{}).
Compared with the observed \lya{} EW (20--334~\AA{}), the larger intrinsic EW implies that only a fraction of the \lya{} photons are able to escape from the host galaxies (on the scale of the NIRSpec microshutter) and through the IGM, which we will discuss in the following section.

\section{Redshift Evolution of \lya{} Escape Fraction }\label{sec:fesc-z}

\begin{figure}
    \centering
    \includegraphics[width=\columnwidth]{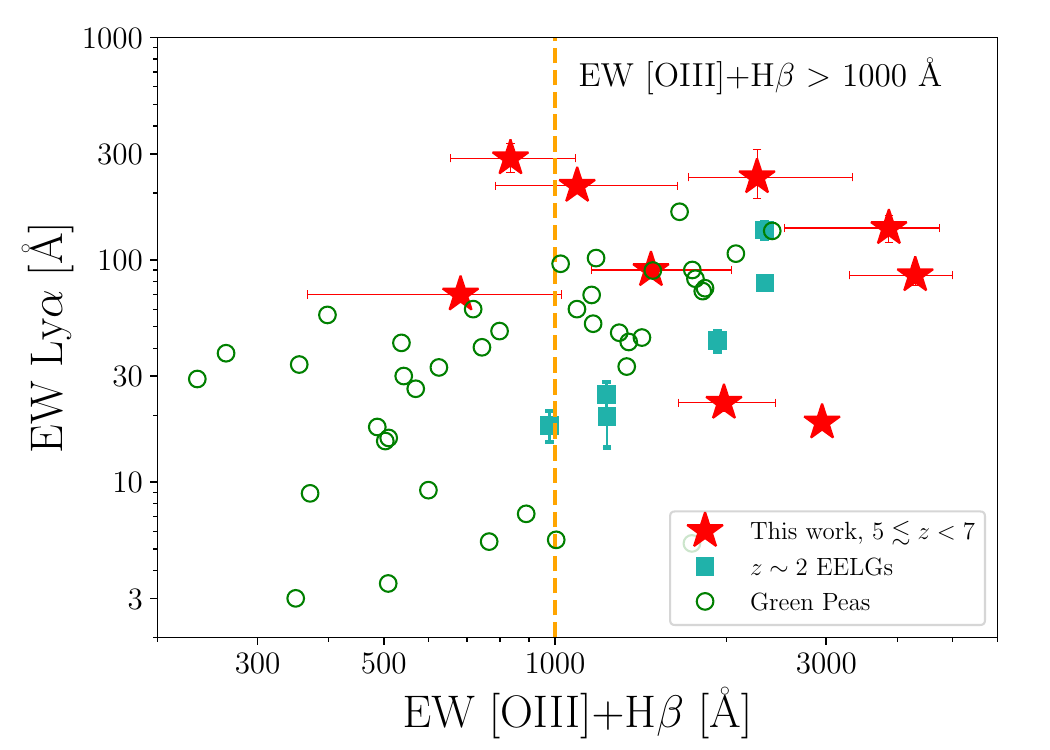}
    \caption{The observed \lya{} EW as a function of EW [\oiii{}]+\hb{}.
    The red stars correspond to the $5\leq z < 7$ \lya{} emitting galaxies presented in this work.
    For comparison, we also plot the $z\sim2$ EELGs \citep{Tang2021} as well as the $z\sim 0$ Green Peas \citep{Yang2017}.
    In general, we find galaxies with high \lya{} EWs also preferentially show high [\oiii{}]+\hb{} EWs.
    We show a dashed line at EW [\oiii{}]+\hb{} = 1000A to denote the threshold we adopt for our \lya{} analysis in \S \ref{sec:fesc-z}.}
    \label{fig:ew_ew_o3hb}
\end{figure}

\begin{figure*}
    \centering
    \includegraphics[width=0.8\textwidth]{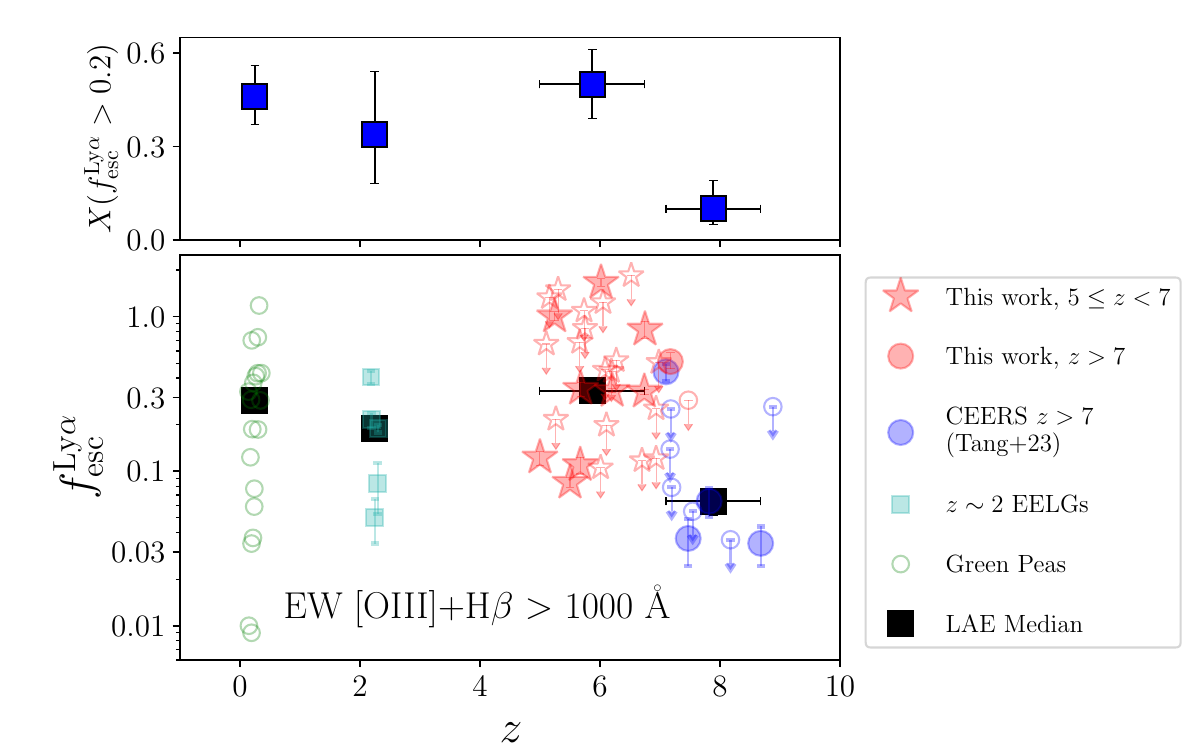}
    \caption{The \lya{} escape fraction (\fesc{}) as a function of redshift.
    Here, we only consider galaxies with large EW [\oiii{}]+\hb{} that are most efficient in producing ionizing and thus \lya{} photons, and with less dust attenuation (V-band optical depth $\tau_V < 0.7$).
    We adopt the results by assuming the case B recombination as commonly done in the literature \citep[e.g.,][]{Tang2023}.
    Shown are the $5\leq z<7$ galaxies (red stars) and the $z>7$ ones from both this work  (red circles) and those reported in \protect\cite{Tang2023} (blue circles).
    We plot the \lya{} emitting galaxies with filled symbols and those lacking \lya{} detections with open ones.
    We also show the low-redshift measurements from literature samples, including the $z\sim0.3$ Green Peas \citep{Yang2017} and the $z\sim2$ EELGs \citep{Tang2021}.
    For sources with prism spectra from \protect\cite{Tang2023} and this work, the \fesc{} values have been derived from the \lya{} flux corrected for prism observations (see \S \ref{sec:lya_det}).
    The black squares indicate median \fesc{} measured at each redshift.
    While the median \fesc{} is uniformly high at $z<7$ ($\approx$ 0.19--0.33) for sources detected in \lya{}, a significant drop for the median \fesc{} (0.06) is found among the $z\sim$ 7--9 sample. 
    This redshift evolution is also clear from the fraction of high-\fesc{} sources, the observed $X$(\fesc{}$>0.2$), shown as blue squares in the top panel, with $X$(\fesc{}$>0.2$) = 0.34--0.46 at $z<3$, $\sim$ 0.50 at $5 \leq z < 7$, and decreasing to 0.10 at $z>7$.
    }
    \label{fig:fesc-z}
\end{figure*}

\begin{table*}
    \centering
    \begin{threeparttable}[t]
\begin{tabular}{cccccccccc}
    \hline
    ID & $z_{\rm spec}$ & RA & Dec & $M_{\rm UV}$ & EW [\oiii{}]+\hb{} & $\tau_V$ & EW \lya{} & \fesc{} & \fesc{} \\
    & & (deg) & (deg) & (mag) & (\AA{}) &  & (\AA{}) & (Case B) & (Case B, corrected)\tnote{a}\\ 
    \hline
CEERS-1912   & 5.11  & 215.0108337 &  53.0133278 & $-19.8_{-0.1}^{+0.1}$  & $1484_{- 197}^{+ 202}$  & $0.153_{-0.052}^{+0.029}$  & $<285$  & $<0.62$ & $<0.67$\\[1pt]
CEERS-82052  & 5.16  & 214.7665608 &  52.7822694 & $-17.4_{-1.3}^{+0.6}$  & $1338_{- 385}^{+ 356}$  & $0.009_{-0.007}^{+0.030}$  & $<311$  & $<1.31$ & $<1.33$\\[1pt]
CEERS-4210   & 5.26  & 215.2372075 &  53.0610861 & $-20.5_{-0.5}^{+0.4}$  & $1071_{-  52}^{+  60}$  &                       --   & $< 64$  & $<0.16$ & $<0.22$\\[1pt]
CEERS-81022  & 5.30  & 214.8004492 &  52.7488889 & $-18.7_{-0.1}^{+0.1}$  & $1007_{- 451}^{+ 449}$  & $0.013_{-0.010}^{+0.037}$  & $<128$  & $<1.36$ & $<1.50$\\[1pt]
CEERS-2168   & 5.66  & 215.1526021 &  53.0570611 & $-20.3_{-0.2}^{+0.1}$  & $1265_{- 116}^{+ 130}$  &                       --   & $< 84$  & $<0.53$ & $<0.68$\\[1pt]
CEERS-83592  & 5.73  & 214.9566929 &  52.8337806 & $-18.8_{-0.1}^{+0.1}$  & $1041_{- 334}^{+ 452}$  & $0.007_{-0.005}^{+0.021}$  & $<180$  & $<0.96$ & $<1.09$\\[1pt]
DDT-2315     & 5.75  & 214.9191721 &  52.8935000 & $-19.8_{-0.0}^{+0.0}$  & $1090_{- 213}^{+ 272}$  & $0.073_{-0.062}^{+0.043}$  & $<227$  & $<0.84$ & $<0.84$\\[1pt]
CEERS-397    & 6.01  & 214.8361971 &  52.8826917 & $-21.0_{-0.0}^{+0.0}$  & $1968_{- 243}^{+ 269}$  & $0.008_{-0.006}^{+0.023}$  & $< 18$  & $<0.05$ & $<0.11$\\[1pt]
CEERS-362    & 6.05  & 214.8126892 &  52.8815361 & $-18.5_{-0.1}^{+0.1}$  & $1198_{- 384}^{+ 394}$  & $0.008_{-0.006}^{+0.024}$  & $<174$  & $<1.08$ & $<1.23$\\[1pt]
CEERS-81063  & 6.08  & 214.7991100 &  52.7251194 & $-19.1_{-0.1}^{+0.1}$  & $1311_{- 329}^{+ 371}$  & $0.050_{-0.043}^{+0.044}$  & $< 88$  & $<0.36$ & $<0.45$\\[1pt]
CEERS-1518   & 6.11  & 215.0068021 &  52.9650417 & $-21.2_{-0.1}^{+0.1}$  & $1446_{-  63}^{+  70}$  &                       --   & $< 19$  & $<0.10$ & $<0.20$\\[1pt]
CEERS-1065   & 6.19  & 215.1168542 &  53.0010806 & $-20.0_{-0.4}^{+0.3}$  & $1688_{- 199}^{+ 238}$  &                       --   & $<157$  & $<0.37$ & $<0.44$\\[1pt]
CEERS-81068  & 6.27  & 214.8205071 &  52.7371472 & $-18.9_{-0.1}^{+0.1}$  & $3137_{- 813}^{+ 460}$  & $0.006_{-0.004}^{+0.015}$  & $<170$  & $<0.47$ & $<0.52$\\[1pt]
DDT-663      & 6.52  & 214.8789692 &  52.8967472 & $-20.1_{-0.0}^{+0.0}$  & $1027_{- 242}^{+ 191}$  & $0.004_{-0.003}^{+0.009}$  & $<101$  & $<1.85$ & $<1.85$\\[1pt]
CEERS-1414   & 6.70  & 215.1280287 &  52.9849361 & $-20.9_{-0.0}^{+0.0}$  & $2657_{- 270}^{+ 268}$  & $0.010_{-0.008}^{+0.114}$  & $< 71$  & $<0.08$ & $<0.12$\\[1pt]
CEERS-717    & 6.93  & 215.0814058 &  52.9721806 & $-21.5_{-0.1}^{+0.1}$  & $1178_{-  69}^{+  74}$  &                       --   & $< 15$  & $<0.12$ & $<0.25$\\[1pt]
CEERS-1143   & 6.93  & 215.0770062 &  52.9695056 & $-20.2_{-0.3}^{+0.2}$  & $3223_{- 229}^{+ 269}$  &                       --   & $< 74$  & $<0.09$ & $<0.12$\\[1pt]
DDT-445      & 6.98  & 214.9416108 &  52.9291306 & $-19.3_{-0.1}^{+0.1}$  & $2061_{- 564}^{+1450}$  & $0.021_{-0.017}^{+0.062}$  & $<214$  & $<0.51$ & $<0.51$\\[1pt]
CEERS-80432  & 7.47  & 214.8120558 &  52.7467472 & $-20.0_{-0.1}^{+0.1}$  & $1959_{- 320}^{+1250}$  & $0.009_{-0.007}^{+0.025}$  & $< 68$  & $<0.20$ & $<0.29$\\[1pt]
    \hline
\end{tabular}

\begin{tablenotes}
\item[a] The \fesc{} with prism \lya{} flux corrections, where the correction factors are computed using the \lya{} mock observations as shown in Figure~\ref{fig:ew_simulation}. Specifically, for each source without a \lya{} detection, we adopt the correction factor derived from the mock spectra at the same continuum magnitude ($M_{\rm UV}$) that has the recovered EW the same as the measured EW upper limit.

\end{tablenotes}
\end{threeparttable}
    \caption{The galaxies lacking \lya{} detections from our CEERS spectroscopic sample and with EW [\oiii{}]+\hb{} $>$ 1000~\AA{} that are included in our escape fraction analysis.
    Here, the EW [\oiii{}]+\hb{} are derived by SED fitting to the NIRCam photometry adopting our default \beagle{} models described in \S \ref{sec:sed} for sources with NIRCam coverage, and directly from prism for those not observed with NIRCam but with rest-frame optical continuum detected.
    We report the source IDs, spectroscopic redshifts, coordinates, absolute UV magnitudes, the V-band optical depths ($\tau_{V}$) and the EW [\oiii{}]+\hb{}  derived from SED fitting, and the $7\sigma$ upper limits for \lya{} EW and \fesc{} (assuming case B recombination).}
    \label{tb:noLAE}
\end{table*}

\jwst{} has recently opened a new window on \lya{} emitting galaxies in the reionization era (e.g., \citealt{Bunker2023,Jones2023,Jung2023,Saxena2023,Saxena2023b,Tang2023}). 
Initial investigations have demonstrated that many of the strongest known \lya{} emitters at $z\gtrsim 7$ are not effective leakers of \lya{} radiation, with typical \lya{} escape fractions  (0.073) that suggest significant attenuation relative to similar systems at $z\simeq 0.3-3$ \citep{Bunker2023,Tang2023}.  
It is not clear whether the line photons are mostly scattered by \hi{} in the galaxy or in the surrounding IGM. 
More recent work has begun to uncover the first $z\gtrsim 7$ galaxies with very large \lya{} escape fractions ($\gtrsim 0.3$, \citealt{Jung2023,Saxena2023,Saxena2023b,Tang2023}), leaving little room for attenuation from the IGM. 
These \lya{} emitters potentially provide signposts of large ionized bubbles expected around overdensities of faint galaxies. 

Future NIRSpec observations promise to build on the early work described above, delivering the distribution of \lya{} escape fractions (as a function of galaxy properties) at $z\gtrsim 7$. 
If we are to robustly link these measurements to quantitative constraints on the IGM ionization state, we need knowledge of the distribution of \lya{} escape fractions in similar galaxies just after reionization, where the impact of the damping wing from the neutral IGM is less important.  
Our primary goal in this section is to use our CEERS NIRSpec prism database to begin investigating the \lya{} escape fractions at the tail end of reionization. 
We seek to determine what fraction of galaxies have very high \lya{} escape fractions at $z\simeq 5-7$ (similar to the galaxy reported in \citealt{Saxena2023}) and to quantify how this fraction evolves with redshift into the reionization era.

Our CEERS prism database consists of 69  galaxies with confident rest-optical emission line redshifts at $5\leq z<7$. 
As reported in \S\ref{sec:data_sample},  10 of these systems have \lya{} detections in the prism spectra (Figure~\ref{fig:spec}). 
The implied EWs are often quite large,  with 7 in excess of 70~\AA\ and a further 3 in the range 100--300~\AA, comparable to the most extreme systems uncovered in ground-based surveys \citep[e.g.,][]{Trainor2015,Hashimoto2017,Hashimoto2017_muse,Maseda2018,Ning2020,Vanzella2020,Matthee2021,Kerutt2022,Torralba-Torregrosa2023}. 
These values approach the maximum intrinsic \lya{} EWs expected from stellar population synthesis models for very young and metal poor stellar populations (200 to $>$600~\AA; Figure~\ref{fig:ew_model}),  suggesting a significant fraction of the line is likely to have been transmitted through both the galaxy and IGM.  
The majority of strong \lya{} emitters have intense [\oiii{}]+\hb{} emission (Figure~\ref{fig:ew_ew_o3hb}), with values largely following the trend between the two quantities seen in lower redshift samples \citep{Du2020,Tang2021}. 
Here we see that the typical \lya{} EWs start to increase above [\oiii{}]+\hb{} EW = 1000~\AA, with the majority having \lya{} EWs between 50 and 300~\AA\ at all redshifts considered.

To infer the \lya{} escape fractions in the $z\simeq 5-7$ galaxies, we first quantify the intrinsic \lya{} luminosity passing through the NIRSpec microshutter. 
We calculate the intrinsic \lya{} luminosity from the dust-corrected H$\alpha$ flux and basic recombination assumptions.
In CEERS-1334, we use the dust-corrected \hb{} flux instead owing to its anomalously high \ha{} flux.
We also use the dust-corrected \hb{} flux for the two $z > 7$ galaxies in our sample,  because the \ha{} emission line falls outside the prism wavelength coverage. 
We note that this definition of the \lya{} escape fraction will focus on the transmission through the microshutter, as is common when \lya{} escape fractions are calculated with slit-based spectrographs \citep[e.g.,][]{Begley2023,Bunker2023,Roy2023,Saxena2023b,Tang2023}.
In this paper, we will primarily focus on the evolution of \lya{} escape fractions derived self-consistently from NIRSpec MSA observations at $z\gtrsim 5$.
While potential corrections for \lya{} scattered into a diffuse halo may be somewhat more important when comparing to ground based observations at lower redshifts (given the slightly larger slits),  such comparisons are not central to this paper.

We first consider case B recombination where the intrinsic line ratio \lya{}/\ha{} (\lya{}/\hb{}) = 8.6 (24.6) at gas temperature $T_e$ = $10^4$ K and electron density $n_e$ = 250 cm$^{-3}$ using the {\sc Pyneb} \citep{Luridiana2015} package with the \cite{Storey1995} atomic data. 
Alternative assumptions on gas properties ($T_e$ =5, 000--20,000 K and $n_e$ = 100--1000 cm$^{-3}$) only change the \lya{}/\ha{} (\lya{}/\hb{}) ratios by $\sim$ 10 percent, with larger ratios at higher $T_e$ and larger $n_e$. 
However, we note that the assumption of case B recombination with optically thick \hii{} regions is not always valid, in particular for galaxies with substantial leakage of \lya{} or Lyman continuum photons \citep[e.g.,][]{Gazagnes2020}. 
Given the large \lya{} EWs quoted above, this may be relevant for a subset of our sample.
Here the escape fraction may be better estimated by case A recombination \citep{Osterbrock2006}, where the intrinsic  \lya{}/\ha{} (\lya{}/\hb{}) ratio is 12.0 (33.7) for $T_e$ = $10^4$ K and $n_e$ = 250 cm$^{-3}$ gas. 
This is a factor of 1.4 larger than that of case B, resulting in smaller \lya{} escape fractions for a given \lya{} and H$\alpha$ flux when case A is adopted. 
Unless otherwise stated, we will focus on the case B escape fractions in the following.
For galaxies lacking \lya{} detections, we place 7$\sigma$ upper limits on \fesc{},  combining the \lya{} flux upper limits derived in \S \ref{sec:lya_det} and the dust-corrected \ha{} or \hb{} flux.
 
The derived \lya{} escape fractions are reported in Table~\ref{tb:lya} for the sources with \lya{} detections.
As expected, the results confirm the presence of large \lya{} escape fractions in galaxies with large EWs. 
The median value in our $z\simeq 5-7$ \lya{} emitting galaxies is \fesc{} = 0.28, well above typical values of UV-selected galaxies at lower redshifts \citep{Hayes2011}. 
If we instead adopt the case A value, we find \fesc{} decreases by about 30\% (median \fesc{} decreases to 0.20), but the values still indicate that significant transmission of \lya{} photons is fairly common at $z\simeq 5-7$. 
One of our goals was identifying galaxies similar to the strong \lya{} emitter with near unity  \fesc{} identified by \citet{Saxena2023} at $z\simeq 7.3$.
Our sample contains four \lya{} emitting galaxies at $5\leq z < 7$ with $\gtrsim$ 50\% \lya{} escape fractions. 
Two of these systems (CEERS-82069 and 476) have escape fractions (and EWs) that are similar to the $z=7.3$ galaxy in \citet{Saxena2023}. 
Each of these has prominent \lya{} coupled with weak \ha{} lines in the rest-optical (Figure~\ref{fig:spec}), implying \lya{}/\ha{} ratios approaching the theoretical limit.
The  \lya{} escape fractions quoted above do not account for the impact of the low resolution prism on the recovered line flux (see \S \ref{sec:lya_det}).   To estimate the impact of this effect on our results, we apply the redshift and \muv{}-dependent flux corrections (Figure 4) to our observed \fesc{} values. We find that this typically leads to small increases (median 1.14$\times$) to the  \lya{} escape fractions described above. We provide these corrected values in Table \ref{tb:lya}.

To establish a useful baseline for comparison against $z\gtrsim 7$ samples, we need to consider the distribution of escape fractions in the galaxy population near the end of reionization.
Here, we primarily focus on the \lya{} escape fractions derived at slightly lower redshifts ($5.0\leq z < 6.8$).
The \lya{} escape fractions derived from NIRSpec MSA observations depend on both galaxy properties (i.e., dust, \hi{} covering fraction) and the neutral content of the IGM. 
To reliably link evolution in the escape fractions to the IGM, we must compare samples across cosmic time with matched galaxy properties. 
Here we attempt to do this by considering only galaxies with intense rest-optical nebular emission ([\oiii{}]+\hb{} EW $>$ 1000~\AA{} based on \beagle{} SED modeling or direct measurements from the prism spectra) and lacking significant dust attenuation ($\tau_{\rm V} < 0.7$). This has the effect of isolating sources that are moderately metal poor and dominated by very young stellar populations \citep[e.g.,][]{Tang2019,Curti2023,Endsley2023,Nakajima2023,Sanders2023_directZ,Boyett2024}. 
There is one source in our sample (CEERS-82069) that appears to have extremely young stellar populations (CSFH age = $6_{-4}^{+48}$ Myr) coupled with a much lower [\oiii{}]+\hb{} EW. As we discussed in \S\ref{sec:sed}, this source has a very blue UV slope that may suggest the emission lines may be weakened by the leakage of ionizing photons. 
To ensure we include all the young and metal-poor sources, we include this source in our sample for calculating the distribution of \lya{} escape fractions.
This leaves us 24 $5.0\leq z<6.8$ galaxies satisfying our selection criteria, including 9 \lya{} emitting galaxies (median \fesc{} = 0.25) and an additional 15 systems with \lya{} non-detections (see Table~\ref{tb:noLAE}).

To infer the \fesc{} distribution, we adopt a Bayesian approach following the method outlined in \cite{Schenker2014} and \cite{Boyett2022}.
Briefly, we assume a lognormal distribution for \fesc{} with parameters ($\mu, \sigma$) in the following functional form
\begin{equation}\label{eq:pdf}
    f(x | \mu, \sigma) = 
    \begin{cases}
    A \bigl(2\pi\sigma^2 x^2\bigr)^{-\frac{1}{2}} e^{ -\frac{(\ln (x) - \mu)^2}{2\sigma^2} }, & \text{$0<x\leq1$}.\\
    0, & \text{otherwise}.
    \end{cases}
\end{equation}
where $A= 2 / \Bigl(1 + {\rm erf} \bigl(\frac{- \mu}{\sigma\sqrt{2}}\bigr) \Bigr)$ is the normalization factor, and $x=$ \fesc{}.
Here, we have adopted a \fesc{} upper limit of 1.
For systems with derived \fesc{} (or the upper limits) $>1$, we will assume \fesc{} (or the upper limits) = 1 in deriving the distribution.
We use Bayes' Theorem to derive the posterior probability distribution of model parameters ($\mu, \sigma$) given the observed data ($D$)
\begin{equation}
    P(\mu,\sigma | D) \propto P(D | \mu, \sigma) P(\mu, \sigma), 
\end{equation}
where $P(\mu,\sigma | D)$ is the likelihood of observing the data given the model parameters, and $P(\mu, \sigma)$ are the priors.
We adopt a uniform prior on the lognormal location parameter $\mu$ from $-9.2$--0 (corresponding to median $x$ = 0.0001--1 in the linear space for the probability density function in Equation \ref{eq:pdf}), and Gaussian prior on the scale parameter $\sigma$ (centered at 0.6 with standard deviation 0.3).

For a given set of model parameters ($\mu$, $\sigma$), we compute the likelihood of observing the data considering both \fesc{} measurements and those with upper/lower limits.
For each system with \fesc{} measurements (detected both in \lya{} and Balmer emission lines), the Gaussian measurement uncertainty of \fesc{} is given by 
\begin{equation}
    P(x)_{{\scriptscriptstyle{D}}_{i}} = 
    B\bigl(2\pi \sigma_{{\scriptscriptstyle{D}}_{i}}^2\bigr)^{-\frac{1}{2}} e^{ - \frac{ (x-\mu_{{\scriptscriptstyle{D}}_{i}})^2}{2\sigma_{{\scriptscriptstyle{D}}_{i}}^2} },
\end{equation}
where $\mu_{{\scriptscriptstyle{D}}_{i}}$ and $\sigma_{{\scriptscriptstyle{D}}_{i}}$ are the measured \fesc{} and uncertainty for the $i$-th source.
The individual likelihood for each  source is thus
\begin{equation}
    P(D_i|\mu,\sigma)_{\rm det}  = \int_{0}^{1} P(x)_{{\scriptscriptstyle{D}}_{i}} \cdot P(x|\mu,\sigma) {\rm d}x.
\end{equation}
For sources not detected in \lya{} and with \fesc{} upper limits, the individual likelihood is given by
\begin{equation}
    P(D_i|\mu,\sigma)_{\rm upper\  limit} = P( x < x_{7\sigma} | \mu,\sigma),
\end{equation}
where the $x_{7\sigma}$ corresponds to the observed \fesc{} $7\sigma$ upper limit.
Similarly, in the case of \fesc{} lower limits due to non-detection of Balmer emission lines (e.g., CEERS-80239), the individual likelihood is \begin{equation}
    P(D_i|\mu,\sigma)_{\rm lower\  limit} = P( x > x_{7\sigma} | \mu,\sigma),
\end{equation}
where the $x_{7\sigma}$ corresponds to the observed \fesc{} $7\sigma$ lower limit. 
The likelihood over the entire dataset is thus taken as the product of the individual likelihood of each source, including both \fesc{} measurements and those with \fesc{} upper/lower limits.

To derive the \lya{} escape fraction distribution at $5.0\leq z<6.8$, we adopt the Markov chain Monte Carlo approach to sample the posteriors of the model parameters ($\mu$, $\sigma$) with the {\sc Emcee} package \citep{Foreman-Mackey2013}.
We first consider the raw \lya{} escape fractions derived from the prism spectra, and then we will consider the distribution changes if we make flux corrections accounting for the impact of the low resolution prism on the line recovery (Figure \ref{fig:ew_simulation}).
Considering the raw escape fractions, we compute the marginalized posterior distributions of lognormal parameters, finding $\mu = -1.92_{-0.36}^{+0.37}$ and $\sigma = 1.10_{-0.16}^{+0.21}$.
In what follows, we define the percentage of the galaxies in this sample with moderately high escape fractions of \lya{} photons ( \fesc{} $>$ 0.2) as $X$(\fesc{}$>0.2$). 
This analysis demonstrates that $36_{-10}^{+11}\%$ of the high-EW ([\oiii{}]+\hb{} $>$ 1000~\AA{}) $z\simeq 6$ systems have \fesc{}$>0.2$.  If we apply the flux corrections to the prism 
measurements described above (see Figure \ref{fig:ew_simulation}), we find a slightly larger fraction of the sample ($50_{-11}^{+11}\%$) has  \lya{} escape fractions in excess of 0.2 (lognormal parameters $\mu = -1.53_{-0.30}^{+0.34}$, $\sigma = 1.02_{-0.17}^{+0.20}$).

Lower redshift samples of similarly selected galaxies also show large escape fractions of \lya{} emission \citep[e.g.,][]{Henry2015,Izotov2016,Yang2017,Tang2021,Flury2022_II}.
Applying a similar selection ([\oiii{}]+\hb{} EW $>$ 1000~\AA) to the $z\sim0.3$ Green Pea sample of \citet{Yang2017}, we derive the \fesc{} distribution for the resulting sample following our method above.
We obtain a lognormal location parameter $\mu = -1.50_{-0.36}^{+0.44}$ and scale parameter $\sigma = 1.30_{-0.15}^{+0.18}$, which suggests significant leakage of \lya{} photons (\fesc{} $>$ 0.2) in  $46_{-9}^{+10}\%$ of the $z\sim0.3$ sample. 
In addition, at $z\sim$ 2, we find \fesc{} lognormal distribution with $\mu = -1.89_{-0.41}^{+0.42}$ and $\sigma = 0.78_{-0.20}^{+0.22}$ with the \cite{Tang2021} EELGs satisfying the same [\oiii{}]+\hb{} EW cut.
The modeled \fesc{} distribution translates into a fraction of  $34_{-16}^{+20}\%$ of the EELGs that show \fesc{} $>$ 0.2. 
This suggests that the fractions of large escape fraction systems in $z\simeq 0.3-2$ samples are broadly comparable to what we have found using NIRSpec at $z=$ 5.0--6.8 (see Figure \ref{fig:fesc-z}).

We now seek to determine if existing constraints on the \lya{} escape fraction at $z\gtrsim 7$ reveal evolution with respect to our benchmark measurement at $z\simeq 5.0$--6.8.
Here we build on the $z>7$ spectroscopic catalog assembled in \citet{Tang2023} (see also \citealt{Jung2023,Larson2023}), adding the 10 new galaxies we confirmed at $z>7$ in this paper (see \S \ref{sec:lya_det}). 
The $z>7$ database contains 10 galaxies with [\oiii{}]+\hb{} EW $>$ 1000~\AA{}.
We also include two galaxies that lack NIRCam SEDs but have very large O32 ratios ($>5$) that are often linked to large [\oiii{}]+H$\beta$ EWs (see \citealt{Tang2023}).
(We note that our main $z>7$ results do not change by including these sources.)
This results in 12 $z>7$ galaxies that satisfy our selection.
We note that the \fesc{} derived in \citet{Tang2023}  assumed zero dust attenuation based on Balmer decrements from composite spectra.
For consistency with our $z\simeq 5-6$ measurements, we apply small dust corrections to the $z\gtrsim 7$ values following the same method described in \S \ref{sec:sed}.
This results in a median 14\% decrease of the original \lya{} escape fractions, as the predicted intrinsic \lya{} luminosity is modestly larger following the dust correction. 
The derived lognormal \fesc{} distribution ($\mu = -2.96_{-0.43}^{+0.40}$ and $\sigma = 1.06_{-0.19}^{+0.21}$) is markedly different from that at $z\simeq 5-6$.

The results demonstrate that large escape fractions (\fesc{} $>$ 0.2) are only found in $9_{-5}^{+8}$\% of the $z>7$ systems in CEERS.
Considering the corrections to the prism-based \fesc{} values, we find this high-\fesc{} fraction is $10_{-5}^{+9}$\% (lognormal parameters $\mu = -2.94_{-0.46}^{+0.40}$, $\sigma = 1.07_{-0.18}^{+0.21}$).
This fraction is lower than that at $z\simeq 0.3$, $z\simeq 2$, and $z\simeq 5-6$ (see Figure \ref{fig:fesc-z}), as would be expected given the inferred neutral fraction of the IGM at $z>7$ \citep{Mesinger2015,Zheng2017,Mason2018,Mason2019,Hoag2019,Whitler2020,Bolan2022}. 
Comparison between the $z\sim 6$ and $z>7$ samples admittedly faces limited statistics and uncertainties in the prism-based EWs, but the data are nonetheless consistent with a significant decline in \fesc{} toward earlier times (Figure \ref{fig:fesc-z}). 
Furthermore, in \S\ref{sec:discussion}, we will show that the CEERS EGS field may have large ionized sightlines that amplify \lya{} relative to the global average at $z\simeq 7$.  
In this sense, the downturn in the \lya{} escape fractions is likely to be greater in other deep fields.
As sample sizes increase at $z\gtrsim 7$, it will be possible to build up the statistics while also considering sources matched in a wider range of galaxy properties.

\section{\lya{} Associations at \texorpdfstring{$\lowercase{z}>7$}{z>7} in the EGS Field}\label{sec:bubble}

\begin{figure*}
    \centering
    \includegraphics[width=\columnwidth]{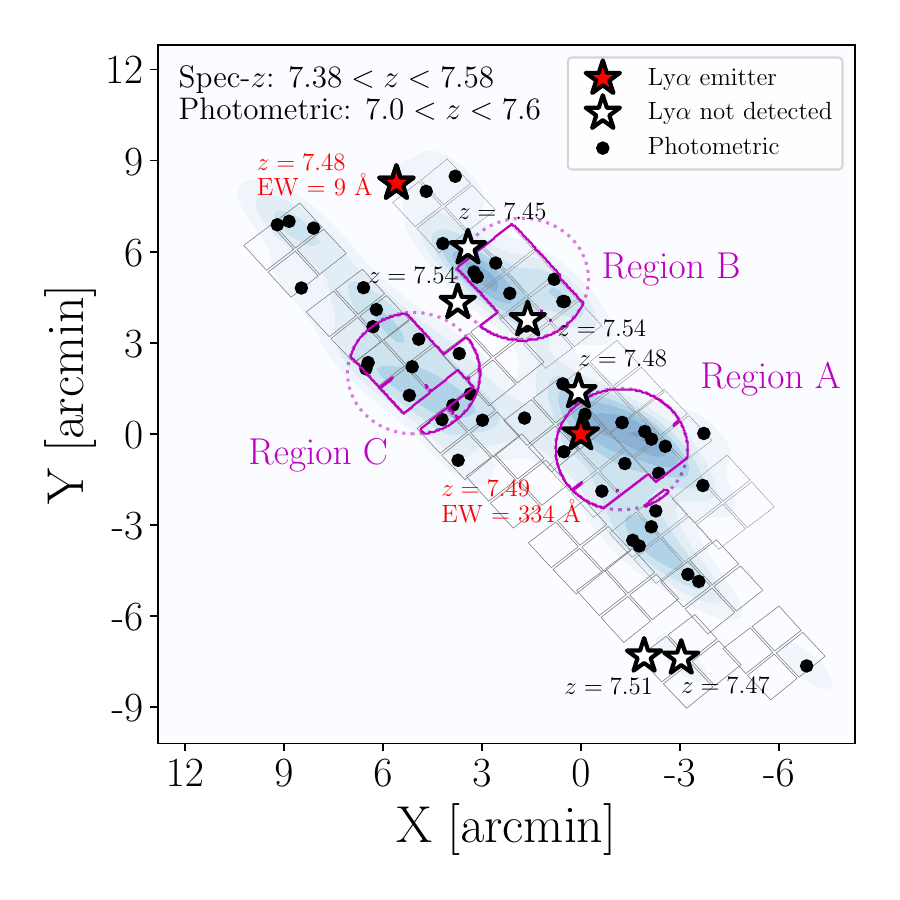}
    \includegraphics[width=\columnwidth]{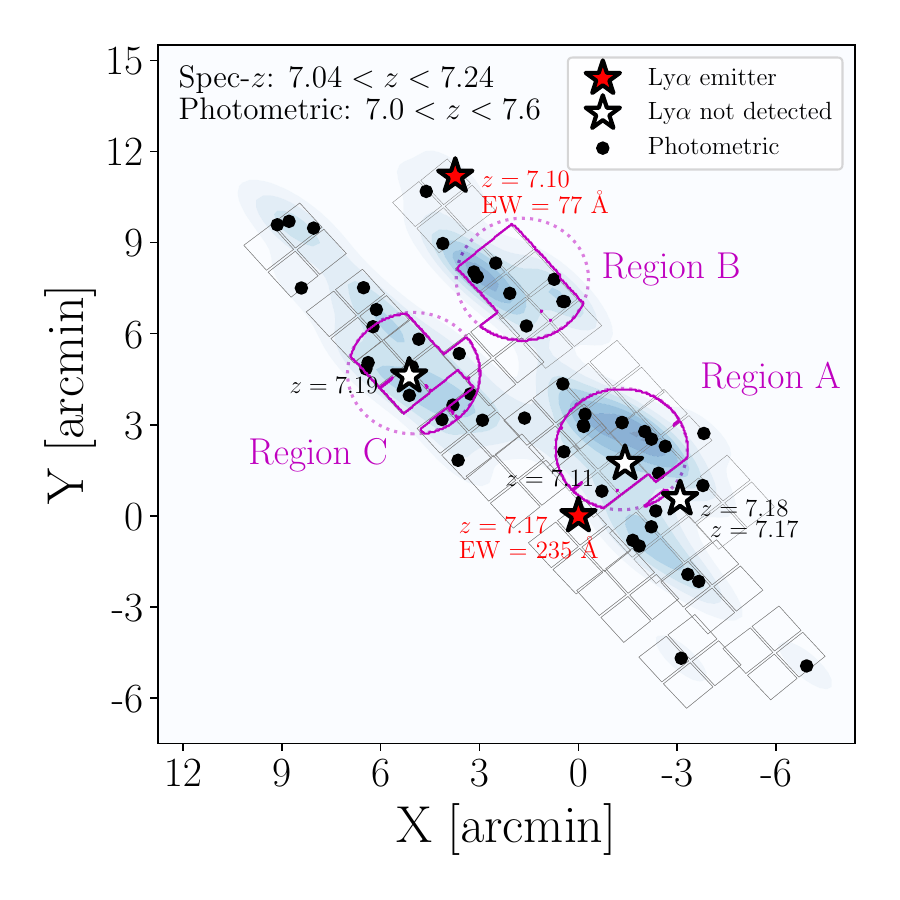}
    \caption{Spatial distribution of galaxies around the two \lya{} emitting galaxies at $z=7.17$ and $z=7.49$ in our sample.
    We place each \lya{} emitting galaxy at the origin (0,0).
    We include the spectroscopically confirmed galaxies from CEERS observations (this work and \citealt{Tang2023}) and ground-based observations \citep{Roberts-Borsani2016,Stark2017}.
    The \lya{} emitting galaxies and galaxies not detected in \lya{} are shown as filled red stars and open stars, respectively.
    We overplot the NIRCam-selected galaxies with similar photometric redshifts (based on F410M color excess) in black dots over the entire CEERS NIRCam footprint (gray), with the blue scale colors in the background indicating the corresponding Gaussian kernel density (higher densities at deeper blue colors).
    Regions A, B, and C (magenta circles with 2~arcmin in radius) correspond to photometric overdensities at 3.1--3.7 $\sigma$ identified near the \lya{} emitting galaxies.}
    \label{fig:spatial}
\end{figure*}

\begin{figure*}
    \centering
    \includegraphics[width=1\textwidth]{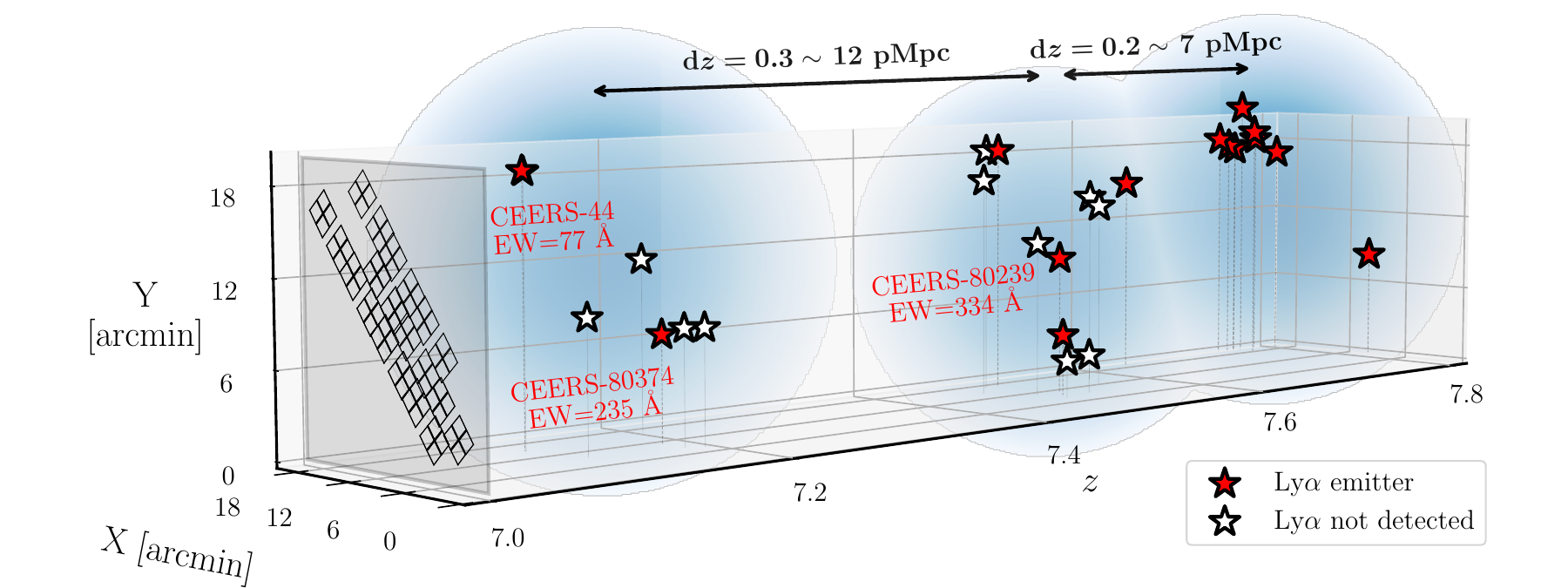}
    \caption{Spectroscopically-confirmed galaxies in the CEERS EGS field at $z=$ 7.1--7.8. The presence of numerous \lya{} emitting galaxies (red stars) in this field suggests overlapping ionized bubbles along the line of sight (shaded blue regions), but limited (mostly) to bright ground-based samples it has been challenging to robustly interpret the observations. Deep \jwst{} spectroscopy of fainter  NIRCam-selected galaxies in this field would allow a tomographic map of ionized bubbles to be mapped around the spatial distribution of galaxies in the field.    }
    \label{fig:3dbubbles}
\end{figure*}

We have demonstrated in \S\ref{sec:fesc-z} that large \lya{} escape fractions and \lya{} EWs are relatively common in faint $z\simeq 6$ galaxies, with \fesc{}$>$0.2  presented in 50\%  of those with [\oiii{}]+\hb{} EW$>$1000~\AA{}.
At $z\gtrsim 7$, we expect high \fesc{} systems to become less common as the IGM grows more neutral, with detections mostly limited to galaxies in large ionized regions of the IGM. 
The discovery of \lya{} emission in numerous  $z\simeq 7.5-9$ galaxies in the EGS \citep{Oesch2015,Zitrin2015,Roberts-Borsani2016,Stark2017,Tilvi2020,Jung2022,Larson2022,Cooper2023} with ground-based telescopes has long hinted at the possible presence of several ionized bubbles in the field. 
The first epoch of CEERS NIRSpec spectroscopy targeted a number of these previously-known \lya{} emitters, generally revealing very low escape fractions \citep{Jung2023,Tang2023} that indicate  \lya{} photons face significant attenuation. 
This may suggest that these systems are situated in relatively small ionized bubbles (with significant IGM attenuation) or it may reflect large column densities of \hi{} on galaxy scales. 
As these first galaxies targeted tend to be very luminous (and fairly massive) and often with large \lya{} velocity offset \citep[e.g.,][]{Tang2023,Bunker2023} implying significant scattering of \lya{} photons, it is perhaps likely that the \hi{} within the galaxy is largely responsible for reducing the measured escape fraction through the NIRSpec microshutter. 
In this case, the \lya{} emitters could still trace large ionized structures where the IGM attenuation is minimal.  

The more recent CEERS NIRSpec observations have targeted fainter $z\gtrsim 7$ galaxies identified in NIRCam images of the EGS field \citep[e.g.,][]{ArrabalHaro2023b}. 
In this paper, we present likely detections of \lya{} in two of these galaxies, CEERS-80239  ($z=7.49$) and CEERS-80374 ($z=7.10$).
Because of the faint continuum of both galaxies (\muv{} = $-18.2$ and $-18.5$, respectively), both detections require extremely large EW  \lya{} emission ($334_{-62}^{+109}$~\AA{} and $205_{-27}^{+48}$~\AA{}). 
These large \lya{} EWs suggest these galaxies are situated in ionized regions that facilitate significant transmission of \lya{} through the IGM (also see \citealt{Witstok2023}). 
In one of these systems (CEERS-80374), we are able to measure the \lya{} escape fraction (\fesc{} = $0.472_{-0.048}^{+0.066}$) via detection of \hb, indicating that at least $\sim 50$\% of the \lya{} photons are transmitted through the ISM and IGM (with a larger IGM transmission implied if some of the attenuation occurs in the ISM and CGM of the galaxy).\footnote{In the other galaxy, CEERS-80239, we do not detect \hb{} given its faintness (see Figure~\ref{fig:spec}). Nonetheless, a large escape fraction (\fesc{} = $0.56_{-0.17}^{0.31}$) is implied by comparing the observed \lya{} EW with the value derived from \beagle{} photoionization modeling (\S \ref{sec:sed}) .} 
To achieve this large \fesc{} may require that the nearest patch of mostly neutral IGM be located at least 1--2 physical~Mpc (pMpc) away. 
Here, we follow the prescription detailed in \cite{Mason2020_bubble} to compute the \lya{} transmission assuming a range of ionized bubble sizes and fully neutral IGM outside the bubble.
Specifically, for a source observed at redshift $z_{\rm s}$ with photons observed at wavelength $\lambda_{\rm obs} = \lambda_{\rm em}(1+z_{\rm obs})$, the corresponding \lya{} optical depth is
\begin{equation}\label{eq:tau_lya}
    \tau_\alpha(\lambda_{\rm obs}) = \int_{z_{\rm obs}}^{z_{\rm s}} {\rm d}z c \frac{{\rm d}t}{{\rm d}z} x_{\rm HI}(z) n_{H}(z) \sigma_{\alpha}(\frac{\lambda_{\rm obs}}{1+z},T),
\end{equation}
where $n_{\rm H}$ is the total hydrogen number density, $x_{\rm HI}$ is the hydrogen neutral fraction, and $\sigma_\alpha$ is the \lya{} scattering cross-section through an ensemble of hydrogen atoms.
To estimate the bubble size required for a given \lya{} escape fraction, we have assumed the \lya{} velocity offset is 100--200 km~s$^{-1}$, a typical range for faint galaxies with strong \lya{} \citep{Saxena2023}.\footnote{While both sources have rest-optical emission line detections which establish the systemic redshift, they are observed with the $R=100$ prism which precludes robust \lya{} velocity offset measurements. 
High-resolution ($R>1000$) NIRSpec spectroscopy would be required to measure velocity offset $<$ 130 \kms{}.} 
Adopting slightly different assumptions will alter the values quoted above, but it is unlikely to change the fact that these two galaxies likely must sit in large ionized regions to facilitate observation of 
\lya{} with EW in excess of 200-300~\AA.\footnote{Very large velocity offsets could conceivably allow \lya{} to escape in smaller bubbles with less IGM attenuation. 
But such large velocity offsets are not expected in galaxies with such large \lya{} EWs \citep[e.g.,][]{Erb2014,Nakajima2018,Cassata2020,Tang2021,Prieto-Lyon2023}.}

The next step is identifying the galaxy population responsible for carving out the ionized regions that are facilitating such efficient \lya{} escape. 
Given that both CEERS-80374 and CEERS-80239 are very faint continuum sources (and unlikely to create significant bubbles on their own), it is probable that they are part of overdense structures that are responsible for creating the ionized region that is enabling efficient \lya{} escape. 
We first search for spectroscopically-confirmed galaxies at the redshifts of both galaxies. 
A map is shown in Figure~\ref{fig:spatial}.  
We note that the new \lya{} emitter CEERS-80239 ($z=7.49$) is at a similar redshift as CEERS-698 ($z=7.469$), an extremely UV-luminous (\muv{} = $-21.86$) galaxy shown previously to have \lya{} emission with Keck \citep{Roberts-Borsani2016,Stark2017}. 
Taking the systemic redshifts of both, we find the two are separated by only 3.1~pMpc. 
CEERS NIRSpec observations also confirm an additional six $z=7.45-7.54$ galaxies between CEERS-698 and CEERS-80239, hinting at an extended structure that could be contributing to the ionized bubble. 
The other new \lya{} emitter we identify (CEERS-80374) is at a  similar redshift as CEERS-44 (separated by 4.7~pMpc), another strong \lya{} emitter (EW=77\AA) in the same field \citep{Tang2023}.
CEERS-44 also shows a large \lya{} escape fraction (\fesc{} = $0.339\pm0.044$), perhaps suggesting reduced IGM attenuation to \lya{} photons at $z\simeq 7.1-7.2$ across the field. 
Another four galaxies with very similar spectroscopic redshifts and close separations (0.7--3.9~pMpc) from CEERS-80374 and CEERS-44 have been identified in the CEERS program (see \citealt{Tang2023} and Figure~\ref{fig:spatial}).
We present these spectroscopically confirmed galaxies at each \lya{} emitting galaxy redshift in Table~\ref{tb:tb_neighbor_specz}.
Unfortunately, their \lya{} line either falls outside the detector or has upper limits on \lya{} EW ($<$185~\AA{}), making them also potential \lya{} emitting galaxies within the same ionized regions.

The NIRCam imaging obtained with CEERS provides a more complete census of the galaxies likely to be at the same redshift as 
the two $z\gtrsim 7$ \lya{} emitters identified in this study. 
In the remainder of this section, we characterize color-selected galaxies likely to lie at $z\simeq 7.0-7.6$ and assess implications for overdense structures in the vicinity of CEERS-80374 and CEERS-80239. 
We use an updated catalog of F814W-dropout galaxies spanning the entire footprint of CEERS NIRCam observations, expanding from the catalog published in \cite{Endsley2023_ceers}.
These dropout galaxies are selected with the same color selection as in \cite{Endsley2023_ceers} but include the full database of CEERS imaging (whereas the earlier paper only included the first epoch). 
Photometric redshifts are inferred from \beagle{} SED fitting to the \hst{}/ACS+\jwst{}/NIRCam multi-band photometry.
The reader is directed \cite{Endsley2023_ceers} for more details.
In total, our selection yields 269 galaxies with F200W$<$28 mag at $z\sim$ 6.5--8 across the 92.1~arcmin$^{2}$ CEERS footprint with overlapping ACS imaging.

Our goal is to identify galaxies in this sample that are likely to be at similar redshifts as CEERS-80374 ($z\simeq 7.17$) and CEERS-80239 ($z\simeq 7.49$). 
We adopt an empirically-motivated approach utilizing the F410M medium band photometry to isolate galaxies at $z\sim$ 7.0--7.6 (see \citealt{Endsley2023} for full details).  
In this redshift range, the [\oiii{}]+\hb{} emission lines fall in the F410M filter, leading to red colors in F356W - F410M. 
By selecting 
dropouts in our sample that are red in F356W - F410M ($>0.6$), we are likely to pick out those galaxies with large EW [\oiii{}]+\hb{} ($>$ 400~\AA{}).  
Given the [\oiii{}]+\hb{} EW distribution in CEERS \citep{Endsley2023_ceers}, this corresponds to 82\% of the population at these redshifts. 
This color cut results in a sample of 54 galaxies likely to lie at $z\sim$ 7.0--7.6 based on the F410M excesses (see Table \ref{tb:tb_f410m}). 
The \beagle{} SED fits described above indicate that the photometric redshifts of all 54 sources lie in the desired redshift range.  

We show the distribution of photometric targets in Figure~\ref{fig:spatial}. 
We identify three regions (region A, B, and C) in the vicinity of CEERS-80374 and CEERS-80239 with high surface densities of $z\sim$ 7.0--7.6 galaxies.
Each of the two regions contains 9--12 photometric galaxies within a radius of 2~arcmin (0.6 projected~pMpc), corresponding to a surface area of 7.5--10.9~arcmin$^{2}$ with overlapping NIRCam imaging.
The derived surface density (1.1--1.3 galaxies /~arcmin$^2$) is $\approx 2 \times$ greater than the average over the full footprint.
In particular, region A overlaps with CEERS-80239 and is found only 0.5~arcmin away from CEERS-80374, suggesting that one (or both) of the \lya{} emitters may be associated with this overdensity. 
The other regions (B \& C) are both found $\approx$ 5.5~arcmin away from CEERS-80239, which could be expected if the structures extend to larger distances over the field.

In order to quantify the strength of these photometric overdensities, we compare the observed galaxy number with that predicted from the UV luminosity function and our selection efficiency.
Our method to compute the expected number of galaxies ($\langle N \rangle$) broadly follows \cite{Endsley2023}. 
We summarize the approach below (also see \citealt{Endsley2021,Whitler2023c}).
As a first step, we consider the selection efficiency imposed by both the dropout selection and the F356W - F410M color cut before estimating the predicted number of observable galaxies.
We start by modeling the probability of a galaxy being selected by both criteria as a function of both redshift and magnitude using a model SED with a flat rest-UV continuum in $f_\nu$ ($\beta = -2.0$).
At each redshift and magnitude, we add [\oiii{}]+\hb{} emission lines with the equivalent widths randomly drawn from the \cite{Endsley2023_ceers} distribution at $z\sim7$, assuming a fixed [\oiii{}]$\lambda$5007/\hb{} ratio = 6 (e.g., \citealt{Tang2019}).
We compute the mock photometry in each ACS+NIRCam filter, taking into account the IGM transmission model of \cite{Inoue2014} and perturbing the photometry with the typical noise estimated from the observed values for our sample.
The selection efficiency at each redshift ($z$ = 6.5--8.0 in $\Delta z$ = 0.02 unit spacing) and magnitude (F150W = 25.0--28.0 in 0.1 mag unit spacing) is then estimated as the fraction of 1000 realizations passing both our dropout criteria for $z \sim 6.5 - 8$ galaxies and the  F356W - F410M color cut.

We combine the selection efficiency with the redshift-dependent UV luminosity function of \cite{Bouwens2022} to derive the strength of the photometric overdensities.
To compute the cosmic mean galaxy number surface density, we convolve the UV luminosity function with the selection efficiency and integrate it over $z=$ 7.0--7.6 and magnitudes brighter than $M_{\rm UV} = -19$, the $M_{\rm UV}$ limit of our sample (also see \citealt{Endsley2023_ceers}).
Comparing the resulting mean surface density (0.4 galaxies /~arcmin$^2$) with what we observe in Region A, B, and C, we found the three regions correspond to overdensities at $3.1_{-0.9}^{+0.9}$, $3.1_{-1.0}^{+1.0}$, and $3.7_{-1.2}^{+1.1}$ $\sigma$, respectively.
We note that our selection includes candidate galaxies over a relatively wide redshift range ($z=7.0$--7.6). 
It is plausible each of the regions could be considerably more overdense if the redshifts of the candidate galaxies are more narrowly distributed around a given redshift.

The photometric data suggest that overdensities (in region A) may surround the two intense \lya{} emitters identified in this study. 
The potential overdensities in regions B and C suggest that there may be extended structures across the field at the redshifts of one or both of the \lya{} emitters. 
A single overdensity of this amplitude ($\sim 3 \sigma$) over a radius of $R \sim 0.6$~pMpc alone can create moderately-sized bubbles ($R_{\ion{H}{ii}} \simeq$ 0.8~pMpc) with nominal assumptions on source properties (see \citealt{Endsley2022_overdensity,Whitler2023c}).
Given the presence of three such overdensities in this field, it is conceivable that collectively they create a much larger ionized region ($R_{\ion{H}{ii}}\gtrsim$ 1--2~pMpc) as is likely required for explaining the observed escape fractions of the \lya{} emitting galaxies.
Given that the galaxies in this sample are selected on F356W-F410M colors that indicate strong [\oiii{}]+H$\beta$ emission, it should be trivial in the future to map out rest-optical emission line redshifts (with NIRSpec or NIRCam grism) and better characterize whether the galaxies in the overdense regions are associated with the $z\simeq 7.2$ or $z\simeq 7.5$ extreme \lya{} emitters.

\section{Discussion}\label{sec:discussion}

The spectroscopic capabilities of \jwst{} are ushering in a new era of investigating reionization.
With rest-optical emission line redshifts now easily accessible at $z\simeq 6-9$, it is becoming possible to fully map out the spatial distribution of galaxies in 3D across known deep fields, identifying regions that are overdense and those that are underdense. 
Deep \lya{} follow-up will in turn offer insight into which of these regions have been able to carve out large ($\gtrsim 1$~pMpc) ionized bubbles. 
In so doing, these observations provide a {\it local} view of reionization, taking a new step beyond the global reionization studies of the past two decades. 
In this section, we contrast observations of two $z\gtrsim 7$ overdensities with \lya{} measurements.

The EGS field presents one of the most interesting cases of a likely ionized region, with 16 robust \lya{} detections at 
$7<z<8$ \citep{Oesch2015,Roberts-Borsani2016,Stark2017,Tilvi2020,Jung2022,Jung2023,Tang2023,Witten2023}.
No other deep field has proven as rich in $z\gtrsim 7$ \lya{} emitters, in spite of considerable observational investment. 
We described new \lya{} detections in the EGS field in \S\ref{sec:bubble}, building the case for three associations of \lya{} emitters along the line of sight: at $z\simeq 7.2$, $z\simeq 7.5$, and $z\simeq 7.7$ (see Figure \ref{fig:3dbubbles}). 
The radial distances between these associations are 7~pMpc ($z\simeq 7.2$ and $z\simeq 7.5$) and 12~pMpc ($z\simeq 7.5$ and $z\simeq 7.7$), comparable to the diameters of large ionized bubbles expected at $z\simeq 7$ \citep{Lu2023}. 
The new discovery of \lya{} with EW$>$200~\AA\ in these associations confirms that there is likely to be large ionized sightlines ($>$1--2~pMpc) at $z\simeq 7.1-7.7$ in the EGS. 
Both photometric and spectroscopic samples are now revealing significant overdensities of galaxies in this redshift range (see \S\ref{sec:discussion}).

These observations raise the possibility that the EGS contains an extended structure oriented radially along the line of sight over part of the region spanning $z\simeq 7.2$ to $z\simeq  7.7$. 
Such filament-like structures are expected to be common topological features in the reionization era (e.g., \citealt{Elbers2019,Elbers2023}). 
If viewed along the line of sight, ionized filaments may create the optimal conditions for detecting \lya{} at $z\gtrsim 7$, providing large ionized sightlines for galaxies throughout the structure. 
In particular, galaxies at the far-end of the structure can have their \lya{} transmission significantly boosted. 
For example, we find that galaxies viewed at the far side of large ($>10$~pMpc) filaments will transmit 2.4--2.9$\times$  more \lya{} flux than if they were situated in relatively small ($R = 0.5$~pMpc) ionized bubbles. 
Here, we follow the procedures described in \citet{Mason2020_bubble} to estimate the \lya{} transmission at different sizes of the ionized region using Equation \ref{eq:tau_lya}, with the range of factors dictated by our assumed distribution of \lya{} velocity offsets (50--200 \kms{}). 
Clearly, radial-extended ionized filaments will amplify many \lya{} lines that otherwise would not be visible. 

The future prospects for characterizing the ionized regions in the EGS are good.
To date, the majority of existing \lya{} detections have come from brighter galaxies identified in  \hst{} imaging. 
The NIRCam photometric samples extend significantly fainter, providing much-improved statistics of the galaxy distribution. 
Thus far, only a very small fraction of the NIRCam samples have been followed up with spectroscopy (fewer than 6\% of the galaxies reported in \S\ref{sec:bubble} have NIRSpec observations, and none of them have $R\geq 1000$ spectra required for bubble size inferences). 
Once a larger subset of them have rest-optical spectra, it will be possible to map the 3D spatial distribution of galaxies through the field. 
Such a dataset will reveal whether overdensities are peaked near the known \lya{} emitters (as may be expected if we are observing a series of bubbles along the line of sight) or whether the entire redshift range spanning $z\simeq 7.1-7.7$ is uniformly overdense (as may be expected if there is a single extended ionized sightline). 
Deep $R\geq 1000$ \lya{} spectroscopy of the NIRCam-identified galaxies will enable direct tests of these scenarios. Here the fainter magnitudes of the NIRCam database are critical. 
As shown in \S\ref{sec:fesc-z}, \lya{} EWs often reach extremely large values ($>$100~\AA) in UV-faint galaxies (M$_{\rm{UV}}>-18$).
When such objects are identified at $z\gtrsim 7$ with small \lya{} velocity offsets ($<$100 km~s$^{-1}$), it immediately points to large radial ionized sightlines \citep[e.g.,][]{Mason2020_bubble,Saxena2023}. 
Comprehensive high resolution \lya{} spectroscopy of the NIRCam-selected galaxies in Figure \ref{fig:spatial} will allow these ionized regions to be mapped across the EGS.

\begin{figure}
    \centering
    \includegraphics[width=\columnwidth]{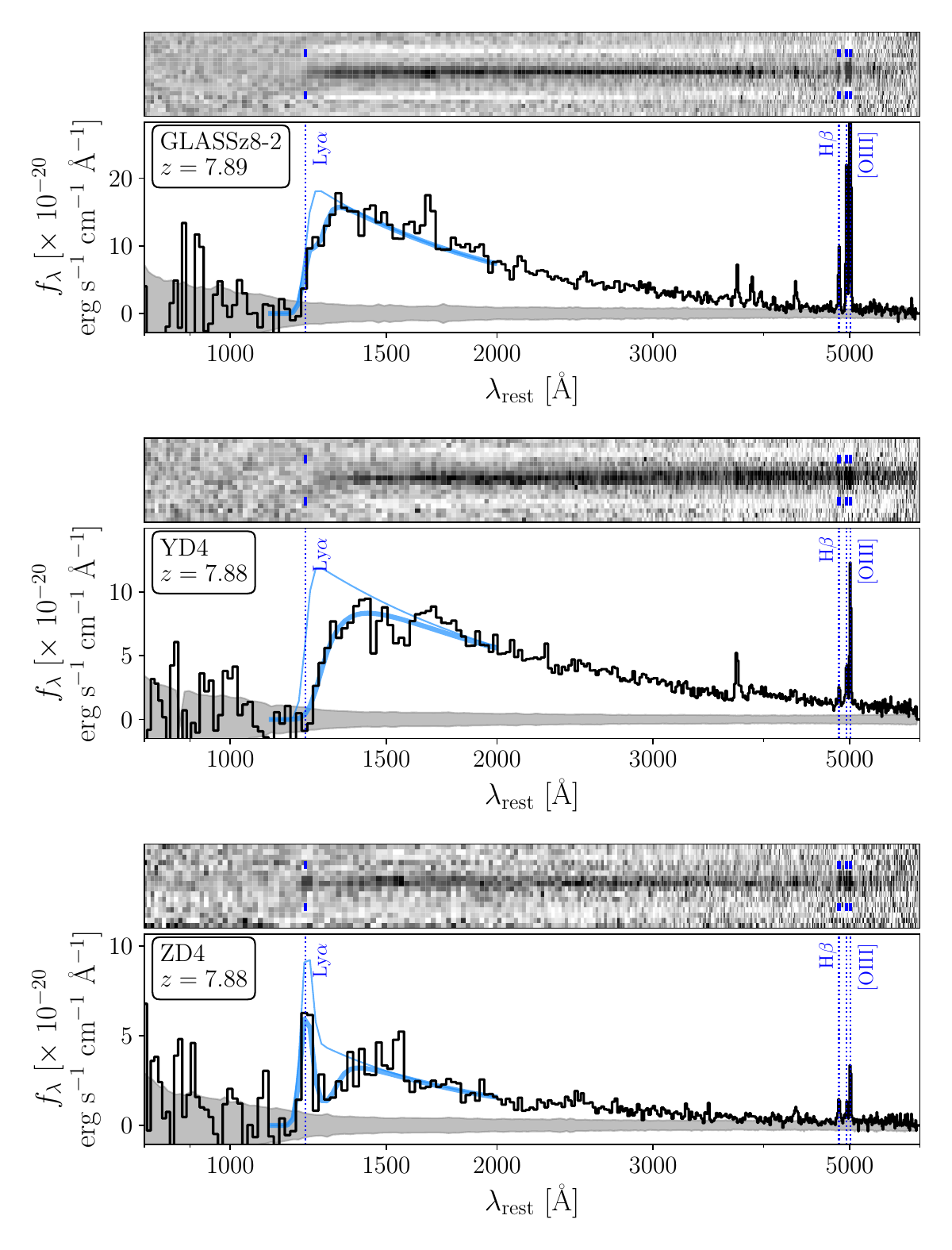}
    \caption{New \jwst{} prism spectra of three galaxies in the  $z=7.88$ overdensity confirmed in Abell 2744 by \protect\cite{Morishita2023}.
   Previous spectra have not shown \lya{} emission, suggesting the overdensity may have yet to ionize its surroundings.  
   The new spectra (from UNCOVER; \citealt{Bezanson2022}) reveal strong \lya{} in source ZD4 (bottom panel), suggesting that \lya{} is not completely absent in the region. 
   We also see evidence for damped \lya{} absorption in all sources, likely attenuating \lya{}.
   We show the intrinsic spectra prior to damped Lyman-$\alpha$ absorption (DLA) as a thin blue curve, and the damped \lya{} fit, convolved with the instrumental resolution, as a thick blue curve.
   We suggest that DLAs are likely to be common in small-scale overdensities (within projected radius $\lesssim$ 60~pkpc) like that seen in Abell 2744 given the large number of foreground galaxies. }
    \label{fig:uncover}
\end{figure}

\begin{figure*}
    \centering
    \includegraphics[width=0.8\textwidth]{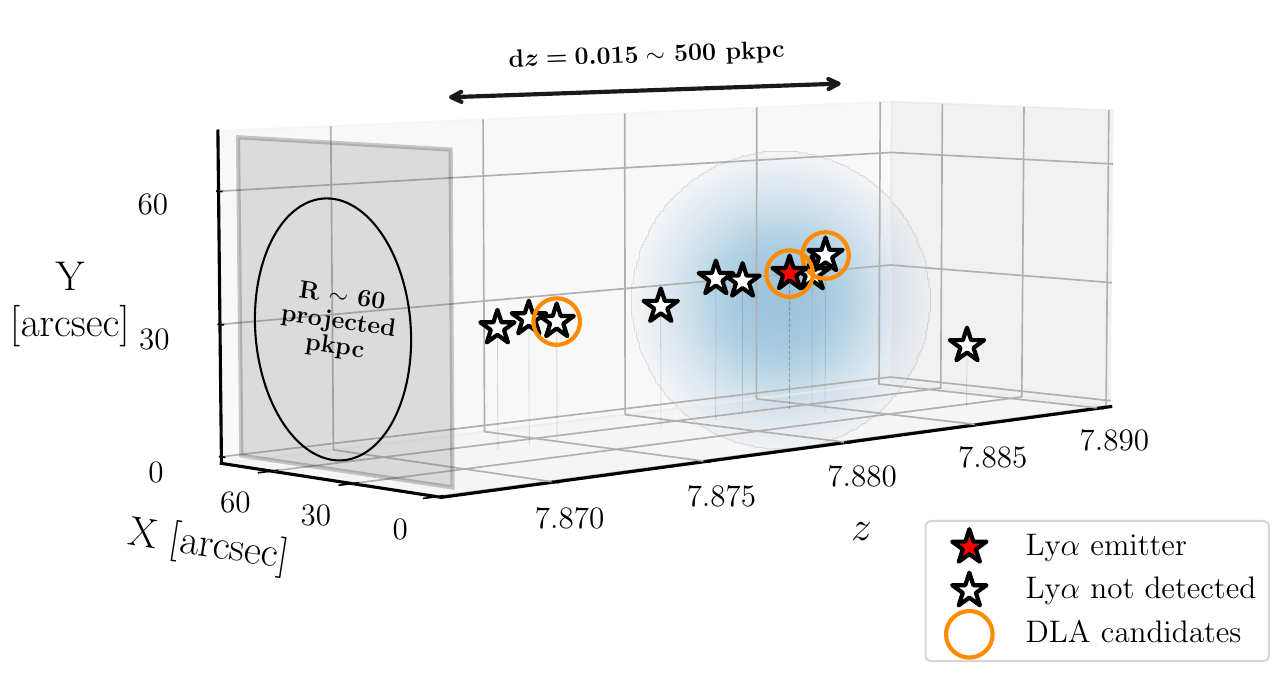}
    \caption{The 3D distribution of spectroscopically confirmed galaxies in the $z=7.88$ overdensity in the Abell 2744 field.
    We show the newly identified \lya{} emitting galaxy from  UNCOVER (ZD4) with a red filled star and the previously-confirmed galaxies lacking \lya{} detection from \protect\cite{Morishita2023} and \protect\cite{Hashimoto2023} with open stars. 
    The three galaxies observed by UNCOVER with damped \lya{} absorption are marked with large orange circles.  
    Note the close proximity of many of the galaxies within a projected radius of 60~pkpc, likely contributing to the prevalence of DLAs in the field. 
    }
    \label{fig:3dbubbles_uncover}
\end{figure*}

The connection between $z\gtrsim 7$ \lya{} visibility and galaxy overdensities is now starting to be seen in several other fields \citep[e.g.,][]{Hu2021,Endsley2023,Witstok2023,Witten2023}. 
However, one recently-confirmed overdensity at $z=7.88$ in the Abell 2744 field presents a conflicting picture \citep{Morishita2023,Hashimoto2023}.  
NIRSpec observations of this field from the GLASS (ERS-1324, PI: Treu; \citealt{Treu2022}), DDT-2756 (PI: W. Chen; \citealt{Roberts-Borsani2023}), and GO-1840 (PI: Hashimoto; \citealt{Hashimoto2023}) programs have confirmed redshifts for 8 galaxies clustered within a small region with a projected radius of 60 physical kpc (pkpc). 
The density of galaxies in this small area implies a strong galaxy overdensity ($24_{-8}^{+12} \sigma$; \citealt{Roberts-Borsani2023}), albeit over much smaller physical scales than the CEERS overdensities discussed above.
Five of the eight galaxies have \lya{} constraints from NIRSpec. None show \lya{} emission, with small reported $2\sigma$ \lya{} EW upper limits ranging from 15.9--27.6~\AA{}. 
The \lya{} non-detections raise the question of whether this dense region has been able to ionize its surroundings.

The UNCOVER program (GO-2561, PIs: I. Labb\'e \& R. Bezanson; \citealt{Bezanson2022}) has recently obtained NIRSpec prism spectra of two of the seven spectroscopically-confirmed galaxies (GLASSz8-2 and YD4) and one of eight photometric candidates (ZD4) associated with the $z=7.88$ overdensity \citep{Morishita2023}.
Following the procedures described in \S \ref{sec:spec}, we obtained  reduced spectra for the three galaxies, which are shown in Figure \ref{fig:uncover}.
The  photometric candidate ZD4 is confirmed to lie at $z=7.88$, placing it as part of the same overdensity.
In addition to strong rest-optical emission lines ([\oiii{}]+H$\beta$ EW = $794_{-220}^{+310}$~\AA), we identify an emission line feature (SNR = 5.9) at a wavelength consistent with  \lya{} in the galaxy.
We measure a relatively large \lya{} EW ($68_{-9}^{+10}$~\AA{}), suggesting the overdensity may indeed have carved out an ionized region. 
For the other two galaxies where we do not detect \lya{}, we obtain $7\sigma$ upper limits on the equivalent widths (EW $<$ 85~\AA{} for GLASSz8-2 and EW $<$ 91~\AA{} for YD4) that are consistent with the previous constraints \citep{Morishita2023}.

The UNCOVER spectra also hint at why \lya{} is not seen in many galaxies of the \citet{Morishita2023} overdensity. 
In each of the three spectra, we see evidence for damped \lya{} absorption (see Figure \ref{fig:uncover}) that will significantly attenuate any \lya{} emission that is present. 
Strong \hi{} absorption is perhaps not surprising given the close separations (within a projected radius of 60~pkpc) of many of the galaxies in the protocluster structure (see Figure \ref{fig:3dbubbles_uncover}). 
In contrast, the EGS field presents a galaxy overdensity on much larger scales ($>$1~pMpc), where the obscuring effect of DLAs is likely to be less common.
To estimate the viability of the damped \lya{} interpretation for the continuua of the UNCOVER spectra, we perform a simple analysis fitting a DLA transmission profile to the spectra over the window of 1100--2000~\AA{} in the rest frame. 
We compute the \lya{} optical depth using the cross-section approximation by \citet{Tasitsiomi2006} and assume a  continuum spectrum  (prior to absorption by the damped \lya{} system) obtained by extrapolating the power-law fit to the observed continuum over 1400--3000~\AA{} in the rest-frame.
We perform the fit in the observed frame, convolving the model spectrum with the instrumental resolution of the prism.
We fit using the \hi{} column density, covering fraction, and peculiar velocity of the DLA, and the \lya{} EW (fixing a \lya{} FWHM of 200 \kms{}) prior to attenuation by the DLA as free parameters. 
As this region is a dense candidate protocluster, redshifted DLA systems may arise from foreground protocluster members with line-of-sight motion towards the observed galaxies, motivating our inclusion of peculiar velocity as a free parameter.
We find degeneracies between the peculiar velocity of the absorber, the \hi{} covering fraction, and the \lya{} EW (prior to attenuation by the DLA). The parameters also are sensitive to different assumptions about the local ionization state of the IGM in the vicinity of the galaxies. 
A detailed characterization of these properties is outside the scope of the current study given these degeneracies and the resolution of the spectrum ($\sim$ 4000 \kms{} per pixel).
Higher resolution spectroscopy is required for a more robust view of the DLA properties.

Nevertheless, in all cases, we find large \hi{} column densities ($N_{\rm HI} \approx  10^{21.5}$--$10^{23.5}$ cm$^{-2}$)\footnote{We note that given the prism resolution and SNR, each spectrum can be fit with a column density of $N_{\rm HI}< 10^{23}$ cm$^{-2}$. } are required to reproduce the observed spectra (see Figure \ref{fig:uncover}). 
The inferred column densities are high but similar to that inferred for other $z>8$ galaxies with recent NIRSpec prism observations \citep[e.g.,][]{Heintz2023}. 
We suggest such column densities may be more commonly seen in protocluster environments at $z\gtrsim 7$ where a large number of galaxies are closely separated.
Such dense neutral \hi{} in the local surroundings could heavily attenuate the \lya{} photons, contributing to the apparent lack of \lya{} detection in the majority (5/6) of the member galaxies associated with the \cite{Morishita2023} overdensity. 
We finally note that the UNCOVER spectrum of ZD4 has recently been presented in \cite{Cameron2023}. 
They suggest nebular continuum may provide an alternative explanation to damped \lya{} absorption. Higher resolution data will ultimately be required to fully disentangle the relative contribution of these effects. 
However, the weak rest-optical emission lines in ZD4 (H$\beta$ EW = $217_{-85}^{+122}$~\AA) suggest that such strong nebular emission is not likely present in this system. 
The ionizing photon production efficiency implied by the H$\beta$ detection in ZD4 (log ($\xi_{\rm ion}$ / erg$^{-1}$ Hz) = $25.5_{-0.2}^{+0.1}$ ) is well below that required for the nebular continuum to dominate the far-UV continuum  \citep[e.g.,][]{Topping2022_ceers}. 
Also, given the presence of such apparent absorption features redwards of \lya{} wavelength in all three galaxies observed with UNCOVER in this $z=7.88$ overdensity, we suggest that the damped \lya{} absorption remains a very likely explanation for ZD4 and the other two systems.  

Strong galaxy overdensities that span small physical size scales  (within a projected radius of 60~pkpc) are more likely to have small ionized bubbles. 
In particular, for the Abell 2744 overdensity at $z=7.88$, \citet{Lu2023} demonstrated that the range of bubble sizes was likely to be $\sim0.5$~pMpc (with a range between 0.2 and 1.8~Mpc), lower than that expected in the CEERS field (1.0~pMpc, with a range between 0.2--2.7~pMpc). 
The small bubble sizes potentially explain the weak \lya{} emission reported in \citet{Morishita2023}. 
The observations here suggest an additional factor regulating \lya{} visibility in this region may be damped \lya{} absorption. 
This is likely to be most impactful in attempts to characterize \lya{} in large overdensities spanning small scales, where the close galaxy separations ($<$ 50--390~pkpc) lead to the common presence of foreground absorbers. 
Efforts to characterize ionized bubbles may be best focused on regions that are overdense over larger scales ($>$ 1~pMpc). 
Systematic surveys of Lyman alpha visibility as a function of galaxy overdensity will enable us to directly connect ionized regions to both their local environment and galaxy properties, paving the way for a detailed understanding of how galaxies achieved reionization.

\section{Summary}\label{sec:summary}

We investigate \lya{} properties of galaxies at 
$5 \leq z < 7$ in \jwst{}/NIRSpec prism spectroscopy taken as part of  CEERS \citep{Finkelstein2022,Finkelstein2023} and DDT-2750 \citep{ArrabalHaro2023a,ArrabalHaro2023b} observations of 
the EGS field. We summarize our main findings below.

\begin{enumerate}
    \item We have assembled a spectroscopic catalog of 69 galaxies at $5.00<z<6.98$ confirmed via rest-optical emission line detections in \jwst{} prism spectroscopy. 
    Many of the galaxies are faint continuum sources (F150W = 27--29 mag) selected in NIRCam imaging, allowing characterization of \lya{} output in very low mass star-forming systems. 
    We identify likely \lya{} emission (SNR = 5.6--26.9) in 10 of the 69 galaxies. 
    Given the faint continuum flux densities, the implied \lya{} EWs extend to very large values (ranging up to 286~\AA{} with a median of 134~\AA{}). 
    These are comparable to the \lya{} EWs found in faint galaxies with MUSE \citep[e.g.,][]{Hashimoto2017_muse,Maseda2018,Vanzella2020,Kerutt2022} and other narrowband surveys \citep[e.g.,][]{Trainor2015,Hashimoto2017,Harikane2018,Ning2020,Matthee2021}.

    \item We present likely detections of \lya{} (SNR = 7.8 and 8.2) in two $z>7$ galaxies, CEERS-80374 ($z=7.49$) and CEERS-80239 ($z=7.10$). 
    These prism spectra were obtained in the second epoch of CEERS observations, after the first papers describing $z\gtrsim 7$ \lya{} spectroscopy were released \citep{Jung2023,Tang2023}. Both galaxies are very faint in the continuum (F150W = 28.7 and 28.9 mag), with the \lya{} detections implying extremely large \lya{} EWs ($334_{-62}^{+109}$~\AA{} and $205_{-27}^{+48}$~\AA{}). 
    Such intense \lya{} likely requires near-unity transmission through the IGM, possibly indicating the two galaxies are situated in large ionized regions, as recently found in the faint \lya{} emitter JADES-GS-z7-LA \citep{Saxena2023}.
    
    \item We explore the physical properties of our sample using NIRCam SEDs and NIRSpec rest-optical prism spectroscopy. 
    The galaxies in our sample with \lya{} emission tend to be dominated by extremely young stellar populations (median age = 6 Myr for constant star formation), as is common for galaxies that have undergone a recent burst of star formation. 
    This results in very efficient ionizing photon production (median $\xi_{\rm ion}$ = $10^{25.8}$ erg$^{-1}$ Hz) that translates into intense intrinsic \lya{} emission (median EW = 256~\AA). 
    As expected for such young systems, we see intense [\oiii{}]+H$\beta$ emission (median EW = 1982~\AA), well above the average for $z\simeq 5-7$ galaxies \citep{Endsley2023}.
    
    \item We compute \lya{} escape fractions for our sample using flux ratios of the \lya{} and the Balmer emission lines (\ha{} or \hb{}) in the NIRSpec data. 
    We find that large escape fractions are still common in the galaxy population at $z\simeq 5-7$, with a Case B median value of \fesc{} = 0.28. 
    This is well above what is common in the first $z\gtrsim 7$ spectroscopic samples (median \fesc{} = 0.073; \citealt{Tang2023}). 
    We find two galaxies in our $z\simeq 5-7$ sample that show \lya{}/\ha{} ratios suggesting close-to-unity ($>$ 80\%) escape fractions of \lya{}. 
    These efficient \lya{} leakers appear similar to the extreme \lya{} emitters now being found at $z\gtrsim 7$ (e.g., \citealt{Saxena2023}).
    
    \item  The distribution of \lya{} escape fractions at $z\simeq 5-7$ in our sample provides a baseline to compare against emerging $z>7$ samples \citep[e.g.,][]{Tang2023,Jung2023}, allowing insight into the impact of the neutral IGM on \lya{}. 
    We find that  $50_{-11}^{+11}$\% of the $z\sim 6$ galaxies with large EW [\oiii{}]+\hb{} ($>$ 1000~\AA{}) have moderately large \fesc{} ($>0.2$). 
    We show that lower redshift ($z\simeq 0-2$) galaxies with similar properties show comparable \lya{} escape fractions. If the IGM significantly attenuates $z\gtrsim 7$ \lya{} emission, we expect this percentage to go down. 
    While current statistics are limited, we find only $10_{-5}^{+9}$\% of the CEERS $z>7$ galaxies (from both \citealt{Tang2023} and this work) satisfying the EW [\oiii{}]+\hb{} selection have such large escape fractions.
    As \jwst{} builds larger $z\gtrsim 7$ spectroscopic samples, it will become possible to reduce the uncertainties on such measurements, improving efforts to link measurements of the \lya{} escape fraction to the IGM. 
    
    \item The two $z\gtrsim 7$ galaxies with likely detections of strong \lya{} emission are found to be located at the same redshifts as other \lya{} emitters in the CEERS EGS field, potentially hinting at extended ionized structures at both redshifts. 
    We characterize the distribution of faint galaxies around both \lya{} emitters using NIRCam imaging. 
    We find evidence for photometric overdensities of $z\simeq 7.0$--7.6 galaxies (at the 3.1--3.7 $\sigma$ level), as would be expected if there are large ionized bubbles in the field. 
    The detection of a large number of high-EW \lya{} emitting galaxies combined with the presence of significant photometric overdensities also suggest the presence of long ionized sightlines that boost the \lya{} emission in this field.
    Future spectroscopic observations are required to better characterize the overdensity structures and ionized region sizes at the redshifts of these \lya{} emitters.

    \item We further investigate the dependence of $z>7$ \lya{} transmission on the local environment, considering the $z=7.88$ galaxy overdensity with 7 previously confirmed member galaxies, z7p9OD, in the Abell 2744 field. 
    Utilizing the new UNCOVER deep NIRSpec prism observations covering this overdensity, we confirm one more photometrically selected member to be associated with the overdensity, which is also the only source detected in \lya{} (with EW = $69_{-9}^{+10}$~\AA{}) in this region. 
    We also identify damped \lya{} absorption in all three member galaxies with deep UNCOVER spectra, suggesting the presence of high column density \hi{} in the local surroundings of the overdensity. 
    Such \hi{} gas may obscure much of the \lya{} photons from the member galaxy, leading to the low fraction (1/6) of \lya{} detection in this overdensity. 
    \jwst{} NIRSpec observations mapping a larger sample of $z>7$ overdensities will be required to statistically investigate how \lya{} visibility may be affected by overdense environments spanning different physical scales.

\end{enumerate}

\begin{table*}
    \centering
    \begin{threeparttable}[t]
\begin{tabular}{ccccccccc}
    \hline
    ID & RA & Dec. & $z$ & $M_{\rm UV}$ & EW \lya{} & Separation from & Separation from & Ref.\tnote{a} \\ 
    & (deg) & (deg) & & (mag) & (\AA{}) &  CEERS-80239 (\arcmin{}) & CEERS-80374 (\arcmin{}) &\\
    \hline 

\noalign{\vskip 2.5pt} 
\multicolumn{9}{l}{Spectroscopically confirmed galaxies near CEERS-80239}\\[2.5pt]
\hline
    
DDT-434       & 214.8980100 & 52.8929639  & $7.47 $  & $-19.5_{- 0.1}^{+ 0.1}$  & --                &  1.39  &   --   & This work  \\[1pt]
DDT-449       & 214.9404890 & 52.9325555  & $7.54 $  & $-18.9_{- 0.2}^{+ 0.1}$  & $<185$            &  4.09  &   --   & This work  \\[1pt]
CEERS-689     & 214.9990525 & 52.9419778  & $7.545$  & $-22.14$                   & $<26.1$           &  5.71  &   --   & T23, N23  \\[1pt]
CEERS-1163    & 214.9904679 & 52.9719889  & $7.448$  & $-20.24$                   & --                &  7.02  &   --   & T23, N23  \\[1pt]
CEERS-80445   & 214.8431150 & 52.7478861  & $7.50 $  & $-19.6_{-0.1}^{+0.1}$    & --                &  7.57  &   --   & This work, N23  \\[1pt]
CEERS-80432   & 214.8120558 & 52.7467472  & $7.47 $  & $-20.0_{-0.1}^{+0.1}$    & $<68$             &  7.99  &   --   & This work, N23  \\[1pt]
CEERS-698     & 215.0503167 & 53.0074417  & $7.470$  & $-21.86$                   & $9.5\pm3.1$       &  9.96  &   --   & RB16, S17, T23, N23, S23  \\[1pt]

\hline

\noalign{\vskip 2.5pt} 
\multicolumn{9}{l}{Spectroscopically confirmed galaxies near CEERS-80374}\\[2.5pt]
\hline

CEERS-534     & 214.8591171 & 52.8536389  & $7.114$  & $-20.6_{- 0.1}^{+ 0.1}$  & $<161 $           &   --   &  2.23  & This work  \\[1pt]
CEERS-498     & 214.8130450 & 52.8342500  & $7.18 $  & $-20.21$                   & $<53$             &   --   &  3.13  & T23  \\[1pt]
CEERS-499     & 214.8130042 & 52.8341694  & $7.168$  & $-16.97$                   & --                &   --   &  3.13  & T23  \\[1pt]
CEERS-1038    & 215.0396971 & 52.9015972  & $7.194$  & $-19.25$                   & --                &   --   &  6.89  & T23, N23  \\[1pt]
CEERS-44      & 215.0011150 & 53.0112694  & $7.100$  & $-19.37$                   & $77.6\pm5.5$      &   --   & 11.79  & T23, N23  \\[1pt]

\hline    
\end{tabular}

\begin{tablenotes}
\item[a] References: T23 -- \cite{Tang2023}; N23 -- \cite{Nakajima2023}; RB16 --- \cite{Roberts-Borsani2016}; S17 -- \cite{Stark2017}; S23 -- \cite{Sanders2023}.
\end{tablenotes}
    
\end{threeparttable}
    \caption{List of spectroscopically confirmed galaxies near the two new $z>7$ \lya{} emitting galaxies presented in this work, CEERS-80239 and CEERS-80374.
    We present the galaxies within ${\rm} z$ = 0.1 from each \lya{} emitting galaxy separately, each part sorted by increasing separations from the corresponding \lya{} emitting galaxy.
    We also report their coordinates, spectroscopic redshifts, absolute UV magnitudes, and the \lya{} EWs ($7\sigma$ upper limits for non-detections).}
    \label{tb:tb_neighbor_specz}
\end{table*}

\begin{table*}
    \centering
    \begin{threeparttable}[t]
\begin{tabular}{ccccccccc}
    \hline
    ID & RA & Dec. & F356W - F410M & $z$ & $M_{\rm UV}$ & Separation from & Separation from & Ref.\tnote{a} \\ 
    & (deg) & (deg) & (mag) & & (mag) &  CEERS-80239 (\arcmin{}) & CEERS-80374 (\arcmin{}) &\\
    \hline

z7\_F410M\_ 1  & 214.8936194 & 52.8740006  & $1.1_{-0.1}^{+0.1}$  & $7.14_{-0.11}^{+0.29}$  & $-19.9_{-0.1}^{+0.1}$  &  0.26  &  2.95  &  This work  \\[1pt]
z7\_F410M\_ 2  & 214.8939244 & 52.8745818  & $1.0_{-0.0}^{+0.0}$  & $7.28_{-0.17}^{+0.19}$  & $-19.9_{-0.1}^{+0.1}$  &  0.29  &  2.99  &  This work  \\[1pt]
z7\_F410M\_ 3  & 214.8925124 & 52.8806457  & $0.8_{-0.0}^{+0.0}$  & $7.29_{-0.17}^{+0.16}$  & $-20.1_{-0.1}^{+0.1}$  &  0.66  &  3.35  &  This work  \\[1pt]
z7\_F410M\_ 4  & 214.9102765 & 52.8600728  & $1.4_{-0.0}^{+0.0}$  & $7.13_{-0.03}^{+0.03}$  & $-19.6_{-0.0}^{+0.1}$  &  0.78  &  2.16  &  This work  \\[1pt]
z7\_F410M\_ 5  & 214.8615723 & 52.8760338  & $0.8_{-0.1}^{+0.1}$  & $7.26_{-0.14}^{+0.13}$  & $-19.6_{-0.1}^{+0.1}$  &  1.30  &  3.34  &        E23  \\[1pt]
z7\_F410M\_ 6  & 214.8615875 & 52.8761597  & $1.2_{-0.1}^{+0.1}$  & $7.14_{-0.07}^{+0.09}$  & $-19.3_{-0.0}^{+0.0}$  &  1.30  &  3.35  &        E23  \\[1pt]
z7\_F410M\_ 7  & 214.8591919 & 52.8535843  & $1.0_{-0.0}^{+0.0}$  & $7.14_{-0.08}^{+0.15}$  & $-20.8_{-0.1}^{+0.1}$  &  1.65  &  2.22  &        E23  \\[1pt]
z7\_F410M\_ 8  & 214.9110902 & 52.8973152  & $1.4_{-0.1}^{+0.1}$  & $7.41_{-0.18}^{+0.09}$  & $-19.1_{-0.1}^{+0.1}$  &  1.74  &  4.37  &  This work  \\[1pt]
z7\_F410M\_ 9  & 214.9431503 & 52.8785455  & $0.8_{-0.2}^{+0.1}$  & $7.25_{-0.28}^{+0.26}$  & $-19.2_{-0.1}^{+0.1}$  &  1.78  &  3.61  &  This work  \\[1pt]
z7\_F410M\_10  & 214.8425293 & 52.8711929  & $1.0_{-0.0}^{+0.0}$  & $7.41_{-0.09}^{+0.08}$  & $-20.0_{-0.0}^{+0.0}$  &  1.94  &  3.43  &        E23  \\[1pt]
z7\_F410M\_11  & 214.8784485 & 52.8384552  & $0.7_{-0.0}^{+0.0}$  & $7.00_{-0.05}^{+0.12}$  & $-19.7_{-0.1}^{+0.1}$  &  1.99  &  1.08  &        E23  \\[1pt]
z7\_F410M\_12  & 214.8369751 & 52.8670006  & $1.4_{-0.0}^{+0.0}$  & $7.37_{-0.10}^{+0.09}$  & $-19.7_{-0.0}^{+0.1}$  &  2.15  &  3.36  &        E23  \\[1pt]
z7\_F410M\_13  & 214.8253632 & 52.8630600  & $1.4_{-0.1}^{+0.1}$  & $7.26_{-0.10}^{+0.09}$  & $-19.5_{-0.1}^{+0.1}$  &  2.59  &  3.49  &        E23  \\[1pt]
z7\_F410M\_14  & 214.8309479 & 52.8482819  & $0.9_{-0.1}^{+0.1}$  & $7.41_{-0.14}^{+0.11}$  & $-19.5_{-0.1}^{+0.1}$  &  2.69  &  2.81  &        E23  \\[1pt]
z7\_F410M\_15  & 214.9783644 & 52.8773967  & $1.0_{-0.0}^{+0.0}$  & $7.42_{-0.17}^{+0.10}$  & $-19.6_{-0.1}^{+0.1}$  &  3.01  &  4.29  &  This work  \\[1pt]
z7\_F410M\_16  & 214.8333282 & 52.8275871  & $0.7_{-0.1}^{+0.1}$  & $7.24_{-0.16}^{+0.17}$  & $-19.6_{-0.1}^{+0.1}$  &  3.41  &  2.35  &        E23  \\[1pt]
z7\_F410M\_17  & 214.9885098 & 52.8918060  & $1.0_{-0.1}^{+0.1}$  & $7.18_{-0.20}^{+0.26}$  & $-19.1_{-0.1}^{+0.1}$  &  3.60  &  5.18  &  This work  \\[1pt]
z7\_F410M\_18  & 214.7931213 & 52.8701210  & $0.6_{-0.2}^{+0.1}$  & $7.17_{-0.16}^{+0.17}$  & $-19.3_{-0.1}^{+0.1}$  &  3.73  &  4.67  &        E23  \\[1pt]
z7\_F410M\_19  & 214.8370667 & 52.8188782  & $0.9_{-0.1}^{+0.1}$  & $7.37_{-0.17}^{+0.12}$  & $-19.3_{-0.1}^{+0.1}$  &  3.73  &  2.24  &        E23  \\[1pt]
z7\_F410M\_20  & 214.9987301 & 52.8553770  & $1.2_{-0.0}^{+0.0}$  & $7.31_{-0.15}^{+0.12}$  & $-20.8_{-0.1}^{+0.1}$  &  3.82  &  4.08  &  This work  \\[1pt]
z7\_F410M\_21  & 214.8526001 & 52.8114624  & $1.0_{-0.1}^{+0.1}$  & $7.21_{-0.11}^{+0.11}$  & $-19.5_{-0.1}^{+0.1}$  &  3.84  &  1.84  &        E23  \\[1pt]
z7\_F410M\_22  & 214.9415894 & 52.9291191  & $1.0_{-0.0}^{+0.0}$  & $7.00_{-0.03}^{+0.06}$  & $-19.0_{-0.1}^{+0.1}$  &  3.92  &  6.45  &        E23  \\[1pt]
z7\_F410M\_23  & 215.0030565 & 52.8856697  & $1.4_{-0.0}^{+0.0}$  & $7.32_{-0.14}^{+0.14}$  & $-19.1_{-0.1}^{+0.1}$  &  3.99  &  5.27  &  This work  \\[1pt]
z7\_F410M\_24  & 214.7939301 & 52.8415337  & $0.8_{-0.0}^{+0.0}$  & $7.60_{-0.02}^{+0.01}$  & $-20.7_{-0.0}^{+0.0}$  &  4.07  &  3.90  &        E23  \\[1pt]
z7\_F410M\_25  & 214.8471985 & 52.8083382  & $1.3_{-0.0}^{+0.0}$  & $7.27_{-0.12}^{+0.13}$  & $-19.5_{-0.1}^{+0.1}$  &  4.09  &  2.10  &        E23  \\[1pt]
z7\_F410M\_26  & 215.0122205 & 52.8777771  & $0.8_{-0.1}^{+0.1}$  & $7.19_{-0.13}^{+0.17}$  & $-19.9_{-0.1}^{+0.1}$  &  4.23  &  5.21  &  This work  \\[1pt]
z7\_F410M\_27  & 214.9098511 & 52.9425430  & $0.7_{-0.1}^{+0.1}$  & $7.38_{-0.14}^{+0.15}$  & $-19.5_{-0.1}^{+0.1}$  &  4.39  &  7.07  &        E23  \\[1pt]
z7\_F410M\_28  & 214.9110565 & 52.9425240  & $1.2_{-0.0}^{+0.0}$  & $7.03_{-0.04}^{+0.11}$  & $-20.3_{-0.1}^{+0.1}$  &  4.39  &  7.07  &        E23  \\[1pt]
z7\_F410M\_29  & 214.9977908 & 52.9139550  & $0.9_{-0.1}^{+0.1}$  & $7.13_{-0.17}^{+0.29}$  & $-19.4_{-0.1}^{+0.1}$  &  4.53  &  6.45  &  This work  \\[1pt]
z7\_F410M\_30  & 214.9554901 & 52.9470100  & $0.9_{-0.1}^{+0.1}$  & $7.29_{-0.18}^{+0.16}$  & $-19.6_{-0.1}^{+0.1}$  &  5.10  &  7.62  &        E23  \\[1pt]
z7\_F410M\_31  & 214.9183197 & 52.9547272  & $1.2_{-0.1}^{+0.1}$  & $7.29_{-0.07}^{+0.09}$  & $-19.6_{-0.1}^{+0.1}$  &  5.16  &  7.82  &        E23  \\[1pt]
z7\_F410M\_32  & 215.0395352 & 52.8910261  & $0.8_{-0.1}^{+0.1}$  & $7.23_{-0.18}^{+0.21}$  & $-18.9_{-0.1}^{+0.1}$  &  5.35  &  6.48  &  This work  \\[1pt]
z7\_F410M\_33  & 215.0371798 & 52.9067190  & $1.7_{-0.0}^{+0.0}$  & $7.28_{-0.13}^{+0.15}$  & $-20.2_{-0.1}^{+0.1}$  &  5.57  &  7.03  &  This work  \\[1pt]
z7\_F410M\_34  & 214.8065082 & 52.7927667  & $0.8_{-0.0}^{+0.0}$  & $7.27_{-0.18}^{+0.18}$  & $-19.6_{-0.1}^{+0.1}$  &  5.65  &  3.84  &  This work  \\[1pt]
z7\_F410M\_35  & 215.0316755 & 52.9218393  & $1.3_{-0.1}^{+0.1}$  & $7.42_{-0.33}^{+0.10}$  & $-19.0_{-0.2}^{+0.5}$  &  5.82  &  7.57  &  This work  \\[1pt]
z7\_F410M\_36  & 214.7973715 & 52.7888976  & $0.7_{-0.1}^{+0.1}$  & $7.23_{-0.15}^{+0.20}$  & $-20.3_{-0.1}^{+0.1}$  &  6.03  &  4.24  &  This work  \\[1pt]
z7\_F410M\_37  & 214.9827271 & 52.9560814  & $0.9_{-0.0}^{+0.0}$  & $7.42_{-0.30}^{+0.06}$  & $-20.3_{-0.1}^{+0.1}$  &  6.05  &  8.45  &        E23  \\[1pt]
z7\_F410M\_38  & 214.9830170 & 52.9560051  & $0.7_{-0.0}^{+0.0}$  & $6.95_{-0.01}^{+0.01}$  & $-19.7_{-0.0}^{+0.0}$  &  6.05  &  8.45  &        E23  \\[1pt]
z7\_F410M\_39  & 214.9672394 & 52.9636269  & $1.0_{-0.0}^{+0.0}$  & $7.37_{-0.23}^{+0.11}$  & $-19.9_{-0.1}^{+0.1}$  &  6.19  &  8.69  &        E23  \\[1pt]
z7\_F410M\_40  & 214.9854889 & 52.9587479  & $0.9_{-0.0}^{+0.0}$  & $7.26_{-0.17}^{+0.20}$  & $-20.0_{-0.1}^{+0.1}$  &  6.24  &  8.63  &        E23  \\[1pt]

z7\_F410M\_41  & 215.0758595 & 52.9055814  & $1.0_{-0.1}^{+0.1}$  & $7.32_{-0.13}^{+0.15}$  & $-19.4_{-0.1}^{+0.1}$  &  6.85  &  8.06  &  This work  \\[1pt]
z7\_F410M\_42  & 215.0740322 & 52.9090114  & $0.6_{-0.1}^{+0.1}$  & $7.49_{-0.27}^{+0.10}$  & $-19.5_{-0.1}^{+0.1}$  &  6.86  &  8.13  &  This work  \\[1pt]
z7\_F410M\_43  & 215.0699407 & 52.9287066  & $1.3_{-0.1}^{+0.1}$  & $7.31_{-0.10}^{+0.14}$  & $-19.5_{-0.1}^{+0.1}$  &  7.22  &  8.81  &  This work  \\[1pt]
z7\_F410M\_44  & 215.0671249 & 52.9381049  & $0.7_{-0.1}^{+0.1}$  & $7.26_{-0.23}^{+0.25}$  & $-19.5_{-0.1}^{+0.1}$  &  7.42  &  9.14  &  This work  \\[1pt]
z7\_F410M\_45  & 215.0115204 & 52.9743690  & $0.8_{-0.1}^{+0.0}$  & $7.18_{-0.13}^{+0.28}$  & $-19.7_{-0.1}^{+0.1}$  &  7.53  &  9.86  &        E23  \\[1pt]
z7\_F410M\_46  & 214.8120564 & 52.7467451  & $1.4_{-0.0}^{+0.0}$  & $7.32_{-0.16}^{+0.13}$  & $-19.7_{-0.1}^{+0.1}$  &  7.99  &  5.63  &  This work  \\[1pt]
z7\_F410M\_47  & 215.0778539 & 52.9501237  & $0.8_{-0.1}^{+0.1}$  & $7.32_{-0.12}^{+0.14}$  & $-20.4_{-0.1}^{+0.1}$  &  8.15  &  9.94  &  This work  \\[1pt]
z7\_F410M\_48  & 215.0254059 & 53.0029793  & $1.1_{-0.1}^{+0.1}$  & $7.43_{-0.11}^{+0.09}$  & $-19.5_{-0.1}^{+0.1}$  &  9.26  & 11.64  &        E23  \\[1pt]
z7\_F410M\_49  & 215.0011139 & 53.0112724  & $1.2_{-0.0}^{+0.0}$  & $7.04_{-0.02}^{+0.03}$  & $-19.5_{-0.0}^{+0.0}$  &  9.30  & 11.79  &        E23  \\[1pt]
z7\_F410M\_50  & 215.1298690 & 52.9499481  & $1.7_{-0.1}^{+0.1}$  & $7.23_{-0.13}^{+0.19}$  & $-19.8_{-0.1}^{+0.1}$  &  9.73  & 11.26  &  This work  \\[1pt]
z7\_F410M\_51  & 214.7071113 & 52.7425158  & $1.2_{-0.0}^{+0.0}$  & $7.34_{-0.14}^{+0.11}$  & $-20.0_{-0.1}^{+0.1}$  & 10.26  &  8.51  &  This work  \\[1pt]
z7\_F410M\_52  & 215.1196409 & 52.9828335  & $0.7_{-0.0}^{+0.0}$  & $7.45_{-0.27}^{+0.07}$  & $-20.6_{-0.1}^{+0.1}$  & 10.55  & 12.41  &  This work  \\[1pt]
z7\_F410M\_53  & 215.1401974 & 52.9865438  & $0.8_{-0.1}^{+0.1}$  & $7.21_{-0.15}^{+0.21}$  & $-20.0_{-0.1}^{+0.1}$  & 11.27  & 13.07  &  This work  \\[1pt]
z7\_F410M\_54  & 215.1500154 & 52.9846551  & $1.3_{-0.0}^{+0.0}$  & $7.32_{-0.10}^{+0.11}$  & $-20.0_{-0.1}^{+0.1}$  & 11.48  & 13.23  &  This work  \\[1pt]
\hline    
\end{tabular}

\begin{tablenotes}
\item[a] References: E23 -- \cite{Endsley2023_ceers}
\end{tablenotes}

\end{threeparttable}
    \caption{List of photometric samples selected with CEERS NIRCam imaging at redshifts similar to the two $z>7$ \lya{} emitting galaxies, CEERS-80239 ($z=7.49$) and CEERS-80374 ($z=7.17$).
    Each galaxy has F356W - F410M $>$ 0.6 which indicates a redshift of $z =$ 7.0--7.6.
    We sort them by increasing separations from CEERS-80239.
    We report their coordinates, F356W - F410M colors, \beagle{} photometric redshifts, absolute UV magnitudes, and their separation from each of the \lya{} emitting galaxies.}
    \label{tb:tb_f410m}
\end{table*}

\section*{Acknowledgements}

The authors thank the referee, Dr. Takuya Hashimoto, for helpful comments that improved the quality of this paper.
We thank Jacopo Chevallard for the use of the \beagle{} tool used for much of our SED fitting analysis, Gabe Brammer for providing the optical imaging of the EGS as part of CHArGE program, John Chisholm for providing data of LzLCS sample, and Ting-Yi Lu for helpful discussion on ionized bubbles.
This work is based in part on observations made with the NASA/ESA/CSA JWST. The data were obtained from the Mikulski Archive for Space Telescopes at the Space Telescope Science Institute, which is operated by the Association of Universities for Research in Astronomy, Inc., under NASA contract NAS 5-03127 for JWST. 
These observations are associated with programs \#ERS-1345 (CEERS), DDT-2750, and GO-2561 (UNCOVER).
The authors acknowledge the CEERS, DDT-2750, and UNCOVER teams led by Steven L. Finkelstein, Pablo Arrabal Haro, and I. Labb\'e \& R. Bezanson for developing their observing program with a zero-exclusive-access period.

DPS acknowledges support from the National Science Foundation through the grant AST-2109066.
CM acknowledges support by the VILLUM FONDEN under grant 37459 and the Carlsberg Foundation under grant CF22-1322. The Cosmic Dawn Center (DAWN) is funded by the Danish National Research Foundation under grant DNRF140.
LW acknowledges support from the National Science Foundation Graduate Research Fellowship under Grant No. DGE-2137419.
M. Tang acknowledges funding from the JWST Arizona/Steward Postdoc in Early galaxies and Reionization (JASPER) Scholar contract at the University of Arizona.
This material is based in part upon High Performance Computing
(HPC) resources supported by the University of Arizona TRIF, UITS,
and Research, Innovation, and Impact (RII) and maintained by the
UArizona Research Technologies department.

This work made use of {\sc Astropy}:\footnote{\url{http://www.astropy.org}} a community-developed core Python package and an ecosystem of tools and resources for astronomy \citep{astropy:2013, astropy:2018, astropy:2022}; \beagle{} \citep{Chevallard2016}; {\sc Emcee} \citep{Foreman-Mackey2013}; {\sc Jupyter} \citep{Kluyver2016}; {\sc Matplotlib} \citep{Hunter:2007}; {\sc Numpy} \citep{harris2020array}; {\sc Photutils}, an {\sc Astropy} package for detection and photometry of astronomical sources \citep{Bradley2022}; {\sc Scikit-image} \citep{vanderWalt2014}; {\sc Scipy} \citep{2020SciPy-NMeth}; and {\sc Sedpy} \citep{Johnson2021_sedpy}.

%%%%%%%%%%%%%%%%%%%%%%%%%%%%%%%%%%%%%%%%%%%%%%%%%%
\section*{Data Availability}

The JWST and HST imaging data used in this work are
available through the MAST Portal (\url{https://mast.stsci.edu/portal/Mashup/Clients/Mast/Portal.html}).
The data product and analysis code used in this work will be shared upon reasonable request to the corresponding author.

%%%%%%%%%%%%%%%%%%%% REFERENCES %%%%%%%%%%%%%%%%%%

% The best way to enter references is to use BibTeX:

\bibliographystyle{mnras}
\bibliography{main} % if your bibtex file is called example.bib

\begin{thebibliography}{}
\makeatletter
\relax
\def\mn@urlcharsother{\let\do\@makeother \do\$\do\&\do\#\do\^\do\_\do\%\do\~}
\def\mn@doi{\begingroup\mn@urlcharsother \@ifnextchar [ {\mn@doi@} {\mn@doi@[]}}
\def\mn@doi@[#1]#2{\def\@tempa{#1}\ifx\@tempa\@empty \href {http://dx.doi.org/#2} {doi:#2}\else \href {http://dx.doi.org/#2} {#1}\fi \endgroup}
\def\mn@eprint#1#2{\mn@eprint@#1:#2::\@nil}
\def\mn@eprint@arXiv#1{\href {http://arxiv.org/abs/#1} {{\tt arXiv:#1}}}
\def\mn@eprint@dblp#1{\href {http://dblp.uni-trier.de/rec/bibtex/#1.xml} {dblp:#1}}
\def\mn@eprint@#1:#2:#3:#4\@nil{\def\@tempa {#1}\def\@tempb {#2}\def\@tempc {#3}\ifx \@tempc \@empty \let \@tempc \@tempb \let \@tempb \@tempa \fi \ifx \@tempb \@empty \def\@tempb {arXiv}\fi \@ifundefined {mn@eprint@\@tempb}{\@tempb:\@tempc}{\expandafter \expandafter \csname mn@eprint@\@tempb\endcsname \expandafter{\@tempc}}}

\bibitem[\protect\citeauthoryear{{Arrabal Haro} et~al.,}{{Arrabal Haro} et~al.}{2023a}]{ArrabalHaro2023a}
{Arrabal Haro} P.,  et~al., 2023a, \mn@doi [\nat] {10.1038/s41586-023-06521-7}, \href {https://ui.adsabs.harvard.edu/abs/2023Natur.622..707A} {622, 707}

\bibitem[\protect\citeauthoryear{{Arrabal Haro} et~al.,}{{Arrabal Haro} et~al.}{2023b}]{ArrabalHaro2023b}
{Arrabal Haro} P.,  et~al., 2023b, \mn@doi [\apjl] {10.3847/2041-8213/acdd54}, \href {https://ui.adsabs.harvard.edu/abs/2023ApJ...951L..22A} {951, L22}

\bibitem[\protect\citeauthoryear{{Astropy Collaboration} et~al.,}{{Astropy Collaboration} et~al.}{2013}]{astropy:2013}
{Astropy Collaboration} et~al., 2013, \mn@doi [\aap] {10.1051/0004-6361/201322068}, \href {http://adsabs.harvard.edu/abs/2013A%26A...558A..33A} {558, A33}

\bibitem[\protect\citeauthoryear{{Astropy Collaboration} et~al.,}{{Astropy Collaboration} et~al.}{2018}]{astropy:2018}
{Astropy Collaboration} et~al., 2018, \mn@doi [\aj] {10.3847/1538-3881/aabc4f}, \href {https://ui.adsabs.harvard.edu/abs/2018AJ....156..123A} {156, 123}

\bibitem[\protect\citeauthoryear{{Astropy Collaboration} et~al.,}{{Astropy Collaboration} et~al.}{2022}]{astropy:2022}
{Astropy Collaboration} et~al., 2022, \mn@doi [apj] {10.3847/1538-4357/ac7c74}, \href {https://ui.adsabs.harvard.edu/abs/2022ApJ...935..167A} {935, 167}

\bibitem[\protect\citeauthoryear{{Bacon} et~al.,}{{Bacon} et~al.}{2017}]{Bacon2017}
{Bacon} R.,  et~al., 2017, \mn@doi [\aap] {10.1051/0004-6361/201730833}, \href {https://ui.adsabs.harvard.edu/abs/2017A&A...608A...1B} {608, A1}

\bibitem[\protect\citeauthoryear{{Bacon} et~al.,}{{Bacon} et~al.}{2023}]{Bacon2023}
{Bacon} R.,  et~al., 2023, \mn@doi [\aap] {10.1051/0004-6361/202244187}, \href {https://ui.adsabs.harvard.edu/abs/2023A&A...670A...4B} {670, A4}

\bibitem[\protect\citeauthoryear{{Bagley} et~al.,}{{Bagley} et~al.}{2023}]{Bagley2023}
{Bagley} M.~B.,  et~al., 2023, \mn@doi [\apjl] {10.3847/2041-8213/acbb08}, \href {https://ui.adsabs.harvard.edu/abs/2023ApJ...946L..12B} {946, L12}

\bibitem[\protect\citeauthoryear{{Barkana} \& {Loeb}}{{Barkana} \& {Loeb}}{2004}]{Barkana2004}
{Barkana} R.,  {Loeb} A.,  2004, \mn@doi [\apj] {10.1086/421079}, \href {https://ui.adsabs.harvard.edu/abs/2004ApJ...609..474B} {609, 474}

\bibitem[\protect\citeauthoryear{{Begley}, {Cullen}, {McLure}, {Shapley}, {Dunlop}, {Carnall}, {McLeod}  \& {Donnan}}{{Begley} et~al.}{2023}]{Begley2023}
{Begley} R.,  {Cullen} F.,  {McLure} R.~J.,  {Shapley} A.~E.,  {Dunlop} J.~S.,  {Carnall} A.~C.,  {McLeod} D.~J.,   {Donnan} C.~T.,  2023, \mn@doi [\mnras] {10.1093/mnras/stad3417}, \href {https://ui.adsabs.harvard.edu/abs/2023MNRAS.tmp.3237B} {}

\bibitem[\protect\citeauthoryear{{Bezanson} et~al.,}{{Bezanson} et~al.}{2022}]{Bezanson2022}
{Bezanson} R.,  et~al., 2022, \mn@doi [arXiv e-prints] {10.48550/arXiv.2212.04026}, \href {https://ui.adsabs.harvard.edu/abs/2022arXiv221204026B} {p. arXiv:2212.04026}

\bibitem[\protect\citeauthoryear{{B{\"o}ker} et~al.,}{{B{\"o}ker} et~al.}{2023}]{Boker2023}
{B{\"o}ker} T.,  et~al., 2023, \mn@doi [\pasp] {10.1088/1538-3873/acb846}, \href {https://ui.adsabs.harvard.edu/abs/2023PASP..135c8001B} {135, 038001}

\bibitem[\protect\citeauthoryear{{Bolan} et~al.,}{{Bolan} et~al.}{2022}]{Bolan2022}
{Bolan} P.,  et~al., 2022, \mn@doi [\mnras] {10.1093/mnras/stac1963}, \href {https://ui.adsabs.harvard.edu/abs/2022MNRAS.517.3263B} {517, 3263}

\bibitem[\protect\citeauthoryear{{Bouwens} et~al.,}{{Bouwens} et~al.}{2010}]{Bouwens2010}
{Bouwens} R.~J.,  et~al., 2010, \mn@doi [\apjl] {10.1088/2041-8205/708/2/L69}, \href {https://ui.adsabs.harvard.edu/abs/2010ApJ...708L..69B} {708, L69}

\bibitem[\protect\citeauthoryear{{Bouwens}, {Illingworth}, {Ellis}, {Oesch}  \& {Stefanon}}{{Bouwens} et~al.}{2022}]{Bouwens2022}
{Bouwens} R.~J.,  {Illingworth} G.,  {Ellis} R.~S.,  {Oesch} P.,   {Stefanon} M.,  2022, \mn@doi [\apj] {10.3847/1538-4357/ac86d1}, \href {https://ui.adsabs.harvard.edu/abs/2022ApJ...940...55B} {940, 55}

\bibitem[\protect\citeauthoryear{{Boyett}, {Stark}, {Bunker}, {Tang}  \& {Maseda}}{{Boyett} et~al.}{2022}]{Boyett2022}
{Boyett} K. N.~K.,  {Stark} D.~P.,  {Bunker} A.~J.,  {Tang} M.,   {Maseda} M.~V.,  2022, \mn@doi [\mnras] {10.1093/mnras/stac1109}, \href {https://ui.adsabs.harvard.edu/abs/2022MNRAS.513.4451B} {513, 4451}

\bibitem[\protect\citeauthoryear{{Boyett} et~al.,}{{Boyett} et~al.}{2024}]{Boyett2024}
{Boyett} K.,  et~al., 2024, \mn@doi [arXiv e-prints] {10.48550/arXiv.2401.16934}, \href {https://ui.adsabs.harvard.edu/abs/2024arXiv240116934B} {p. arXiv:2401.16934}

\bibitem[\protect\citeauthoryear{Bradley et~al.,}{Bradley et~al.}{2022}]{Bradley2022}
Bradley L.,  et~al., 2022, astropy/photutils: 1.5.0, \mn@doi{10.5281/zenodo.6825092}, \url {https://doi.org/10.5281/zenodo.6825092}

\bibitem[\protect\citeauthoryear{{Brammer}, {Strait}, {Matharu}  \& {Momcheva}}{{Brammer} et~al.}{2022}]{Brammer2022}
{Brammer} G.,  {Strait} V.,  {Matharu} J.,   {Momcheva} I.,  2022, {grizli}, Zenodo, \mn@doi{10.5281/zenodo.6672538}

\bibitem[\protect\citeauthoryear{{Bruzual} \& {Charlot}}{{Bruzual} \& {Charlot}}{2003}]{Bruzual2003}
{Bruzual} G.,  {Charlot} S.,  2003, \mn@doi [\mnras] {10.1046/j.1365-8711.2003.06897.x}, \href {https://ui.adsabs.harvard.edu/abs/2003MNRAS.344.1000B} {344, 1000}

\bibitem[\protect\citeauthoryear{{Bunker} et~al.,}{{Bunker} et~al.}{2023}]{Bunker2023}
{Bunker} A.~J.,  et~al., 2023, \mn@doi [\aap] {10.1051/0004-6361/202346159}, \href {https://ui.adsabs.harvard.edu/abs/2023A&A...677A..88B} {677, A88}

\bibitem[\protect\citeauthoryear{{Calzetti}, {Armus}, {Bohlin}, {Kinney}, {Koornneef}  \& {Storchi-Bergmann}}{{Calzetti} et~al.}{2000}]{Calzetti2000}
{Calzetti} D.,  {Armus} L.,  {Bohlin} R.~C.,  {Kinney} A.~L.,  {Koornneef} J.,   {Storchi-Bergmann} T.,  2000, \mn@doi [\apj] {10.1086/308692}, \href {https://ui.adsabs.harvard.edu/abs/2000ApJ...533..682C} {533, 682}

\bibitem[\protect\citeauthoryear{{Cameron}, {Katzm}, {Witten}, {Saxena}, {Laporte}  \& {Bunker}}{{Cameron} et~al.}{2023}]{Cameron2023}
{Cameron} A.~J.,  {Katzm} H.,  {Witten} C.,  {Saxena} A.,  {Laporte} N.,   {Bunker} A.~J.,  2023, arXiv e-prints, \href {https://ui.adsabs.harvard.edu/abs/2023arXiv231102051C} {p. arXiv:2311.02051}

\bibitem[\protect\citeauthoryear{{Caruana}, {Bunker}, {Wilkins}, {Stanway}, {Lorenzoni}, {Jarvis}  \& {Ebert}}{{Caruana} et~al.}{2014}]{Caruana2014}
{Caruana} J.,  {Bunker} A.~J.,  {Wilkins} S.~M.,  {Stanway} E.~R.,  {Lorenzoni} S.,  {Jarvis} M.~J.,   {Ebert} H.,  2014, \mn@doi [\mnras] {10.1093/mnras/stu1341}, \href {https://ui.adsabs.harvard.edu/abs/2014MNRAS.443.2831C} {443, 2831}

\bibitem[\protect\citeauthoryear{{Cassata} et~al.,}{{Cassata} et~al.}{2015}]{Cassata2015}
{Cassata} P.,  et~al., 2015, \mn@doi [\aap] {10.1051/0004-6361/201423824}, \href {https://ui.adsabs.harvard.edu/abs/2015A&A...573A..24C} {573, A24}

\bibitem[\protect\citeauthoryear{{Cassata} et~al.,}{{Cassata} et~al.}{2020}]{Cassata2020}
{Cassata} P.,  et~al., 2020, \mn@doi [\aap] {10.1051/0004-6361/202037517}, \href {https://ui.adsabs.harvard.edu/abs/2020A&A...643A...6C} {643, A6}

\bibitem[\protect\citeauthoryear{{Chabrier}}{{Chabrier}}{2003}]{Chabrier2003}
{Chabrier} G.,  2003, \mn@doi [\pasp] {10.1086/376392}, \href {https://ui.adsabs.harvard.edu/abs/2003PASP..115..763C} {115, 763}

\bibitem[\protect\citeauthoryear{{Charlot} \& {Longhetti}}{{Charlot} \& {Longhetti}}{2001}]{Charlot2001}
{Charlot} S.,  {Longhetti} M.,  2001, \mn@doi [\mnras] {10.1046/j.1365-8711.2001.04260.x}, \href {https://ui.adsabs.harvard.edu/abs/2001MNRAS.323..887C} {323, 887}

\bibitem[\protect\citeauthoryear{{Chevallard} \& {Charlot}}{{Chevallard} \& {Charlot}}{2016}]{Chevallard2016}
{Chevallard} J.,  {Charlot} S.,  2016, \mn@doi [\mnras] {10.1093/mnras/stw1756}, \href {https://ui.adsabs.harvard.edu/abs/2016MNRAS.462.1415C} {462, 1415}

\bibitem[\protect\citeauthoryear{{Chevallard} et~al.,}{{Chevallard} et~al.}{2018}]{Chevallard2018}
{Chevallard} J.,  et~al., 2018, \mn@doi [\mnras] {10.1093/mnras/sty1461}, \href {https://ui.adsabs.harvard.edu/abs/2018MNRAS.479.3264C} {479, 3264}

\bibitem[\protect\citeauthoryear{{Chisholm} et~al.,}{{Chisholm} et~al.}{2022}]{Chisholm2022}
{Chisholm} J.,  et~al., 2022, \mn@doi [\mnras] {10.1093/mnras/stac2874}, \href {https://ui.adsabs.harvard.edu/abs/2022MNRAS.517.5104C} {517, 5104}

\bibitem[\protect\citeauthoryear{{Cooper} et~al.,}{{Cooper} et~al.}{2023}]{Cooper2023}
{Cooper} O.~R.,  et~al., 2023, arXiv e-prints, \href {https://ui.adsabs.harvard.edu/abs/2023arXiv230906656C} {p. arXiv:2309.06656}

\bibitem[\protect\citeauthoryear{{Cowie}, {Barger}  \& {Hu}}{{Cowie} et~al.}{2011}]{Cowie2011}
{Cowie} L.~L.,  {Barger} A.~J.,   {Hu} E.~M.,  2011, \mn@doi [\apj] {10.1088/0004-637X/738/2/136}, \href {https://ui.adsabs.harvard.edu/abs/2011ApJ...738..136C} {738, 136}

\bibitem[\protect\citeauthoryear{{Cullen} et~al.,}{{Cullen} et~al.}{2023}]{Cullen2023}
{Cullen} F.,  et~al., 2023, \mn@doi [\mnras] {10.1093/mnras/stad073}, \href {https://ui.adsabs.harvard.edu/abs/2023MNRAS.520...14C} {520, 14}

\bibitem[\protect\citeauthoryear{{Curti} et~al.,}{{Curti} et~al.}{2023}]{Curti2023}
{Curti} M.,  et~al., 2023, \mn@doi [arXiv e-prints] {10.48550/arXiv.2304.08516}, \href {https://ui.adsabs.harvard.edu/abs/2023arXiv230408516C} {p. arXiv:2304.08516}

\bibitem[\protect\citeauthoryear{{Curtis-Lake} et~al.,}{{Curtis-Lake} et~al.}{2012}]{Curtis-Lake2012}
{Curtis-Lake} E.,  et~al., 2012, \mn@doi [\mnras] {10.1111/j.1365-2966.2012.20720.x}, \href {https://ui.adsabs.harvard.edu/abs/2012MNRAS.422.1425C} {422, 1425}

\bibitem[\protect\citeauthoryear{{Davies} et~al.,}{{Davies} et~al.}{2018}]{Davies2018}
{Davies} F.~B.,  et~al., 2018, \mn@doi [\apj] {10.3847/1538-4357/aad6dc}, \href {https://ui.adsabs.harvard.edu/abs/2018ApJ...864..142D} {864, 142}

\bibitem[\protect\citeauthoryear{{Dayal} \& {Ferrara}}{{Dayal} \& {Ferrara}}{2018}]{Dayal2018}
{Dayal} P.,  {Ferrara} A.,  2018, \mn@doi [\physrep] {10.1016/j.physrep.2018.10.002}, \href {https://ui.adsabs.harvard.edu/abs/2018PhR...780....1D} {780, 1}

\bibitem[\protect\citeauthoryear{{De Barros} et~al.,}{{De Barros} et~al.}{2017}]{DeBarros2017}
{De Barros} S.,  et~al., 2017, \mn@doi [\aap] {10.1051/0004-6361/201731476}, \href {https://ui.adsabs.harvard.edu/abs/2017A&A...608A.123D} {608, A123}

\bibitem[\protect\citeauthoryear{{Dijkstra}}{{Dijkstra}}{2014}]{Dijkstra2014}
{Dijkstra} M.,  2014, \mn@doi [\pasa] {10.1017/pasa.2014.33}, \href {https://ui.adsabs.harvard.edu/abs/2014PASA...31...40D} {31, e040}

\bibitem[\protect\citeauthoryear{{Dijkstra}, {Gronke}  \& {Venkatesan}}{{Dijkstra} et~al.}{2016}]{Dijkstra2016}
{Dijkstra} M.,  {Gronke} M.,   {Venkatesan} A.,  2016, \mn@doi [\apj] {10.3847/0004-637X/828/2/71}, \href {https://ui.adsabs.harvard.edu/abs/2016ApJ...828...71D} {828, 71}

\bibitem[\protect\citeauthoryear{{Draine}}{{Draine}}{2011}]{Draine2011}
{Draine} B.~T.,  2011, {Physics of the Interstellar and Intergalactic Medium}

\bibitem[\protect\citeauthoryear{{Du}, {Shapley}, {Tang}, {Stark}, {Martin}, {Mobasher}, {Topping}  \& {Chevallard}}{{Du} et~al.}{2020}]{Du2020}
{Du} X.,  {Shapley} A.~E.,  {Tang} M.,  {Stark} D.~P.,  {Martin} C.~L.,  {Mobasher} B.,  {Topping} M.~W.,   {Chevallard} J.,  2020, \mn@doi [\apj] {10.3847/1538-4357/ab67b8}, \href {https://ui.adsabs.harvard.edu/abs/2020ApJ...890...65D} {890, 65}

\bibitem[\protect\citeauthoryear{{Elbers} \& {van de Weygaert}}{{Elbers} \& {van de Weygaert}}{2019}]{Elbers2019}
{Elbers} W.,  {van de Weygaert} R.,  2019, \mn@doi [\mnras] {10.1093/mnras/stz908}, \href {https://ui.adsabs.harvard.edu/abs/2019MNRAS.486.1523E} {486, 1523}

\bibitem[\protect\citeauthoryear{{Elbers} \& {van de Weygaert}}{{Elbers} \& {van de Weygaert}}{2023}]{Elbers2023}
{Elbers} W.,  {van de Weygaert} R.,  2023, \mn@doi [\mnras] {10.1093/mnras/stad120}, \href {https://ui.adsabs.harvard.edu/abs/2023MNRAS.520.2709E} {520, 2709}

\bibitem[\protect\citeauthoryear{{Endsley} \& {Stark}}{{Endsley} \& {Stark}}{2022}]{Endsley2022_overdensity}
{Endsley} R.,  {Stark} D.~P.,  2022, \mn@doi [\mnras] {10.1093/mnras/stac524}, \href {https://ui.adsabs.harvard.edu/abs/2022MNRAS.511.6042E} {511, 6042}

\bibitem[\protect\citeauthoryear{{Endsley}, {Stark}, {Chevallard}  \& {Charlot}}{{Endsley} et~al.}{2021a}]{Endsley2021}
{Endsley} R.,  {Stark} D.~P.,  {Chevallard} J.,   {Charlot} S.,  2021a, \mn@doi [\mnras] {10.1093/mnras/staa3370}, \href {https://ui.adsabs.harvard.edu/abs/2021MNRAS.500.5229E} {500, 5229}

\bibitem[\protect\citeauthoryear{{Endsley}, {Stark}, {Charlot}, {Chevallard}, {Robertson}, {Bouwens}  \& {Stefanon}}{{Endsley} et~al.}{2021b}]{Endsley2021_mmt}
{Endsley} R.,  {Stark} D.~P.,  {Charlot} S.,  {Chevallard} J.,  {Robertson} B.,  {Bouwens} R.~J.,   {Stefanon} M.,  2021b, \mn@doi [\mnras] {10.1093/mnras/stab432}, \href {https://ui.adsabs.harvard.edu/abs/2021MNRAS.502.6044E} {502, 6044}

\bibitem[\protect\citeauthoryear{{Endsley} et~al.,}{{Endsley} et~al.}{2023a}]{Endsley2023}
{Endsley} R.,  et~al., 2023a, \mn@doi [arXiv e-prints] {10.48550/arXiv.2306.05295}, \href {https://ui.adsabs.harvard.edu/abs/2023arXiv230605295E} {p. arXiv:2306.05295}

\bibitem[\protect\citeauthoryear{{Endsley}, {Stark}, {Whitler}, {Topping}, {Chen}, {Plat}, {Chisholm}  \& {Charlot}}{{Endsley} et~al.}{2023b}]{Endsley2023_ceers}
{Endsley} R.,  {Stark} D.~P.,  {Whitler} L.,  {Topping} M.~W.,  {Chen} Z.,  {Plat} A.,  {Chisholm} J.,   {Charlot} S.,  2023b, \mn@doi [\mnras] {10.1093/mnras/stad1919}, \href {https://ui.adsabs.harvard.edu/abs/2023MNRAS.524.2312E} {524, 2312}

\bibitem[\protect\citeauthoryear{{Erb} et~al.,}{{Erb} et~al.}{2014}]{Erb2014}
{Erb} D.~K.,  et~al., 2014, \mn@doi [\apj] {10.1088/0004-637X/795/1/33}, \href {https://ui.adsabs.harvard.edu/abs/2014ApJ...795...33E} {795, 33}

\bibitem[\protect\citeauthoryear{{Fan} et~al.,}{{Fan} et~al.}{2006}]{Fan2006}
{Fan} X.,  et~al., 2006, \mn@doi [\aj] {10.1086/504836}, \href {https://ui.adsabs.harvard.edu/abs/2006AJ....132..117F} {132, 117}

\bibitem[\protect\citeauthoryear{{Fan}, {Ba{\~n}ados}  \& {Simcoe}}{{Fan} et~al.}{2023}]{fan2023}
{Fan} X.,  {Ba{\~n}ados} E.,   {Simcoe} R.~A.,  2023, \mn@doi [\araa] {10.1146/annurev-astro-052920-102455}, \href {https://ui.adsabs.harvard.edu/abs/2023ARA&A..61..373F} {61, 373}

\bibitem[\protect\citeauthoryear{{Ferland} et~al.,}{{Ferland} et~al.}{2013}]{Ferland2013}
{Ferland} G.~J.,  et~al., 2013, \rmxaa, \href {https://ui.adsabs.harvard.edu/abs/2013RMxAA..49..137F} {49, 137}

\bibitem[\protect\citeauthoryear{{Ferruit} et~al.,}{{Ferruit} et~al.}{2022}]{Ferruit2022}
{Ferruit} P.,  et~al., 2022, \mn@doi [\aap] {10.1051/0004-6361/202142673}, \href {https://ui.adsabs.harvard.edu/abs/2022A&A...661A..81F} {661, A81}

\bibitem[\protect\citeauthoryear{{Finkelstein} et~al.,}{{Finkelstein} et~al.}{2019}]{Finkelstein2019}
{Finkelstein} S.~L.,  et~al., 2019, \mn@doi [\apj] {10.3847/1538-4357/ab1ea8}, \href {https://ui.adsabs.harvard.edu/abs/2019ApJ...879...36F} {879, 36}

\bibitem[\protect\citeauthoryear{{Finkelstein} et~al.,}{{Finkelstein} et~al.}{2022}]{Finkelstein2022}
{Finkelstein} S.~L.,  et~al., 2022, \mn@doi [\apjl] {10.3847/2041-8213/ac966e}, \href {https://ui.adsabs.harvard.edu/abs/2022ApJ...940L..55F} {940, L55}

\bibitem[\protect\citeauthoryear{{Finkelstein} et~al.,}{{Finkelstein} et~al.}{2023}]{Finkelstein2023}
{Finkelstein} S.~L.,  et~al., 2023, \mn@doi [\apjl] {10.3847/2041-8213/acade4}, \href {https://ui.adsabs.harvard.edu/abs/2023ApJ...946L..13F} {946, L13}

\bibitem[\protect\citeauthoryear{{Flury} et~al.,}{{Flury} et~al.}{2022a}]{Flury2022_I}
{Flury} S.~R.,  et~al., 2022a, \mn@doi [\apjs] {10.3847/1538-4365/ac5331}, \href {https://ui.adsabs.harvard.edu/abs/2022ApJS..260....1F} {260, 1}

\bibitem[\protect\citeauthoryear{{Flury} et~al.,}{{Flury} et~al.}{2022b}]{Flury2022_II}
{Flury} S.~R.,  et~al., 2022b, \mn@doi [\apj] {10.3847/1538-4357/ac61e4}, \href {https://ui.adsabs.harvard.edu/abs/2022ApJ...930..126F} {930, 126}

\bibitem[\protect\citeauthoryear{{Fontana} et~al.,}{{Fontana} et~al.}{2010}]{Fontana2010}
{Fontana} A.,  et~al., 2010, \mn@doi [\apjl] {10.1088/2041-8205/725/2/L205}, \href {https://ui.adsabs.harvard.edu/abs/2010ApJ...725L.205F} {725, L205}

\bibitem[\protect\citeauthoryear{{Foreman-Mackey}, {Hogg}, {Lang}  \& {Goodman}}{{Foreman-Mackey} et~al.}{2013}]{Foreman-Mackey2013}
{Foreman-Mackey} D.,  {Hogg} D.~W.,  {Lang} D.,   {Goodman} J.,  2013, \mn@doi [\pasp] {10.1086/670067}, \href {https://ui.adsabs.harvard.edu/abs/2013PASP..125..306F} {125, 306}

\bibitem[\protect\citeauthoryear{{F{\"o}rster Schreiber} et~al.,}{{F{\"o}rster Schreiber} et~al.}{2009}]{ForsterSchreiber2009}
{F{\"o}rster Schreiber} N.~M.,  et~al., 2009, \mn@doi [\apj] {10.1088/0004-637X/706/2/1364}, \href {https://ui.adsabs.harvard.edu/abs/2009ApJ...706.1364F} {706, 1364}

\bibitem[\protect\citeauthoryear{{Fujimoto} et~al.,}{{Fujimoto} et~al.}{2023}]{Fujimoto2023}
{Fujimoto} S.,  et~al., 2023, \mn@doi [\apjl] {10.3847/2041-8213/acd2d9}, \href {https://ui.adsabs.harvard.edu/abs/2023ApJ...949L..25F} {949, L25}

\bibitem[\protect\citeauthoryear{{Furlanetto}, {Zaldarriaga}  \& {Hernquist}}{{Furlanetto} et~al.}{2004}]{Furlanetto2004}
{Furlanetto} S.~R.,  {Zaldarriaga} M.,   {Hernquist} L.,  2004, \mn@doi [\apj] {10.1086/423025}, \href {https://ui.adsabs.harvard.edu/abs/2004ApJ...613....1F} {613, 1}

\bibitem[\protect\citeauthoryear{{Furtak}, {Shuntov}, {Atek}, {Zitrin}, {Richard}, {Lehnert}  \& {Chevallard}}{{Furtak} et~al.}{2023}]{Furtak2023}
{Furtak} L.~J.,  {Shuntov} M.,  {Atek} H.,  {Zitrin} A.,  {Richard} J.,  {Lehnert} M.~D.,   {Chevallard} J.,  2023, \mn@doi [\mnras] {10.1093/mnras/stac3717}, \href {https://ui.adsabs.harvard.edu/abs/2023MNRAS.519.3064F} {519, 3064}

\bibitem[\protect\citeauthoryear{{Gazagnes}, {Chisholm}, {Schaerer}, {Verhamme}  \& {Izotov}}{{Gazagnes} et~al.}{2020}]{Gazagnes2020}
{Gazagnes} S.,  {Chisholm} J.,  {Schaerer} D.,  {Verhamme} A.,   {Izotov} Y.,  2020, \mn@doi [\aap] {10.1051/0004-6361/202038096}, \href {https://ui.adsabs.harvard.edu/abs/2020A&A...639A..85G} {639, A85}

\bibitem[\protect\citeauthoryear{{Goulding} et~al.,}{{Goulding} et~al.}{2023}]{Goulding2023}
{Goulding} A.~D.,  et~al., 2023, \mn@doi [\apjl] {10.3847/2041-8213/acf7c5}, \href {https://ui.adsabs.harvard.edu/abs/2023ApJ...955L..24G} {955, L24}

\bibitem[\protect\citeauthoryear{{Greig}, {Mesinger}, {Haiman}  \& {Simcoe}}{{Greig} et~al.}{2017}]{Greig2017}
{Greig} B.,  {Mesinger} A.,  {Haiman} Z.,   {Simcoe} R.~A.,  2017, \mn@doi [\mnras] {10.1093/mnras/stw3351}, \href {https://ui.adsabs.harvard.edu/abs/2017MNRAS.466.4239G} {466, 4239}

\bibitem[\protect\citeauthoryear{{Gutkin}, {Charlot}  \& {Bruzual}}{{Gutkin} et~al.}{2016}]{Gutkin2016}
{Gutkin} J.,  {Charlot} S.,   {Bruzual} G.,  2016, \mn@doi [\mnras] {10.1093/mnras/stw1716}, \href {https://ui.adsabs.harvard.edu/abs/2016MNRAS.462.1757G} {462, 1757}

\bibitem[\protect\citeauthoryear{{Harikane} et~al.,}{{Harikane} et~al.}{2018}]{Harikane2018}
{Harikane} Y.,  et~al., 2018, \mn@doi [\apj] {10.3847/1538-4357/aabd80}, \href {https://ui.adsabs.harvard.edu/abs/2018ApJ...859...84H} {859, 84}

\bibitem[\protect\citeauthoryear{{Harikane} et~al.,}{{Harikane} et~al.}{2023a}]{Harikane2023_AGN}
{Harikane} Y.,  et~al., 2023a, \mn@doi [arXiv e-prints] {10.48550/arXiv.2303.11946}, \href {https://ui.adsabs.harvard.edu/abs/2023arXiv230311946H} {p. arXiv:2303.11946}

\bibitem[\protect\citeauthoryear{{Harikane}, {Nakajima}, {Ouchi}, {Umeda}, {Isobe}, {Ono}, {Xu}  \& {Zhang}}{{Harikane} et~al.}{2023b}]{Harikane2023}
{Harikane} Y.,  {Nakajima} K.,  {Ouchi} M.,  {Umeda} H.,  {Isobe} Y.,  {Ono} Y.,  {Xu} Y.,   {Zhang} Y.,  2023b, \mn@doi [arXiv e-prints] {10.48550/arXiv.2304.06658}, \href {https://ui.adsabs.harvard.edu/abs/2023arXiv230406658H} {p. arXiv:2304.06658}

\bibitem[\protect\citeauthoryear{Harris et~al.,}{Harris et~al.}{2020}]{harris2020array}
Harris C.~R.,  et~al., 2020, \mn@doi [Nature] {10.1038/s41586-020-2649-2}, 585, 357

\bibitem[\protect\citeauthoryear{{Hashimoto} et~al.,}{{Hashimoto} et~al.}{2017a}]{Hashimoto2017}
{Hashimoto} T.,  et~al., 2017a, \mn@doi [\mnras] {10.1093/mnras/stw2834}, \href {https://ui.adsabs.harvard.edu/abs/2017MNRAS.465.1543H} {465, 1543}

\bibitem[\protect\citeauthoryear{{Hashimoto} et~al.,}{{Hashimoto} et~al.}{2017b}]{Hashimoto2017_muse}
{Hashimoto} T.,  et~al., 2017b, \mn@doi [\aap] {10.1051/0004-6361/201731579}, \href {https://ui.adsabs.harvard.edu/abs/2017A&A...608A..10H} {608, A10}

\bibitem[\protect\citeauthoryear{{Hashimoto} et~al.,}{{Hashimoto} et~al.}{2023}]{Hashimoto2023}
{Hashimoto} T.,  et~al., 2023, \mn@doi [\apjl] {10.3847/2041-8213/acf57c}, \href {https://ui.adsabs.harvard.edu/abs/2023ApJ...955L...2H} {955, L2}

\bibitem[\protect\citeauthoryear{{Hayes}, {Schaerer}, {{\"O}stlin}, {Mas-Hesse}, {Atek}  \& {Kunth}}{{Hayes} et~al.}{2011}]{Hayes2011}
{Hayes} M.,  {Schaerer} D.,  {{\"O}stlin} G.,  {Mas-Hesse} J.~M.,  {Atek} H.,   {Kunth} D.,  2011, \mn@doi [\apj] {10.1088/0004-637X/730/1/8}, \href {https://ui.adsabs.harvard.edu/abs/2011ApJ...730....8H} {730, 8}

\bibitem[\protect\citeauthoryear{{Heckman}, {Sembach}, {Meurer}, {Leitherer}, {Calzetti}  \& {Martin}}{{Heckman} et~al.}{2001}]{Heckman2001}
{Heckman} T.~M.,  {Sembach} K.~R.,  {Meurer} G.~R.,  {Leitherer} C.,  {Calzetti} D.,   {Martin} C.~L.,  2001, \mn@doi [\apj] {10.1086/322475}, \href {https://ui.adsabs.harvard.edu/abs/2001ApJ...558...56H} {558, 56}

\bibitem[\protect\citeauthoryear{{Heckman} et~al.,}{{Heckman} et~al.}{2011}]{Heckman2011}
{Heckman} T.~M.,  et~al., 2011, \mn@doi [\apj] {10.1088/0004-637X/730/1/5}, \href {https://ui.adsabs.harvard.edu/abs/2011ApJ...730....5H} {730, 5}

\bibitem[\protect\citeauthoryear{{Heintz} et~al.,}{{Heintz} et~al.}{2023}]{Heintz2023}
{Heintz} K.~E.,  et~al., 2023, \mn@doi [arXiv e-prints] {10.48550/arXiv.2306.00647}, \href {https://ui.adsabs.harvard.edu/abs/2023arXiv230600647H} {p. arXiv:2306.00647}

\bibitem[\protect\citeauthoryear{{Henry}, {Scarlata}, {Martin}  \& {Erb}}{{Henry} et~al.}{2015}]{Henry2015}
{Henry} A.,  {Scarlata} C.,  {Martin} C.~L.,   {Erb} D.,  2015, \mn@doi [\apj] {10.1088/0004-637X/809/1/19}, \href {https://ui.adsabs.harvard.edu/abs/2015ApJ...809...19H} {809, 19}

\bibitem[\protect\citeauthoryear{{Hoag} et~al.,}{{Hoag} et~al.}{2019}]{Hoag2019}
{Hoag} A.,  et~al., 2019, \mn@doi [\apj] {10.3847/1538-4357/ab1de7}, \href {https://ui.adsabs.harvard.edu/abs/2019ApJ...878...12H} {878, 12}

\bibitem[\protect\citeauthoryear{{Hu} et~al.,}{{Hu} et~al.}{2021}]{Hu2021}
{Hu} W.,  et~al., 2021, \mn@doi [Nature Astronomy] {10.1038/s41550-020-01291-y}, \href {https://ui.adsabs.harvard.edu/abs/2021NatAs...5..485H} {5, 485}

\bibitem[\protect\citeauthoryear{Hunter}{Hunter}{2007}]{Hunter:2007}
Hunter J.~D.,  2007, \mn@doi [Computing in Science \& Engineering] {10.1109/MCSE.2007.55}, 9, 90

\bibitem[\protect\citeauthoryear{{Iliev}, {Mellema}, {Pen}, {Merz}, {Shapiro}  \& {Alvarez}}{{Iliev} et~al.}{2006}]{Iliev2006}
{Iliev} I.~T.,  {Mellema} G.,  {Pen} U.~L.,  {Merz} H.,  {Shapiro} P.~R.,   {Alvarez} M.~A.,  2006, \mn@doi [\mnras] {10.1111/j.1365-2966.2006.10502.x}, \href {https://ui.adsabs.harvard.edu/abs/2006MNRAS.369.1625I} {369, 1625}

\bibitem[\protect\citeauthoryear{{Inoue}, {Shimizu}, {Iwata}  \& {Tanaka}}{{Inoue} et~al.}{2014}]{Inoue2014}
{Inoue} A.~K.,  {Shimizu} I.,  {Iwata} I.,   {Tanaka} M.,  2014, \mn@doi [\mnras] {10.1093/mnras/stu936}, \href {https://ui.adsabs.harvard.edu/abs/2014MNRAS.442.1805I} {442, 1805}

\bibitem[\protect\citeauthoryear{{Izotov}, {Schaerer}, {Thuan}, {Worseck}, {Guseva}, {Orlitov{\'a}}  \& {Verhamme}}{{Izotov} et~al.}{2016}]{Izotov2016}
{Izotov} Y.~I.,  {Schaerer} D.,  {Thuan} T.~X.,  {Worseck} G.,  {Guseva} N.~G.,  {Orlitov{\'a}} I.,   {Verhamme} A.,  2016, \mn@doi [\mnras] {10.1093/mnras/stw1205}, \href {https://ui.adsabs.harvard.edu/abs/2016MNRAS.461.3683I} {461, 3683}

\bibitem[\protect\citeauthoryear{{Izotov}, {Worseck}, {Schaerer}, {Guseva}, {Thuan}, {Fricke}  \& {Orlitov{\'a}}}{{Izotov} et~al.}{2018}]{Izotov2018_higho32}
{Izotov} Y.~I.,  {Worseck} G.,  {Schaerer} D.,  {Guseva} N.~G.,  {Thuan} T.~X.,  {Fricke} Verhamme A.,   {Orlitov{\'a}} I.,  2018, \mn@doi [\mnras] {10.1093/mnras/sty1378}, \href {https://ui.adsabs.harvard.edu/abs/2018MNRAS.478.4851I} {478, 4851}

\bibitem[\protect\citeauthoryear{{Izotov}, {Schaerer}, {Worseck}, {Verhamme}, {Guseva}, {Thuan}, {Orlitov{\'a}}  \& {Fricke}}{{Izotov} et~al.}{2020}]{Izotov2020}
{Izotov} Y.~I.,  {Schaerer} D.,  {Worseck} G.,  {Verhamme} A.,  {Guseva} N.~G.,  {Thuan} T.~X.,  {Orlitov{\'a}} I.,   {Fricke} K.~J.,  2020, \mn@doi [\mnras] {10.1093/mnras/stz3041}, \href {https://ui.adsabs.harvard.edu/abs/2020MNRAS.491..468I} {491, 468}

\bibitem[\protect\citeauthoryear{{Jakobsen} et~al.,}{{Jakobsen} et~al.}{2022}]{Jakobsen2022}
{Jakobsen} P.,  et~al., 2022, \mn@doi [\aap] {10.1051/0004-6361/202142663}, \href {https://ui.adsabs.harvard.edu/abs/2022A&A...661A..80J} {661, A80}

\bibitem[\protect\citeauthoryear{{Jiang} et~al.,}{{Jiang} et~al.}{2022}]{Jiang2022}
{Jiang} L.,  et~al., 2022, \mn@doi [Nature Astronomy] {10.1038/s41550-022-01708-w}, \href {https://ui.adsabs.harvard.edu/abs/2022NatAs...6..850J} {6, 850}

\bibitem[\protect\citeauthoryear{{Jin} et~al.,}{{Jin} et~al.}{2023}]{Jin2023}
{Jin} X.,  et~al., 2023, \mn@doi [\apj] {10.3847/1538-4357/aca678}, \href {https://ui.adsabs.harvard.edu/abs/2023ApJ...942...59J} {942, 59}

\bibitem[\protect\citeauthoryear{{Johnson}}{{Johnson}}{2021}]{Johnson2021_sedpy}
{Johnson} B.~D.,  2021, {bd-j/sedpy: sedpy v0.2.0}, Zenodo, \mn@doi{10.5281/zenodo.4582723}

\bibitem[\protect\citeauthoryear{{Jones} et~al.,}{{Jones} et~al.}{2023}]{Jones2023}
{Jones} G.~C.,  et~al., 2023, \mn@doi [arXiv e-prints] {10.48550/arXiv.2306.02471}, \href {https://ui.adsabs.harvard.edu/abs/2023arXiv230602471J} {p. arXiv:2306.02471}

\bibitem[\protect\citeauthoryear{{Jung} et~al.,}{{Jung} et~al.}{2020}]{Jung2020}
{Jung} I.,  et~al., 2020, \mn@doi [\apj] {10.3847/1538-4357/abbd44}, \href {https://ui.adsabs.harvard.edu/abs/2020ApJ...904..144J} {904, 144}

\bibitem[\protect\citeauthoryear{{Jung} et~al.,}{{Jung} et~al.}{2022}]{Jung2022}
{Jung} I.,  et~al., 2022, \mn@doi [arXiv e-prints] {10.48550/arXiv.2212.09850}, \href {https://ui.adsabs.harvard.edu/abs/2022arXiv221209850J} {p. arXiv:2212.09850}

\bibitem[\protect\citeauthoryear{{Jung} et~al.,}{{Jung} et~al.}{2023}]{Jung2023}
{Jung} I.,  et~al., 2023, \mn@doi [arXiv e-prints] {10.48550/arXiv.2304.05385}, \href {https://ui.adsabs.harvard.edu/abs/2023arXiv230405385J} {p. arXiv:2304.05385}

\bibitem[\protect\citeauthoryear{{Kerutt} et~al.,}{{Kerutt} et~al.}{2022}]{Kerutt2022}
{Kerutt} J.,  et~al., 2022, \mn@doi [\aap] {10.1051/0004-6361/202141900}, \href {https://ui.adsabs.harvard.edu/abs/2022A&A...659A.183K} {659, A183}

\bibitem[\protect\citeauthoryear{{Kim} et~al.,}{{Kim} et~al.}{2023}]{Kim2023}
{Kim} K.~J.,  et~al., 2023, \mn@doi [\apjl] {10.3847/2041-8213/acf0c5}, \href {https://ui.adsabs.harvard.edu/abs/2023ApJ...955L..17K} {955, L17}

\bibitem[\protect\citeauthoryear{{Kluyver} et~al.,}{{Kluyver} et~al.}{2016}]{Kluyver2016}
{Kluyver} T.,  et~al., 2016, in , IOS Press.
pp 87--90, \mn@doi{10.3233/978-1-61499-649-1-87}

\bibitem[\protect\citeauthoryear{{Kocevski} et~al.,}{{Kocevski} et~al.}{2023}]{Kocevski2023}
{Kocevski} D.~D.,  et~al., 2023, \mn@doi [\apjl] {10.3847/2041-8213/ace5a0}, \href {https://ui.adsabs.harvard.edu/abs/2023ApJ...954L...4K} {954, L4}

\bibitem[\protect\citeauthoryear{{Kokorev} et~al.,}{{Kokorev} et~al.}{2022}]{Kokorev2022}
{Kokorev} V.,  et~al., 2022, \mn@doi [\apjs] {10.3847/1538-4365/ac9909}, \href {https://ui.adsabs.harvard.edu/abs/2022ApJS..263...38K} {263, 38}

\bibitem[\protect\citeauthoryear{{Kron}}{{Kron}}{1980}]{Kron1980}
{Kron} R.~G.,  1980, \mn@doi [\apjs] {10.1086/190669}, \href {https://ui.adsabs.harvard.edu/abs/1980ApJS...43..305K} {43, 305}

\bibitem[\protect\citeauthoryear{{Kulkarni}, {Worseck}  \& {Hennawi}}{{Kulkarni} et~al.}{2019}]{Kulkarni2019}
{Kulkarni} G.,  {Worseck} G.,   {Hennawi} J.~F.,  2019, \mn@doi [\mnras] {10.1093/mnras/stz1493}, \href {https://ui.adsabs.harvard.edu/abs/2019MNRAS.488.1035K} {488, 1035}

\bibitem[\protect\citeauthoryear{{Larson} et~al.,}{{Larson} et~al.}{2022}]{Larson2022}
{Larson} R.~L.,  et~al., 2022, \mn@doi [\apj] {10.3847/1538-4357/ac5dbd}, \href {https://ui.adsabs.harvard.edu/abs/2022ApJ...930..104L} {930, 104}

\bibitem[\protect\citeauthoryear{{Larson} et~al.,}{{Larson} et~al.}{2023}]{Larson2023}
{Larson} R.~L.,  et~al., 2023, \mn@doi [\apjl] {10.3847/2041-8213/ace619}, \href {https://ui.adsabs.harvard.edu/abs/2023ApJ...953L..29L} {953, L29}

\bibitem[\protect\citeauthoryear{{Leclercq} et~al.,}{{Leclercq} et~al.}{2017}]{Leclercq2017}
{Leclercq} F.,  et~al., 2017, \mn@doi [\aap] {10.1051/0004-6361/201731480}, \href {https://ui.adsabs.harvard.edu/abs/2017A&A...608A...8L} {608, A8}

\bibitem[\protect\citeauthoryear{{Lin} et~al.,}{{Lin} et~al.}{2024}]{Lin2024}
{Lin} X.,  et~al., 2024, \mn@doi [arXiv e-prints] {10.48550/arXiv.2401.09532}, \href {https://ui.adsabs.harvard.edu/abs/2024arXiv240109532L} {p. arXiv:2401.09532}

\bibitem[\protect\citeauthoryear{{Lu}, {Mason}, {Hutter}, {Mesinger}, {Qin}, {Stark}  \& {Endsley}}{{Lu} et~al.}{2023}]{Lu2023}
{Lu} T.-Y.,  {Mason} C.,  {Hutter} A.,  {Mesinger} A.,  {Qin} Y.,  {Stark} D.~P.,   {Endsley} R.,  2023, \mn@doi [arXiv e-prints] {10.48550/arXiv.2304.11192}, \href {https://ui.adsabs.harvard.edu/abs/2023arXiv230411192L} {p. arXiv:2304.11192}

\bibitem[\protect\citeauthoryear{{Luridiana}, {Morisset}  \& {Shaw}}{{Luridiana} et~al.}{2015}]{Luridiana2015}
{Luridiana} V.,  {Morisset} C.,   {Shaw} R.~A.,  2015, \mn@doi [\aap] {10.1051/0004-6361/201323152}, \href {https://ui.adsabs.harvard.edu/abs/2015A&A...573A..42L} {573, A42}

\bibitem[\protect\citeauthoryear{{Malhotra} \& {Rhoads}}{{Malhotra} \& {Rhoads}}{2002}]{Malhotra2002}
{Malhotra} S.,  {Rhoads} J.~E.,  2002, \mn@doi [\apjl] {10.1086/338980}, \href {https://ui.adsabs.harvard.edu/abs/2002ApJ...565L..71M} {565, L71}

\bibitem[\protect\citeauthoryear{{Mary}, {Bacon}, {Conseil}, {Piqueras}  \& {Schutz}}{{Mary} et~al.}{2020}]{Mary2020}
{Mary} D.,  {Bacon} R.,  {Conseil} S.,  {Piqueras} L.,   {Schutz} A.,  2020, \mn@doi [\aap] {10.1051/0004-6361/201937001}, \href {https://ui.adsabs.harvard.edu/abs/2020A&A...635A.194M} {635, A194}

\bibitem[\protect\citeauthoryear{{Maseda} et~al.,}{{Maseda} et~al.}{2018}]{Maseda2018}
{Maseda} M.~V.,  et~al., 2018, \mn@doi [\apjl] {10.3847/2041-8213/aade4b}, \href {https://ui.adsabs.harvard.edu/abs/2018ApJ...865L...1M} {865, L1}

\bibitem[\protect\citeauthoryear{{Maseda} et~al.,}{{Maseda} et~al.}{2023}]{Maseda2023}
{Maseda} M.~V.,  et~al., 2023, \mn@doi [\apj] {10.3847/1538-4357/acf12b}, \href {https://ui.adsabs.harvard.edu/abs/2023ApJ...956...11M} {956, 11}

\bibitem[\protect\citeauthoryear{{Mason} \& {Gronke}}{{Mason} \& {Gronke}}{2020}]{Mason2020_bubble}
{Mason} C.~A.,  {Gronke} M.,  2020, \mn@doi [\mnras] {10.1093/mnras/staa2910}, \href {https://ui.adsabs.harvard.edu/abs/2020MNRAS.499.1395M} {499, 1395}

\bibitem[\protect\citeauthoryear{{Mason}, {Treu}, {Dijkstra}, {Mesinger}, {Trenti}, {Pentericci}, {de Barros}  \& {Vanzella}}{{Mason} et~al.}{2018a}]{Mason2018}
{Mason} C.~A.,  {Treu} T.,  {Dijkstra} M.,  {Mesinger} A.,  {Trenti} M.,  {Pentericci} L.,  {de Barros} S.,   {Vanzella} E.,  2018a, \mn@doi [\apj] {10.3847/1538-4357/aab0a7}, \href {https://ui.adsabs.harvard.edu/abs/2018ApJ...856....2M} {856, 2}

\bibitem[\protect\citeauthoryear{{Mason} et~al.,}{{Mason} et~al.}{2018b}]{Mason2018_transmission}
{Mason} C.~A.,  et~al., 2018b, \mn@doi [\apjl] {10.3847/2041-8213/aabbab}, \href {https://ui.adsabs.harvard.edu/abs/2018ApJ...857L..11M} {857, L11}

\bibitem[\protect\citeauthoryear{{Mason} et~al.,}{{Mason} et~al.}{2019}]{Mason2019}
{Mason} C.~A.,  et~al., 2019, \mn@doi [\mnras] {10.1093/mnras/stz632}, \href {https://ui.adsabs.harvard.edu/abs/2019MNRAS.485.3947M} {485, 3947}

\bibitem[\protect\citeauthoryear{{Matsuoka} et~al.,}{{Matsuoka} et~al.}{2018}]{Matsuoka2018}
{Matsuoka} Y.,  et~al., 2018, \mn@doi [\apj] {10.3847/1538-4357/aaee7a}, \href {https://ui.adsabs.harvard.edu/abs/2018ApJ...869..150M} {869, 150}

\bibitem[\protect\citeauthoryear{{Matthee} et~al.,}{{Matthee} et~al.}{2021}]{Matthee2021}
{Matthee} J.,  et~al., 2021, \mn@doi [\mnras] {10.1093/mnras/stab1304}, \href {https://ui.adsabs.harvard.edu/abs/2021MNRAS.505.1382M} {505, 1382}

\bibitem[\protect\citeauthoryear{{McGreer}, {Mesinger}  \& {D'Odorico}}{{McGreer} et~al.}{2015}]{McGreer2015}
{McGreer} I.~D.,  {Mesinger} A.,   {D'Odorico} V.,  2015, \mn@doi [\mnras] {10.1093/mnras/stu2449}, \href {https://ui.adsabs.harvard.edu/abs/2015MNRAS.447..499M} {447, 499}

\bibitem[\protect\citeauthoryear{{Mesinger}, {Aykutalp}, {Vanzella}, {Pentericci}, {Ferrara}  \& {Dijkstra}}{{Mesinger} et~al.}{2015}]{Mesinger2015}
{Mesinger} A.,  {Aykutalp} A.,  {Vanzella} E.,  {Pentericci} L.,  {Ferrara} A.,   {Dijkstra} M.,  2015, \mn@doi [\mnras] {10.1093/mnras/stu2089}, \href {https://ui.adsabs.harvard.edu/abs/2015MNRAS.446..566M} {446, 566}

\bibitem[\protect\citeauthoryear{{Morishita} et~al.,}{{Morishita} et~al.}{2023}]{Morishita2023}
{Morishita} T.,  et~al., 2023, \mn@doi [\apjl] {10.3847/2041-8213/acb99e}, \href {https://ui.adsabs.harvard.edu/abs/2023ApJ...947L..24M} {947, L24}

\bibitem[\protect\citeauthoryear{{Naidu}, {Tacchella}, {Mason}, {Bose}, {Oesch}  \& {Conroy}}{{Naidu} et~al.}{2020}]{Naidu2020}
{Naidu} R.~P.,  {Tacchella} S.,  {Mason} C.~A.,  {Bose} S.,  {Oesch} P.~A.,   {Conroy} C.,  2020, \mn@doi [\apj] {10.3847/1538-4357/ab7cc9}, \href {https://ui.adsabs.harvard.edu/abs/2020ApJ...892..109N} {892, 109}

\bibitem[\protect\citeauthoryear{{Naidu} et~al.,}{{Naidu} et~al.}{2022}]{Naidu2022}
{Naidu} R.~P.,  et~al., 2022, \mn@doi [\mnras] {10.1093/mnras/stab3601}, \href {https://ui.adsabs.harvard.edu/abs/2022MNRAS.510.4582N} {510, 4582}

\bibitem[\protect\citeauthoryear{{Nakajima} et~al.,}{{Nakajima} et~al.}{2012}]{Nakajima2012}
{Nakajima} K.,  et~al., 2012, \mn@doi [\apj] {10.1088/0004-637X/745/1/12}, \href {https://ui.adsabs.harvard.edu/abs/2012ApJ...745...12N} {745, 12}

\bibitem[\protect\citeauthoryear{{Nakajima}, {Fletcher}, {Ellis}, {Robertson}  \& {Iwata}}{{Nakajima} et~al.}{2018}]{Nakajima2018}
{Nakajima} K.,  {Fletcher} T.,  {Ellis} R.~S.,  {Robertson} B.~E.,   {Iwata} I.,  2018, \mn@doi [\mnras] {10.1093/mnras/sty750}, \href {https://ui.adsabs.harvard.edu/abs/2018MNRAS.477.2098N} {477, 2098}

\bibitem[\protect\citeauthoryear{{Nakajima}, {Ouchi}, {Isobe}, {Harikane}, {Zhang}, {Ono}, {Umeda}  \& {Oguri}}{{Nakajima} et~al.}{2023}]{Nakajima2023}
{Nakajima} K.,  {Ouchi} M.,  {Isobe} Y.,  {Harikane} Y.,  {Zhang} Y.,  {Ono} Y.,  {Umeda} H.,   {Oguri} M.,  2023, \mn@doi [\apjs] {10.3847/1538-4365/acd556}, \href {https://ui.adsabs.harvard.edu/abs/2023ApJS..269...33N} {269, 33}

\bibitem[\protect\citeauthoryear{{Nakane} et~al.,}{{Nakane} et~al.}{2023}]{Nakane2023}
{Nakane} M.,  et~al., 2023, \mn@doi [arXiv e-prints] {10.48550/arXiv.2312.06804}, \href {https://ui.adsabs.harvard.edu/abs/2023arXiv231206804N} {p. arXiv:2312.06804}

\bibitem[\protect\citeauthoryear{{Napolitano} et~al.,}{{Napolitano} et~al.}{2024}]{Napolitano2024}
{Napolitano} L.,  et~al., 2024, arXiv e-prints, \href {https://ui.adsabs.harvard.edu/abs/2024arXiv240211220N} {p. arXiv:2402.11220}

\bibitem[\protect\citeauthoryear{{Ning} et~al.,}{{Ning} et~al.}{2020}]{Ning2020}
{Ning} Y.,  et~al., 2020, \mn@doi [\apj] {10.3847/1538-4357/abb705}, \href {https://ui.adsabs.harvard.edu/abs/2020ApJ...903....4N} {903, 4}

\bibitem[\protect\citeauthoryear{{Ning}, {Cai}, {Jiang}, {Lin}, {Fu}  \& {Spinoso}}{{Ning} et~al.}{2023}]{Ning2023}
{Ning} Y.,  {Cai} Z.,  {Jiang} L.,  {Lin} X.,  {Fu} S.,   {Spinoso} D.,  2023, \mn@doi [\apjl] {10.3847/2041-8213/acb26b}, \href {https://ui.adsabs.harvard.edu/abs/2023ApJ...944L...1N} {944, L1}

\bibitem[\protect\citeauthoryear{{Oesch} et~al.,}{{Oesch} et~al.}{2015}]{Oesch2015}
{Oesch} P.~A.,  et~al., 2015, \mn@doi [\apjl] {10.1088/2041-8205/804/2/L30}, \href {https://ui.adsabs.harvard.edu/abs/2015ApJ...804L..30O} {804, L30}

\bibitem[\protect\citeauthoryear{{Oke} \& {Gunn}}{{Oke} \& {Gunn}}{1983}]{Oke1983}
{Oke} J.~B.,  {Gunn} J.~E.,  1983, \mn@doi [\apj] {10.1086/160817}, \href {https://ui.adsabs.harvard.edu/abs/1983ApJ...266..713O} {266, 713}

\bibitem[\protect\citeauthoryear{{Ono} et~al.,}{{Ono} et~al.}{2012}]{Ono2012}
{Ono} Y.,  et~al., 2012, \mn@doi [\apj] {10.1088/0004-637X/744/2/83}, \href {https://ui.adsabs.harvard.edu/abs/2012ApJ...744...83O} {744, 83}

\bibitem[\protect\citeauthoryear{{Orlitov{\'a}}, {Verhamme}, {Henry}, {Scarlata}, {Jaskot}, {Oey}  \& {Schaerer}}{{Orlitov{\'a}} et~al.}{2018}]{Orlitova2018}
{Orlitov{\'a}} I.,  {Verhamme} A.,  {Henry} A.,  {Scarlata} C.,  {Jaskot} A.,  {Oey} M.~S.,   {Schaerer} D.,  2018, \mn@doi [\aap] {10.1051/0004-6361/201732478}, \href {https://ui.adsabs.harvard.edu/abs/2018A&A...616A..60O} {616, A60}

\bibitem[\protect\citeauthoryear{{Osterbrock} \& {Ferland}}{{Osterbrock} \& {Ferland}}{2006}]{Osterbrock2006}
{Osterbrock} D.~E.,  {Ferland} G.~J.,  2006, {Astrophysics of gaseous nebulae and active galactic nuclei}

\bibitem[\protect\citeauthoryear{{Ouchi}, {Ono}  \& {Shibuya}}{{Ouchi} et~al.}{2020}]{Ouchi2020}
{Ouchi} M.,  {Ono} Y.,   {Shibuya} T.,  2020, \mn@doi [\araa] {10.1146/annurev-astro-032620-021859}, \href {https://ui.adsabs.harvard.edu/abs/2020ARA&A..58..617O} {58, 617}

\bibitem[\protect\citeauthoryear{{Pahl}, {Shapley}, {Steidel}, {Chen}  \& {Reddy}}{{Pahl} et~al.}{2021}]{Pahl2021}
{Pahl} A.~J.,  {Shapley} A.,  {Steidel} C.~C.,  {Chen} Y.,   {Reddy} N.~A.,  2021, \mn@doi [\mnras] {10.1093/mnras/stab1374}, \href {https://ui.adsabs.harvard.edu/abs/2021MNRAS.505.2447P} {505, 2447}

\bibitem[\protect\citeauthoryear{{Pei}}{{Pei}}{1992}]{Pei1992}
{Pei} Y.~C.,  1992, \mn@doi [\apj] {10.1086/171637}, \href {https://ui.adsabs.harvard.edu/abs/1992ApJ...395..130P} {395, 130}

\bibitem[\protect\citeauthoryear{{Pentericci} et~al.,}{{Pentericci} et~al.}{2014}]{Pentericci2014}
{Pentericci} L.,  et~al., 2014, \mn@doi [\apj] {10.1088/0004-637X/793/2/113}, \href {https://ui.adsabs.harvard.edu/abs/2014ApJ...793..113P} {793, 113}

\bibitem[\protect\citeauthoryear{{Planck Collaboration} et~al.,}{{Planck Collaboration} et~al.}{2016}]{PlanckCollaboration2016}
{Planck Collaboration} et~al., 2016, \mn@doi [\aap] {10.1051/0004-6361/201628897}, \href {https://ui.adsabs.harvard.edu/abs/2016A&A...596A.108P} {596, A108}

\bibitem[\protect\citeauthoryear{{Planck Collaboration} et~al.,}{{Planck Collaboration} et~al.}{2020}]{PlanckCollaboration2020}
{Planck Collaboration} et~al., 2020, \mn@doi [\aap] {10.1051/0004-6361/201833910}, \href {https://ui.adsabs.harvard.edu/abs/2020A&A...641A...6P} {641, A6}

\bibitem[\protect\citeauthoryear{{Price} et~al.,}{{Price} et~al.}{2014}]{Price2014}
{Price} S.~H.,  et~al., 2014, \mn@doi [\apj] {10.1088/0004-637X/788/1/86}, \href {https://ui.adsabs.harvard.edu/abs/2014ApJ...788...86P} {788, 86}

\bibitem[\protect\citeauthoryear{{Prieto-Lyon} et~al.,}{{Prieto-Lyon} et~al.}{2023}]{Prieto-Lyon2023}
{Prieto-Lyon} G.,  et~al., 2023, \mn@doi [\apj] {10.3847/1538-4357/acf715}, \href {https://ui.adsabs.harvard.edu/abs/2023ApJ...956..136P} {956, 136}

\bibitem[\protect\citeauthoryear{{Qin}, {Mesinger}, {Bosman}  \& {Viel}}{{Qin} et~al.}{2021}]{Qin2021}
{Qin} Y.,  {Mesinger} A.,  {Bosman} S. E.~I.,   {Viel} M.,  2021, \mn@doi [\mnras] {10.1093/mnras/stab1833}, \href {https://ui.adsabs.harvard.edu/abs/2021MNRAS.506.2390Q} {506, 2390}

\bibitem[\protect\citeauthoryear{{Raiter}, {Schaerer}  \& {Fosbury}}{{Raiter} et~al.}{2010}]{Raiter2010}
{Raiter} A.,  {Schaerer} D.,   {Fosbury} R.~A.~E.,  2010, \mn@doi [\aap] {10.1051/0004-6361/201015236}, \href {https://ui.adsabs.harvard.edu/abs/2010A&A...523A..64R} {523, A64}

\bibitem[\protect\citeauthoryear{{Reddy} et~al.,}{{Reddy} et~al.}{2015}]{Reddy2015}
{Reddy} N.~A.,  et~al., 2015, \mn@doi [\apj] {10.1088/0004-637X/806/2/259}, \href {https://ui.adsabs.harvard.edu/abs/2015ApJ...806..259R} {806, 259}

\bibitem[\protect\citeauthoryear{{Reddy}, {Steidel}, {Pettini}, {Bogosavljevi{\'c}}  \& {Shapley}}{{Reddy} et~al.}{2016}]{Reddy2016}
{Reddy} N.~A.,  {Steidel} C.~C.,  {Pettini} M.,  {Bogosavljevi{\'c}} M.,   {Shapley} A.~E.,  2016, \mn@doi [\apj] {10.3847/0004-637X/828/2/108}, \href {https://ui.adsabs.harvard.edu/abs/2016ApJ...828..108R} {828, 108}

\bibitem[\protect\citeauthoryear{{Reddy} et~al.,}{{Reddy} et~al.}{2020}]{Reddy2020}
{Reddy} N.~A.,  et~al., 2020, \mn@doi [\apj] {10.3847/1538-4357/abb674}, \href {https://ui.adsabs.harvard.edu/abs/2020ApJ...902..123R} {902, 123}

\bibitem[\protect\citeauthoryear{{Roberts-Borsani} et~al.,}{{Roberts-Borsani} et~al.}{2016}]{Roberts-Borsani2016}
{Roberts-Borsani} G.~W.,  et~al., 2016, \mn@doi [\apj] {10.3847/0004-637X/823/2/143}, \href {https://ui.adsabs.harvard.edu/abs/2016ApJ...823..143R} {823, 143}

\bibitem[\protect\citeauthoryear{{Roberts-Borsani} et~al.,}{{Roberts-Borsani} et~al.}{2023}]{Roberts-Borsani2023}
{Roberts-Borsani} G.,  et~al., 2023, \mn@doi [\nat] {10.1038/s41586-023-05994-w}, \href {https://ui.adsabs.harvard.edu/abs/2023Natur.618..480R} {618, 480}

\bibitem[\protect\citeauthoryear{{Robertson}}{{Robertson}}{2022}]{Robertson2022}
{Robertson} B.~E.,  2022, \mn@doi [\araa] {10.1146/annurev-astro-120221-044656}, \href {https://ui.adsabs.harvard.edu/abs/2022ARA&A..60..121R} {60, 121}

\bibitem[\protect\citeauthoryear{{Robertson}, {Ellis}, {Furlanetto}  \& {Dunlop}}{{Robertson} et~al.}{2015}]{Robertson2015}
{Robertson} B.~E.,  {Ellis} R.~S.,  {Furlanetto} S.~R.,   {Dunlop} J.~S.,  2015, \mn@doi [\apjl] {10.1088/2041-8205/802/2/L19}, \href {https://ui.adsabs.harvard.edu/abs/2015ApJ...802L..19R} {802, L19}

\bibitem[\protect\citeauthoryear{{Roy} et~al.,}{{Roy} et~al.}{2023}]{Roy2023}
{Roy} N.,  et~al., 2023, \mn@doi [\apjl] {10.3847/2041-8213/acdbce}, \href {https://ui.adsabs.harvard.edu/abs/2023ApJ...952L..14R} {952, L14}

\bibitem[\protect\citeauthoryear{{Sanders}, {Shapley}, {Topping}, {Reddy}  \& {Brammer}}{{Sanders} et~al.}{2023a}]{Sanders2023_directZ}
{Sanders} R.~L.,  {Shapley} A.~E.,  {Topping} M.~W.,  {Reddy} N.~A.,   {Brammer} G.~B.,  2023a, \mn@doi [arXiv e-prints] {10.48550/arXiv.2303.08149}, \href {https://ui.adsabs.harvard.edu/abs/2023arXiv230308149S} {p. arXiv:2303.08149}

\bibitem[\protect\citeauthoryear{{Sanders}, {Shapley}, {Topping}, {Reddy}  \& {Brammer}}{{Sanders} et~al.}{2023b}]{Sanders2023}
{Sanders} R.~L.,  {Shapley} A.~E.,  {Topping} M.~W.,  {Reddy} N.~A.,   {Brammer} G.~B.,  2023b, \mn@doi [\apj] {10.3847/1538-4357/acedad}, \href {https://ui.adsabs.harvard.edu/abs/2023ApJ...955...54S} {955, 54}

\bibitem[\protect\citeauthoryear{{Saxena} et~al.,}{{Saxena} et~al.}{2023a}]{Saxena2023b}
{Saxena} A.,  et~al., 2023a, \mn@doi [arXiv e-prints] {10.48550/arXiv.2306.04536}, \href {https://ui.adsabs.harvard.edu/abs/2023arXiv230604536S} {p. arXiv:2306.04536}

\bibitem[\protect\citeauthoryear{{Saxena} et~al.,}{{Saxena} et~al.}{2023b}]{Saxena2023}
{Saxena} A.,  et~al., 2023b, \mn@doi [\aap] {10.1051/0004-6361/202346245}, \href {https://ui.adsabs.harvard.edu/abs/2023A&A...678A..68S} {678, A68}

\bibitem[\protect\citeauthoryear{{Schenker}, {Ellis}, {Konidaris}  \& {Stark}}{{Schenker} et~al.}{2014}]{Schenker2014}
{Schenker} M.~A.,  {Ellis} R.~S.,  {Konidaris} N.~P.,   {Stark} D.~P.,  2014, \mn@doi [\apj] {10.1088/0004-637X/795/1/20}, \href {https://ui.adsabs.harvard.edu/abs/2014ApJ...795...20S} {795, 20}

\bibitem[\protect\citeauthoryear{{Shivaei} et~al.,}{{Shivaei} et~al.}{2020}]{Shivaei2020}
{Shivaei} I.,  et~al., 2020, \mn@doi [\apj] {10.3847/1538-4357/aba35e}, \href {https://ui.adsabs.harvard.edu/abs/2020ApJ...899..117S} {899, 117}

\bibitem[\protect\citeauthoryear{{Stanway}, {Eldridge}  \& {Becker}}{{Stanway} et~al.}{2016}]{Stanway2016}
{Stanway} E.~R.,  {Eldridge} J.~J.,   {Becker} G.~D.,  2016, \mn@doi [\mnras] {10.1093/mnras/stv2661}, \href {https://ui.adsabs.harvard.edu/abs/2016MNRAS.456..485S} {456, 485}

\bibitem[\protect\citeauthoryear{{Stark}}{{Stark}}{2016}]{Stark2016}
{Stark} D.~P.,  2016, \mn@doi [\araa] {10.1146/annurev-astro-081915-023417}, \href {https://ui.adsabs.harvard.edu/abs/2016ARA&A..54..761S} {54, 761}

\bibitem[\protect\citeauthoryear{{Stark}, {Ellis}, {Chiu}, {Ouchi}  \& {Bunker}}{{Stark} et~al.}{2010}]{Stark2010}
{Stark} D.~P.,  {Ellis} R.~S.,  {Chiu} K.,  {Ouchi} M.,   {Bunker} A.,  2010, \mn@doi [\mnras] {10.1111/j.1365-2966.2010.17227.x}, \href {https://ui.adsabs.harvard.edu/abs/2010MNRAS.408.1628S} {408, 1628}

\bibitem[\protect\citeauthoryear{{Stark}, {Ellis}  \& {Ouchi}}{{Stark} et~al.}{2011}]{Stark2011}
{Stark} D.~P.,  {Ellis} R.~S.,   {Ouchi} M.,  2011, \mn@doi [\apjl] {10.1088/2041-8205/728/1/L2}, \href {https://ui.adsabs.harvard.edu/abs/2011ApJ...728L...2S} {728, L2}

\bibitem[\protect\citeauthoryear{{Stark} et~al.,}{{Stark} et~al.}{2017}]{Stark2017}
{Stark} D.~P.,  et~al., 2017, \mn@doi [\mnras] {10.1093/mnras/stw2233}, \href {https://ui.adsabs.harvard.edu/abs/2017MNRAS.464..469S} {464, 469}

\bibitem[\protect\citeauthoryear{{Storey} \& {Hummer}}{{Storey} \& {Hummer}}{1995}]{Storey1995}
{Storey} P.~J.,  {Hummer} D.~G.,  1995, \mn@doi [\mnras] {10.1093/mnras/272.1.41}, \href {https://ui.adsabs.harvard.edu/abs/1995MNRAS.272...41S} {272, 41}

\bibitem[\protect\citeauthoryear{{Tacchella} et~al.,}{{Tacchella} et~al.}{2023}]{Tacchella2023}
{Tacchella} S.,  et~al., 2023, \mn@doi [\mnras] {10.1093/mnras/stad1408}, \href {https://ui.adsabs.harvard.edu/abs/2023MNRAS.522.6236T} {522, 6236}

\bibitem[\protect\citeauthoryear{{Tang}, {Stark}, {Chevallard}  \& {Charlot}}{{Tang} et~al.}{2019}]{Tang2019}
{Tang} M.,  {Stark} D.~P.,  {Chevallard} J.,   {Charlot} S.,  2019, \mn@doi [\mnras] {10.1093/mnras/stz2236}, \href {https://ui.adsabs.harvard.edu/abs/2019MNRAS.489.2572T} {489, 2572}

\bibitem[\protect\citeauthoryear{{Tang}, {Stark}, {Chevallard}, {Charlot}, {Endsley}  \& {Congiu}}{{Tang} et~al.}{2021}]{Tang2021}
{Tang} M.,  {Stark} D.~P.,  {Chevallard} J.,  {Charlot} S.,  {Endsley} R.,   {Congiu} E.,  2021, \mn@doi [\mnras] {10.1093/mnras/stab705}, \href {https://ui.adsabs.harvard.edu/abs/2021MNRAS.503.4105T} {503, 4105}

\bibitem[\protect\citeauthoryear{{Tang} et~al.,}{{Tang} et~al.}{2023}]{Tang2023}
{Tang} M.,  et~al., 2023, \mn@doi [\mnras] {10.1093/mnras/stad2763}, \href {https://ui.adsabs.harvard.edu/abs/2023MNRAS.526.1657T} {526, 1657}

\bibitem[\protect\citeauthoryear{{Tang} et~al.,}{{Tang} et~al.}{2024}]{Tang2024}
{Tang} M.,  et~al., 2024, \mn@doi [arXiv e-prints] {10.48550/arXiv.2402.06070}, \href {https://ui.adsabs.harvard.edu/abs/2024arXiv240206070T} {p. arXiv:2402.06070}

\bibitem[\protect\citeauthoryear{{Tasitsiomi}}{{Tasitsiomi}}{2006}]{Tasitsiomi2006}
{Tasitsiomi} A.,  2006, \mn@doi [\apj] {10.1086/504460}, \href {https://ui.adsabs.harvard.edu/abs/2006ApJ...645..792T} {645, 792}

\bibitem[\protect\citeauthoryear{{Tilvi} et~al.,}{{Tilvi} et~al.}{2014}]{Tilvi2014}
{Tilvi} V.,  et~al., 2014, \mn@doi [\apj] {10.1088/0004-637X/794/1/5}, \href {https://ui.adsabs.harvard.edu/abs/2014ApJ...794....5T} {794, 5}

\bibitem[\protect\citeauthoryear{{Tilvi} et~al.,}{{Tilvi} et~al.}{2020}]{Tilvi2020}
{Tilvi} V.,  et~al., 2020, \mn@doi [\apjl] {10.3847/2041-8213/ab75ec}, \href {https://ui.adsabs.harvard.edu/abs/2020ApJ...891L..10T} {891, L10}

\bibitem[\protect\citeauthoryear{{Topping} et~al.,}{{Topping} et~al.}{2022a}]{Topping2022_rebels}
{Topping} M.~W.,  et~al., 2022a, \mn@doi [\mnras] {10.1093/mnras/stac2291}, \href {https://ui.adsabs.harvard.edu/abs/2022MNRAS.516..975T} {516, 975}

\bibitem[\protect\citeauthoryear{{Topping}, {Stark}, {Endsley}, {Plat}, {Whitler}, {Chen}  \& {Charlot}}{{Topping} et~al.}{2022b}]{Topping2022_ceers}
{Topping} M.~W.,  {Stark} D.~P.,  {Endsley} R.,  {Plat} A.,  {Whitler} L.,  {Chen} Z.,   {Charlot} S.,  2022b, \mn@doi [\apj] {10.3847/1538-4357/aca522}, \href {https://ui.adsabs.harvard.edu/abs/2022ApJ...941..153T} {941, 153}

\bibitem[\protect\citeauthoryear{{Topping} et~al.,}{{Topping} et~al.}{2023}]{Topping2023}
{Topping} M.~W.,  et~al., 2023, \mn@doi [arXiv e-prints] {10.48550/arXiv.2307.08835}, \href {https://ui.adsabs.harvard.edu/abs/2023arXiv230708835T} {p. arXiv:2307.08835}

\bibitem[\protect\citeauthoryear{{Torralba-Torregrosa} et~al.,}{{Torralba-Torregrosa} et~al.}{2023}]{Torralba-Torregrosa2023}
{Torralba-Torregrosa} A.,  et~al., 2023, \mn@doi [arXiv e-prints] {10.48550/arXiv.2307.10215}, \href {https://ui.adsabs.harvard.edu/abs/2023arXiv230710215T} {p. arXiv:2307.10215}

\bibitem[\protect\citeauthoryear{{Trainor}, {Steidel}, {Strom}  \& {Rudie}}{{Trainor} et~al.}{2015}]{Trainor2015}
{Trainor} R.~F.,  {Steidel} C.~C.,  {Strom} A.~L.,   {Rudie} G.~C.,  2015, \mn@doi [\apj] {10.1088/0004-637X/809/1/89}, \href {https://ui.adsabs.harvard.edu/abs/2015ApJ...809...89T} {809, 89}

\bibitem[\protect\citeauthoryear{{Treu}, {Schmidt}, {Trenti}, {Bradley}  \& {Stiavelli}}{{Treu} et~al.}{2013}]{Treu2013}
{Treu} T.,  {Schmidt} K.~B.,  {Trenti} M.,  {Bradley} L.~D.,   {Stiavelli} M.,  2013, \mn@doi [\apjl] {10.1088/2041-8205/775/1/L29}, \href {https://ui.adsabs.harvard.edu/abs/2013ApJ...775L..29T} {775, L29}

\bibitem[\protect\citeauthoryear{{Treu} et~al.,}{{Treu} et~al.}{2022}]{Treu2022}
{Treu} T.,  et~al., 2022, \mn@doi [\apj] {10.3847/1538-4357/ac8158}, \href {https://ui.adsabs.harvard.edu/abs/2022ApJ...935..110T} {935, 110}

\bibitem[\protect\citeauthoryear{{Vanzella} et~al.,}{{Vanzella} et~al.}{2020}]{Vanzella2020}
{Vanzella} E.,  et~al., 2020, \mn@doi [\mnras] {10.1093/mnrasl/slaa041}, \href {https://ui.adsabs.harvard.edu/abs/2020MNRAS.494L..81V} {494, L81}

\bibitem[\protect\citeauthoryear{{Verhamme}, {Orlitov{\'a}}, {Schaerer}, {Izotov}, {Worseck}, {Thuan}  \& {Guseva}}{{Verhamme} et~al.}{2017}]{Verhamme2017}
{Verhamme} A.,  {Orlitov{\'a}} I.,  {Schaerer} D.,  {Izotov} Y.,  {Worseck} G.,  {Thuan} T.~X.,   {Guseva} N.,  2017, \mn@doi [\aap] {10.1051/0004-6361/201629264}, \href {https://ui.adsabs.harvard.edu/abs/2017A&A...597A..13V} {597, A13}

\bibitem[\protect\citeauthoryear{Virtanen et~al.,}{Virtanen et~al.}{2020}]{2020SciPy-NMeth}
Virtanen P.,  et~al., 2020, \mn@doi [Nature Methods] {10.1038/s41592-019-0686-2}, \href {https://rdcu.be/b08Wh} {17, 261}

\bibitem[\protect\citeauthoryear{{Wang} et~al.,}{{Wang} et~al.}{2020}]{Wang2020}
{Wang} F.,  et~al., 2020, \mn@doi [\apj] {10.3847/1538-4357/ab8c45}, \href {https://ui.adsabs.harvard.edu/abs/2020ApJ...896...23W} {896, 23}

\bibitem[\protect\citeauthoryear{{Wang} et~al.,}{{Wang} et~al.}{2023}]{Wang2023}
{Wang} B.,  et~al., 2023, \mn@doi [\apjl] {10.3847/2041-8213/acfe07}, \href {https://ui.adsabs.harvard.edu/abs/2023ApJ...957L..34W} {957, L34}

\bibitem[\protect\citeauthoryear{{Weinberger}, {Kulkarni}, {Haehnelt}, {Choudhury}  \& {Puchwein}}{{Weinberger} et~al.}{2018}]{Weinberger2018}
{Weinberger} L.~H.,  {Kulkarni} G.,  {Haehnelt} M.~G.,  {Choudhury} T.~R.,   {Puchwein} E.,  2018, \mn@doi [\mnras] {10.1093/mnras/sty1563}, \href {https://ui.adsabs.harvard.edu/abs/2018MNRAS.479.2564W} {479, 2564}

\bibitem[\protect\citeauthoryear{{Whitler}, {Mason}, {Ren}, {Dijkstra}, {Mesinger}, {Pentericci}, {Trenti}  \& {Treu}}{{Whitler} et~al.}{2020}]{Whitler2020}
{Whitler} L.~R.,  {Mason} C.~A.,  {Ren} K.,  {Dijkstra} M.,  {Mesinger} A.,  {Pentericci} L.,  {Trenti} M.,   {Treu} T.,  2020, \mn@doi [\mnras] {10.1093/mnras/staa1178}, \href {https://ui.adsabs.harvard.edu/abs/2020MNRAS.495.3602W} {495, 3602}

\bibitem[\protect\citeauthoryear{{Whitler}, {Stark}, {Endsley}, {Chen}, {Mason}, {Topping}  \& {Charlot}}{{Whitler} et~al.}{2023a}]{Whitler2023c}
{Whitler} L.,  {Stark} D.~P.,  {Endsley} R.,  {Chen} Z.,  {Mason} C.,  {Topping} M.~W.,   {Charlot} S.,  2023a, \mn@doi [arXiv e-prints] {10.48550/arXiv.2305.16670}, \href {https://ui.adsabs.harvard.edu/abs/2023arXiv230516670W} {p. arXiv:2305.16670}

\bibitem[\protect\citeauthoryear{{Whitler}, {Endsley}, {Stark}, {Topping}, {Chen}  \& {Charlot}}{{Whitler} et~al.}{2023b}]{Whitler2023_ceers}
{Whitler} L.,  {Endsley} R.,  {Stark} D.~P.,  {Topping} M.,  {Chen} Z.,   {Charlot} S.,  2023b, \mn@doi [\mnras] {10.1093/mnras/stac3535}, \href {https://ui.adsabs.harvard.edu/abs/2023MNRAS.519..157W} {519, 157}

\bibitem[\protect\citeauthoryear{{Whitler}, {Stark}, {Endsley}, {Leja}, {Charlot}  \& {Chevallard}}{{Whitler} et~al.}{2023c}]{Whitler2023_cosmos}
{Whitler} L.,  {Stark} D.~P.,  {Endsley} R.,  {Leja} J.,  {Charlot} S.,   {Chevallard} J.,  2023c, \mn@doi [\mnras] {10.1093/mnras/stad004}, \href {https://ui.adsabs.harvard.edu/abs/2023MNRAS.519.5859W} {519, 5859}

\bibitem[\protect\citeauthoryear{{Wisotzki} et~al.,}{{Wisotzki} et~al.}{2016}]{Wisotzki2016}
{Wisotzki} L.,  et~al., 2016, \mn@doi [\aap] {10.1051/0004-6361/201527384}, \href {https://ui.adsabs.harvard.edu/abs/2016A&A...587A..98W} {587, A98}

\bibitem[\protect\citeauthoryear{{Witstok} et~al.,}{{Witstok} et~al.}{2023}]{Witstok2023}
{Witstok} J.,  et~al., 2023, \mn@doi [arXiv e-prints] {10.48550/arXiv.2306.04627}, \href {https://ui.adsabs.harvard.edu/abs/2023arXiv230604627W} {p. arXiv:2306.04627}

\bibitem[\protect\citeauthoryear{{Witten} et~al.,}{{Witten} et~al.}{2023}]{Witten2023}
{Witten} C.,  et~al., 2023, \mn@doi [arXiv e-prints] {10.48550/arXiv.2303.16225}, \href {https://ui.adsabs.harvard.edu/abs/2023arXiv230316225W} {p. arXiv:2303.16225}

\bibitem[\protect\citeauthoryear{{Wuyts} et~al.,}{{Wuyts} et~al.}{2011}]{Wuyts2011}
{Wuyts} S.,  et~al., 2011, \mn@doi [\apj] {10.1088/0004-637X/738/1/106}, \href {https://ui.adsabs.harvard.edu/abs/2011ApJ...738..106W} {738, 106}

\bibitem[\protect\citeauthoryear{{Wyithe} \& {Loeb}}{{Wyithe} \& {Loeb}}{2005}]{Wyithe2005}
{Wyithe} J. S.~B.,  {Loeb} A.,  2005, \mn@doi [\apj] {10.1086/429529}, \href {https://ui.adsabs.harvard.edu/abs/2005ApJ...625....1W} {625, 1}

\bibitem[\protect\citeauthoryear{{Yamanaka} et~al.,}{{Yamanaka} et~al.}{2020}]{Yamanaka2020}
{Yamanaka} S.,  et~al., 2020, \mn@doi [\mnras] {10.1093/mnras/staa2507}, \href {https://ui.adsabs.harvard.edu/abs/2020MNRAS.498.3095Y} {498, 3095}

\bibitem[\protect\citeauthoryear{{Yang} et~al.,}{{Yang} et~al.}{2017}]{Yang2017}
{Yang} H.,  et~al., 2017, \mn@doi [\apj] {10.3847/1538-4357/aa7d4d}, \href {https://ui.adsabs.harvard.edu/abs/2017ApJ...844..171Y} {844, 171}

\bibitem[\protect\citeauthoryear{{Yang} et~al.,}{{Yang} et~al.}{2020a}]{Yang2020}
{Yang} J.,  et~al., 2020a, \mn@doi [\apjl] {10.3847/2041-8213/ab9c26}, \href {https://ui.adsabs.harvard.edu/abs/2020ApJ...897L..14Y} {897, L14}

\bibitem[\protect\citeauthoryear{{Yang} et~al.,}{{Yang} et~al.}{2020b}]{Yang2020b}
{Yang} J.,  et~al., 2020b, \mn@doi [\apj] {10.3847/1538-4357/abbc1b}, \href {https://ui.adsabs.harvard.edu/abs/2020ApJ...904...26Y} {904, 26}

\bibitem[\protect\citeauthoryear{{Zackrisson} et~al.,}{{Zackrisson} et~al.}{2017}]{Zackrisson2017}
{Zackrisson} E.,  et~al., 2017, \mn@doi [\apj] {10.3847/1538-4357/836/1/78}, \href {https://ui.adsabs.harvard.edu/abs/2017ApJ...836...78Z} {836, 78}

\bibitem[\protect\citeauthoryear{{Zheng} et~al.,}{{Zheng} et~al.}{2017}]{Zheng2017}
{Zheng} Z.-Y.,  et~al., 2017, \mn@doi [\apjl] {10.3847/2041-8213/aa794f}, \href {https://ui.adsabs.harvard.edu/abs/2017ApJ...842L..22Z} {842, L22}

\bibitem[\protect\citeauthoryear{{Zhu} et~al.,}{{Zhu} et~al.}{2022}]{Zhu2022}
{Zhu} Y.,  et~al., 2022, \mn@doi [\apj] {10.3847/1538-4357/ac6e60}, \href {https://ui.adsabs.harvard.edu/abs/2022ApJ...932...76Z} {932, 76}

\bibitem[\protect\citeauthoryear{{Zitrin} et~al.,}{{Zitrin} et~al.}{2015}]{Zitrin2015}
{Zitrin} A.,  et~al., 2015, \mn@doi [\apjl] {10.1088/2041-8205/810/1/L12}, \href {https://ui.adsabs.harvard.edu/abs/2015ApJ...810L..12Z} {810, L12}

\bibitem[\protect\citeauthoryear{{van der Walt} et~al.,}{{van der Walt} et~al.}{2014}]{vanderWalt2014}
{van der Walt} S.,  et~al., 2014, \mn@doi [PeerJ] {10.7717/peerj.453}, \href {https://ui.adsabs.harvard.edu/abs/2014PeerJ...2..453V} {2, e453}

\makeatother
\end{thebibliography}

% Alternatively you could enter them by hand, like this:
% This method is tedious and prone to error if you have lots of references
%\begin{thebibliography}{99}
%\bibitem[\protect\citeauthoryear{Author}{2012}]{Author2012}
%Author A.~N., 2013, Journal of Improbable Astronomy, 1, 1
%\bibitem[\protect\citeauthoryear{Others}{2013}]{Others2013}
%Others S., 2012, Journal of Interesting Stuff, 17, 198
%\end{thebibliography}

%%%%%%%%%%%%%%%%%%%%%%%%%%%%%%%%%%%%%%%%%%%%%%%%%%

%%%%%%%%%%%%%%%%% APPENDICES %%%%%%%%%%%%%%%%%%%%%

%%%%%%%%%%%%%%%%%%%%%%%%%%%%%%%%%%%%%%%%%%%%%%%%%%

% Don't change these lines
\bsp	% typesetting comment
\label{lastpage}
\end{document}